\documentclass[fleqn,usenatbib]{mnras}

\usepackage[T1]{fontenc}

\DeclareUnicodeCharacter{0301}{\hspace{-1ex}\'{ }}
\DeclareRobustCommand{\VAN}[3]{#2}
\let\VANthebibliography\thebibliography
\def\thebibliography{\DeclareRobustCommand{\VAN}[3]{##3}\VANthebibliography}

\usepackage{graphicx}	
\usepackage{amsmath}	
\usepackage{amssymb}	
\usepackage{multicol}        
\usepackage{bm}		
\usepackage{pdflscape}	
\usepackage{morefloats}
\usepackage{longtable}
\usepackage{geometry} 
\usepackage{booktabs,longtable}
\usepackage{float}
\usepackage{lscape} 
 
\newcommand{\hii}{\,H\,{\small II}}

\usepackage[T1]{fontenc}
\usepackage{ae,aecompl}

\title[Star formation in IRAS 15394$-$5358]{ATOMS: ALMA Three-millimeter Observations of Massive Star-forming
regions $-$ XVII. High-mass star-formation through a large-scale collapse in IRAS 15394$-$5358}

\author[Das et al.]{Swagat R. Das$^{1}$\thanks{Contact e-mail: \href{dasswagat77@gmail.com}{swagat@das.uchile.cl / dasswagat77@gmail.com}.}, Manuel Merello$^{1}$, Leonardo Bronfman$^{1}$, Tie Liu$^{2}$, Guido Garay$^{1,15}$, Amelia Stutz$^{3}$, 
\newauthor
Diego Mardones$^{1}$, Jian-Wen Zhou$^{4}$, Patricio Sanhueza$^{5,6}$, Hong-Li Liu$^{7}$, Enrique Vázquez-Semadeni$^{8}$, 
\newauthor
Gilberto C. Gómez$^{8}$, Aina Palau$^{8}$, Anandmayee Tej$^{9}$, Feng-Wei Xu$^{10,11}$, Tapas Baug$^{12}$, 
\newauthor
Lokesh K. Dewangan$^{13}$, Jinhua He$^{14,15,1}$, Lei Zhu$^{15}$, Shanghuo Li$^{16,17}$, Mika Juvela$^{18}$, Anindya Saha$^{9}$, 
\newauthor
Namitha Issac$^{2}$, Jihye Hwang$^{19}$, Hafiz Nazeer$^{9}$, and L. Viktor Toth$^{20}$
\\ \\
$^{1}$Departamento de Astronomı́a, Universidad de Chile, Las Condes, 7591245 Santiago, Chile \\
$^{2}$ Shanghai Astronomical Observatory, Chinese Academy of Sciences, 80 Nandan Road, Shanghai 200030, China \\
$^{3}$ Departamento de Astronomı́a, Universidad de Concepción, Av. Esteban Iturra s/n, Distrito Universitario, 160-C, Chile \\
$^{4}$ Max-Planck-Institut für Radioastronomie, Auf dem Hügel 69, 53121 Bonn, Germany \\
$^{5}$ National Astronomical Observatory of Japan, National Institutes of Natural Sciences, 2-21-1 Osawa, Mitaka, Tokyo 181-8588, Japan \\
$^{6}$ Astronomical Science Program, The Graduate University for Advanced Studies, SOKENDAI, 2-21-1 Osawa, Mitaka, Tokyo 181-8588, Japan \\
$^{7}$ School of Physics and Astronomy, Yunnan University, Kunming, 650091, People \\
$^{8}$ Instituto de Radioastronomía y Astrofísica, Universidad Nacional Autónoma de México, Apdo. Postal 3-72, Morelia, Michoacán, 58089, México \\
$^{9}$ Indian Institute of Space Science and Technology, Thiruvananthapuram 695 547, Kerala, India \\
$^{10}$ Kavli Institute for Astronomy and Astrophysics, Peking University, Beijing, 100871, People’s Republic of China \\
$^{11}$ Department of Astronomy, School of Physics, Peking University, Beijing, 100871, People’s Republic of China \\
$^{12}$ S. N. Bose National Centre for Basic Sciences, Block-JD, Sector-III, Salt Lake City, Kolkata 700106, India \\
$^{13}$ Physical Research Laboratory, Navrangpura, Ahmedabad-380 009, India \\
$^{14}$ Yunnan Observatories, Chinese Academy of Sciences, 396 Yang-fangwang, Guandu District, Kunming, 650216, P. R. China \\
$^{15}$ Chinese Academy of Sciences South America Center for Astronomy, National Astronomical Observatories, CAS, Beijing 100101,
China \\
$^{16}$ Max Planck Institute of Astronoy, Königstuhl 17, Heidelberg 69117, Germany \\
$^{17}$ SOFIA Science Centre, USRA, NASA Ames Research Centre, MS-12, N232, Moffett Field, CA 94035, USA \\
$^{18}$ Department of Physics, University of Helsinki, PO Box 64, FI-00014 Helsinki, Finland \\
$^{19}$ Korea Astronomy and Space Science Institute, 776 Daedeokdaero, Yuseong-gu, Daejeon 34055, Republic of Korea \\
$^{20}$ Institute of Physics and Astronomy, Eotvos Lorand University, Pazmany Peter setany 1/A, H-1117 Budapest, Hungary \\
}

\date{Accepted XXX. Received YYY; in original form ZZZ}

\pubyear{2015}

\begin{document}
\label{firstpage}
\pagerange{\pageref{firstpage}--\pageref{lastpage}}
\maketitle

\begin{abstract}
Hub-filament systems are considered as natural sites for high-mass star formation. Kinematic analysis of the surroundings of hub-filaments is essential to better understand high-mass star formation within such systems. 
In this work, we present a detailed study of the massive Galactic protocluster IRAS 15394$-$5358, using continuum and molecular line data from the ALMA Three-millimeter Observations of Massive Star-forming Regions (ATOMS) survey. The 3~mm dust continuum map reveals the fragmentation of the massive ($\rm M=843~M_{\odot}$) clump into six cores. The core C-1A is the largest (radius = 0.04~pc), the most massive ($\rm M=157~M_{\odot}$), and lies within the dense central region, along with two smaller cores ($\rm M=7~and~3~M_{\odot}$). The fragmentation process is consistent with the thermal Jeans fragmentation mechanism and virial analysis shows that all the cores have small virial parameter values ($\rm \alpha_{vir}<<2$), suggesting that the cores are gravitationally bound. The mass vs. radius relation indicates that three cores can potentially form at least a single massive star. The integrated intensity map of $\rm H^{13}CO^{+}$ shows that the massive clump is associated with a hub-filament system, where the central hub is linked with four filaments. A sharp velocity gradient is observed towards the hub, suggesting a global collapse where the filaments are actively feeding the hub. We discuss the role of global collapse and the possible driving mechanisms for the massive star formation activity in the protocluster.

\end{abstract}



\section{Introduction}\label{intro}
High-mass ($\rm >8~M_{\odot}$) stars are the powerhouses of galaxies. They play a major role in the formation and evolution of galaxies through their feedback. These high-mass stars also play a central role in synthesizing the chemistry of galaxies. However, the formation mechanism of high-mass stars is still a subject of debate in astrophysics. Formation within the densest regions of molecular clouds, in a complex clustered environment, and the large distances of the massive star-forming regions pose a challenge to studying them observationally.

The ``turbulent core'' \citep{2003ApJ...585..850M}, and ``competitive accretion'' \citep{2001MNRAS.323..785B,2004MNRAS.349..735B} theoretical models are frequently considered for explaining the massive star formation process. The ``turbulent core'' model is a scaled-up version or extension of the low-mass star formation theory \citep{1987ARA&A..25...23S}, proposing a monolithic collapse of high-mass prestellar cores. According to this model, the massive prestellar cores are supported against collapse or fragmentation by a significant degree of turbulence and strong magnetic fields. In the ``competitive accretion'' model, the massive protostars start with initial Jeans fragmentation. The growth in the mass of the objects happens through Bondi-Hoyle accretion \citep{2023arXiv230613846V}. In this model, the massive objects need to be near the centre of the gravitational potential well of the molecular cloud so that it favours more accretion with respect to its neighbouring protostellar objects (e.g., \citealt{2018ApJ...861...14C}). 
Apart from these two models, a few more scenarios of massive star formation are also proposed based on observations. For instance, the global hierarchical collapse (GHC) scenario \citep{2009ApJ...707.1023V,2017MNRAS.467.1313V,2019MNRAS.490.3061V}, where gravitationally driven fragmentation occurs in molecular clouds. This model considers the collapse to be hierarchical, suggesting the presence of small-scale within large-scale collapses, where such large-scale accretion is expected to feed the massive star-forming regions. \citet{2018ARA&A..56...41M} proposed an evolutionary scenario suggesting the high-mass analogues of prestellar cores can be replaced by large-scale ($\rm \sim 0.1 - 1~pc$) gas reservoirs, called starless massive dense cores (MDCs) or starless clumps. During their protostellar phase, these mass reservoirs would concentrate their mass into high-mass cores at the same time that stellar embryos would accrete, skipping the high-mass prestellar core phase. \citet{2020ApJ...900...82P} propose the inertial-inflow (IF) model, a scenario where massive stars are assembled by large-scale, converging, inertial ﬂows that naturally occur in supersonic turbulence. According to this model, the mass distribution of stars is primarily the result of turbulent fragmentation under the hypothesis that the statistical properties of turbulence determine the formation timescale and mass of prestellar cores, posing deﬁnite constraints on the formation mechanism of massive stars.

Observational studies have shown that a global collapse in molecular clouds towards the potential well happens when mass from the surrounding gets transferred through filaments (e.g., \citealt{2013A&A...555A.112P}). The connection of more than three or four filaments with the potential well is usually called a ``hub-filament'' system, where the centre of the potential well is the hub. In such a system, the cores within the hubs grow in mass and size by accreting matter from the surrounding environment. The accretion time can be longer while the matter gets funnelled to the hubs. Several studies have revealed that the hub-filament systems are ideal for the formation of high-mass stars and clusters \citep{2013A&A...555A.112P,2014MNRAS.440.2860H,2015ApJ...804..141Z,2016ApJ...824...31L,2018ApJ...852...12Y,2018ApJ...855....9L,2018ApJ...852..119B,2019MNRAS.487.1259L,2019MNRAS.485.1775I,2020ApJ...903...13D,2020A&A...642A..87K,2022ApJ...941...51H,2023MNRAS.522.3719L,2023ApJ...953...40Y,2023ApJ...957...61H}. Simulations have shown velocity gradients along filaments, resulting from the inflow of matter into the hubs \citep{2010ApJ...709...27W,2014ApJ...791..124G,2016MNRAS.455.3640S,2020ApJ...900...82P,2022ApJ...941...51H}. A recent study utilising ATOMS (ALMA Three-millimetre Observations of Massive Star-forming regions; \citealt{2020MNRAS.496.2790L}) survey data has identified filamentary structures in the ATOMS targets \citep{2022MNRAS.514.6038Z}, and reported ``hub-filament'' systems in many of these targets while conducting a statistical analysis of the accretion flow along the filaments. Another recent work by \citet{2023MNRAS.520.3259X} did a detailed study of core feeding from a steady accretion flow from global collapse, that continues down to 1000~au scales \citep{2021ApJ...909..199O}, in one of the well-known hub-filament systems seen in the protocluster region SDC 335.

Many observational analyses have been carried out to study the accretion flow along the filaments at different scales and mass regimes  \citep{2012ApJ...745...61L,2013ApJ...766..115K,2013A&A...555A.112P,2016A&A...590A...2S,2018ApJ...855....9L,2019MNRAS.487.1259L,2019ApJ...875...24C,2019ApJ...877..114C,2019A&A...629A..81T,2021ApJ...919....3C,2021ApJ...915L..10S,2022ApJ...936..169R,2022ApJ...926..165L,2023ApJ...959L..31O,2023ApJ...953...40Y,2024ApJ...960...76P}. However, the statistics of such analyses is still minimal, and it requires much more attention. High-mass star formation through such a mechanism would be better understood if we probe the kinematics of the surrounding environment of the hub-filament systems. To complement earlier studies of massive star formation in hub-filament systems, in this work, we conduct a detailed kinematic analysis of the protocluster IRAS 15394$-$5358. This massive protocluster hosts a hub-filament system \citep{2022MNRAS.514.6038Z}. Our objective is to explore the massive star formation within the hub-filament system of the protocluster by conducting a comprehensive kinematic analysis. Through detailed examination of the kinematics, including velocity dispersion, line profiles, and spatial distributions, we aim to elucidate the mechanisms driving the formation and evolution of high-mass stars within the complex environment of the hub-filament system. This analysis will contribute to a deeper understanding of the processes involved in massive star formation and the role of hub-filament systems in shaping the stellar population of protoclusters. In the following, we provide a brief detail of the massive protocluster.

\subsection{Details of protocluster IRAS 15394$-$5358}
The object IRAS 15394$-$5358 ($l=326.47$ and $b=0.70$) has been observed in CS \citep{1996A&AS..115...81B} and $\rm ^{13}CO$ \citep{2007A&A...474..891U} line emission. Association with $\rm H_2O$ \citep{1989A&A...221..105S} and 6~GHz $\rm CH_3OH$ \citep{1998MNRAS.297..215C} masers are also observed towards the target. The detection of masers confirms the massive star formation activity in this object.
As part of a survey, \citet{2004A&A...426...97F} detected the 1.2~mm continuum emission from the protocluster. These authors derived several physical parameters for the object using the kinematic distance of 2.8~kpc. 
With 1.2~mm dust continuum observation from SEST-SIMBA, \citet{2007ApJ...666..309G} detected a massive dense clump (G326.474+0.697) with mass of $\rm 1.5\times10^3~M_{\odot}$ and radius of 0.22~pc, adopting distance of 2.8~kpc. Despite being luminous ($\rm 5.4\times10^3~L_{\odot}$; \citealt{2018MNRAS.473.1059U}), no radio emission was detected towards the object \citep{2006ApJ...651..914G}. Probably, the ionizing source(s) is in the very early stage of evolution, deeply embedded and not revealed yet.   
The association of extended green objects (EGOs) with IRAS 15394$-$5358 also indicates an early stage of evolution \citep{2008AJ....136.2391C}. 
\citet{2015A&A...573A..82C} present the NIR imaging ($\rm H_2$ and $K_s$) and spectroscopic ($\rm 0.95 - 2.50~ \mu m$) analysis of IRAS 15394$-$5358 and detect the association of a small cluster of young stellar objects (YSOs). These authors reported the association of multiple outflows from continuum subtracted $\rm H_2$ images. From ATLASGAL 870~$\rm \mu m$ emission, \citet{2018MNRAS.473.1059U} detected a single dust clump towards the IRAS object and derived its physical properties. They reported the clump radius to be of 0.5~pc, dust temperature to be 20.7~K, and $\rm V_{LSR}$ of $\rm-41.6~km~ s^{-1}$. The dust temperature was derived by conducting a greybody fit to the sub-millimetre (sub-mm) dust emission, including from {\it Herschel} 70~$\rm \mu m$ to ATLASGAL 870~$\rm \mu m$. Adopting the distance of 1.8~kpc, they obtained $\rm L_{bol}=5.4\times10^3~L_{\odot}$ and $\rm M=843.3~M_{\odot}$. The higher mass obtained by \citet{2007ApJ...666..309G} could be primarily attributed to the higher distance i.e., 2.8~kpc they adopted. 
In a later study, \citet{2013ApJS..208...11L} provided this object's parallax distance of 1.8~kpc. In this work, we adopt the later and updated distance estimation of 1.8~kpc for this source.

\begin{figure*}
\hspace{-1.5cm}
\includegraphics[scale=0.46]{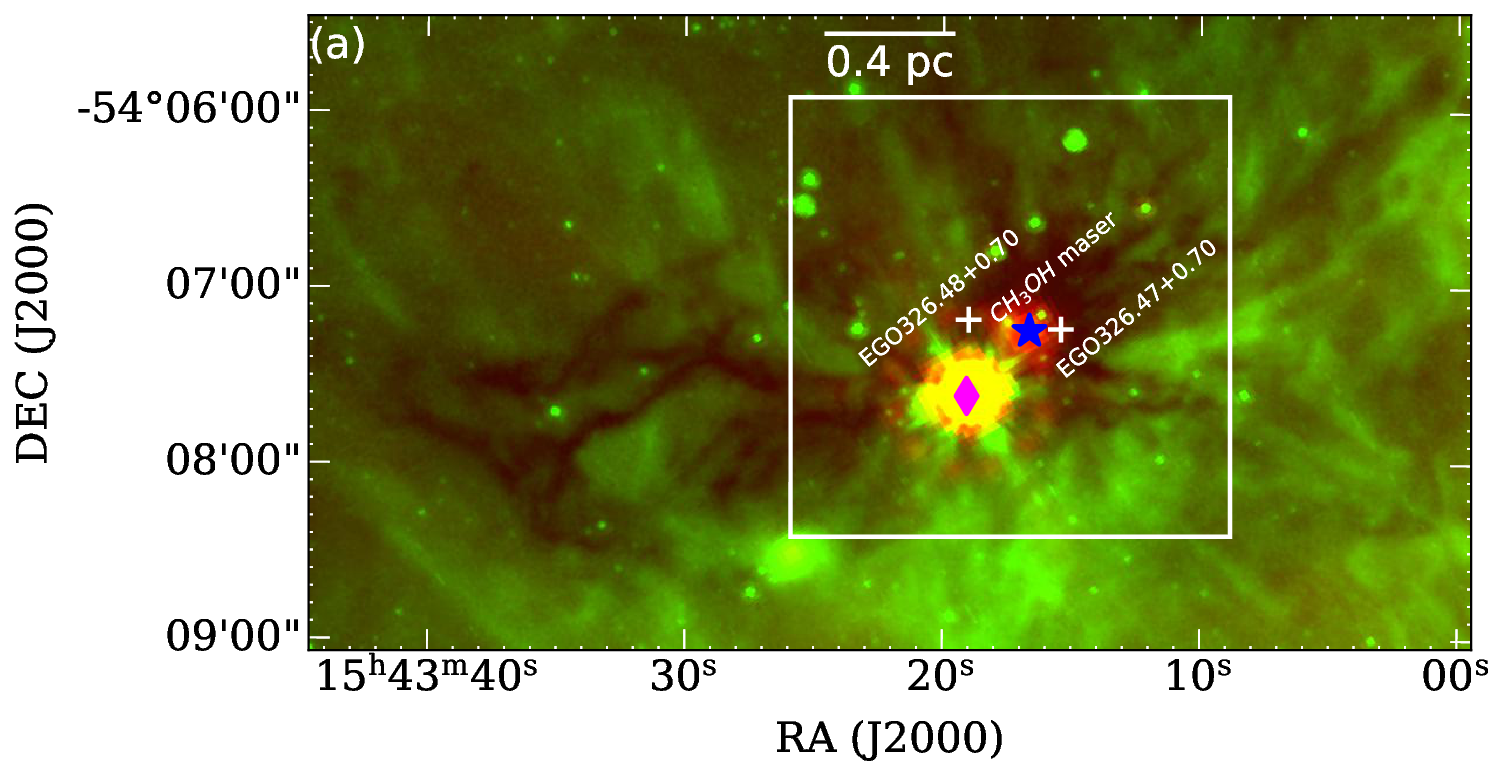}
\includegraphics[scale=0.3]{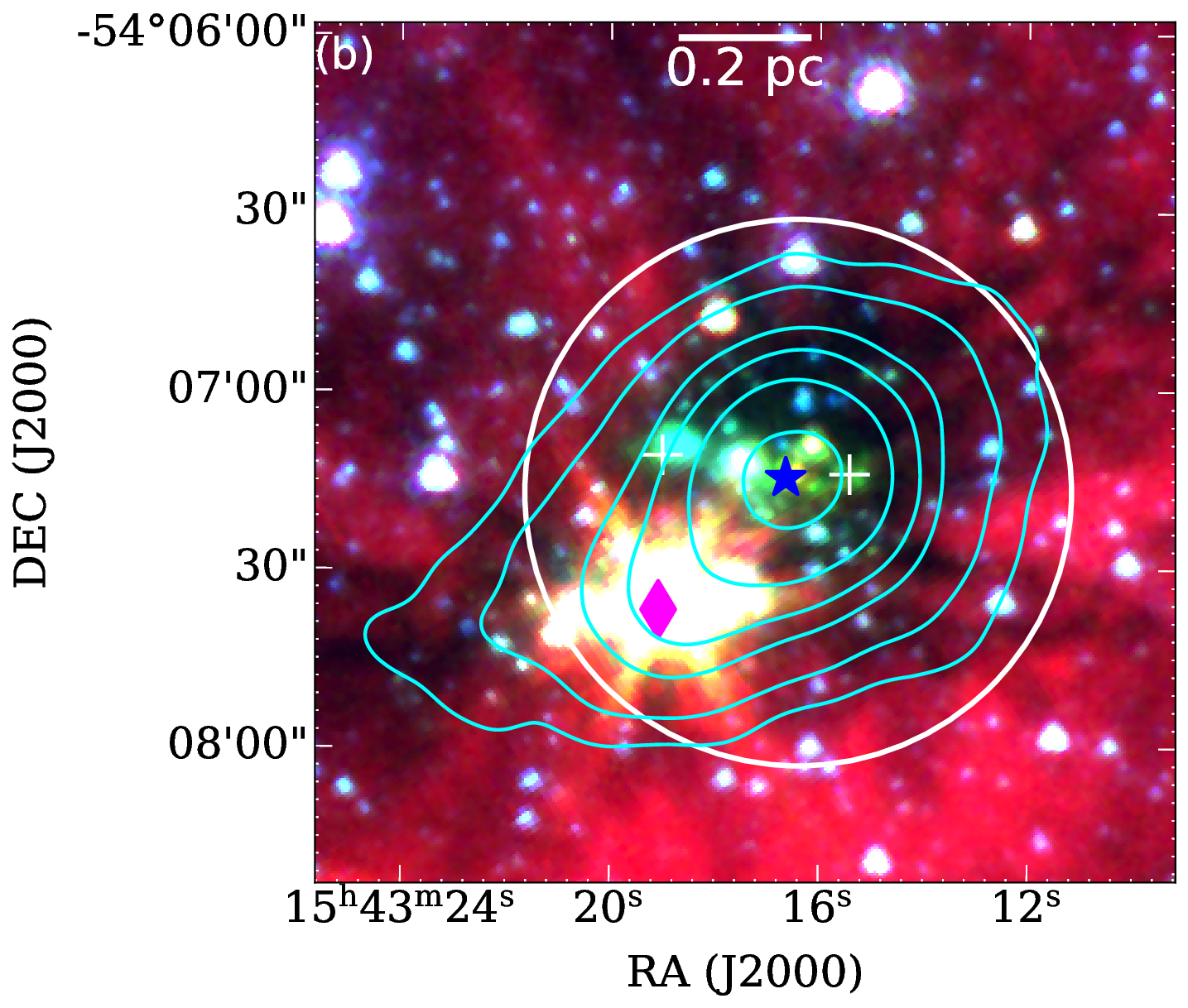}
\caption{(a) Two color composite image (24~$\rm \mu m$ - red and 8~$\rm \mu m$ - green) of the region towards IRAS 15394$-$5358. This image shows a larger field of view of the surrounding area of the star-forming region, which lie on the large IRDC. The white box covers a close region around IRAS 15394$-$5358. (b) Three color-composite image (8~$\rm \mu m$ - red, 4.5~$\rm \mu m$ - green, and 3.6~$\rm \mu m$ - blue) of the white box region shown in (a). The white circle (radius = $0.7^{\prime}$) is the ALMA field of view. The cyan contours are emissions at 870~$\rm \mu m$ from the ATLASGAL survey. The contour levels are 0.6, 1, 2, 3, 5, and 10~Jy/beam. Different objects towards the region found in the literature are shown in different symbols. The white `+' marks are for the EGOs \citep{2008AJ....136.2391C}, the blue star shows the location of 6~GHz $\rm CH_3OH$ maser \citep{1998MNRAS.297..215C}, and the magenta diamond for the position of the 70~$\rm \mu m$ point source. Scale bars of both images are shown on top of each frame.}
\label{mips_irac_map}
\end{figure*}

In Fig. \ref{mips_irac_map}(a), we show the two color-composite image (24~$\rm \mu m$ - red and 8~$\rm \mu m$ - green) of the region towards IRAS 15394$-$5358. This displays a large view of the region around the target. The 8~$\rm \mu m$ reveals a large dark filamentary structure, where the protocluster is located. Two bright blobs appear in 24~$\rm \mu m$ emission. The brighter 24~$\rm \mu m$ blob is associated with the 70~$\rm \mu m$ point source, suggesting it is more evolved. The 6~GHz $\rm CH_3OH$ maser is linked to the less bright 24~$\rm \mu m$ blob and to two EGOs in its close vicinity. This 24~$\rm \mu m$ emitting source is less evolved but associated with massive star formation. Fig. \ref{mips_irac_map}(b) displays the zoom-in view of the white box area of (a). In this close-up view of 8~$\rm \mu m$ emission, we can see the extended structures, which appear dark, seen to be extending from the massive dense core. We show the cold dust emission traced by ATLASGAL 870~$\rm \mu m$ in cyan contours. This cold dust emission reveals a single clump. The $\rm CH_3OH$ maser is located within the clump's peak. The white circle shown in this figure is the ALMA field of view, and the present work is focused on that region.

In this work, we use data from the ATOMS survey and conduct a detailed analysis of the fragmentation and gas kinematics, and probe the high-mass star formation through a large-scale collapse in the massive Galactic clump IRAS 15394$-$5358.
This work is arranged as follows. We describe the details of data sets in Section \ref{data}. This includes the details of ALMA observations and archival data sets. We present results in Section \ref{res}. These includes the results obtained from ALMA 3~mm dust continuum emission and the molecular line emissions. The details of the star formation activity of the protocluster region are provided in Section \ref{star_form_reg}. The nature of mass inflow through the filaments and the massive star formation through a large-scale collapse is discussed in Section \ref{discuss}. We summarize the results in Section \ref{summ}.

\section{Observations and archival data} \label{data}
\subsection{ALMA Observations} \label{alma_obs}
We use the ALMA data from the ATOMS survey (Project ID:
2019.1.00685.S; PI: Tie Liu; \citealt{2020MNRAS.496.2790L}). As part of the ATOMS survey, 146 massive IRAS clumps were observed in the single-pointing mode with both the Atacama Compact 7~m and the 12~m array configurations in band 3. This survey aims to systematically investigate the spatial distribution of various dense gas tracers (e.g., $\rm HCO^{+}$, HCN, and their
isotopologues), high-mass core (HMC) tracers (e.g. $\rm CH_{3}OH$ and $\rm HC_{3}N$), shock tracers (e.g.
SiO and SO), and ionized gas tracers ($\rm H_{40\alpha}$) in a large sample
of Galactic massive clumps. The 146 sources in the ATOMS sample were selected from the CS $\rm J=2-1$ survey of \citet{1996A&AS..115...81B}. In this work, we use data from the 12~m plus 7~m array datasets from this survey. Complete details regarding source selection, observations, and data reduction of the ATOMS survey can be found in \citet{2020MNRAS.496.2790L}. 
The synthesised beam size and $rms$ noise level for the continuum image are $\rm 2.34\arcsec \times 2.1\arcsec$ and $\rm 0.4~mJy~beam^{-1}$, respectively. The synthesised beam sizes for the molecular line datasets range from $\rm 2.4\arcsec \times 2.1\arcsec$ to $\rm 2.7\arcsec \times 2.4\arcsec$ and the $rms$ noise levels range from $\rm 0.12~mJy~beam^{-1}$ to $\rm 9.5~mJy~beam^{-1}$. The $\rm H^{13}CO^+~(1-0)$ line, which is used to study the kinematics of the protocluster has beam size and $rms$ noise level of $\rm 2.7\arcsec \times 2.4\arcsec$ and $\rm 0.12~mJy~beam^{-1}$, respectively. The dataset has a maximum recoverable scale of $\rm 81.8\arcsec$.

\subsection{Archival data}\label{arch_data}
We also use mid-infrared (MIR) to sub-mm archival data sets to study the interaction of the clump with its surrounding environment. We use the MIR images of the GLIMPSE survey \citep{2003PASP..115..953B}, obtained using the $\it Spitzer$ Infrared Array Camera (IRAC) with wavelengths of 3.6, 4.5, 5.8, and 8.0~$\rm \mu m$, with a resolution smaller than $2^{\prime\prime}$. These images trace emission from point sources at shorter wavelengths and warm dust emission at longer wavelengths. The 870~$\rm \mu m$ sub-mm image from the ATLASGAL survey \citep{2009A&A...504..415S} obtained with the Large APEX Bolometer Camera (LABOCA) is also used in this work. The image has a resolution of $19^{\prime\prime}$ and traces emission from cold dust. We also use the target region's column density and dust temperature maps, generated with the PPMAP procedure \citep{2017MNRAS.471.2730M} using the far-infrared (FIR) data obtained from the {\it Herschel} Infrared Galactic plane survey \citep{2010PASP..122..314M, 2016A&A...591A.149M}. The resulting column density and dust temperature maps have $\sim 12^{\prime\prime}$ resolutions and are available in the online archive\footnote{http://www.astro.cardiff.ac.uk/research/ViaLactea/}.

\section{Results} \label{res}

\subsection{Dust continuum emission}\label{dust_cont}
\subsubsection{Spatial distribution}\label{3mm_spatial}
\begin{figure}
\centering
\includegraphics[scale=0.4]{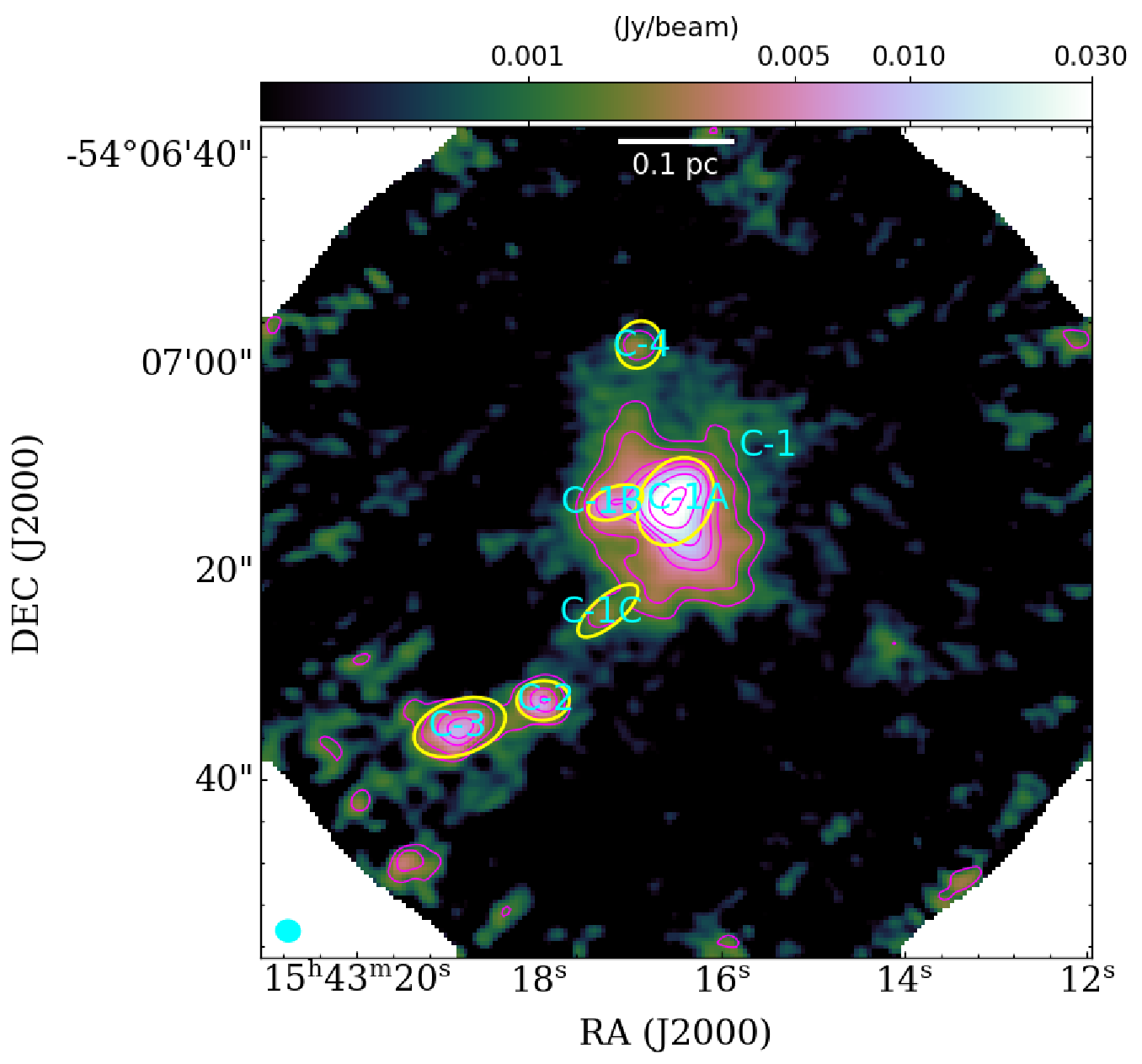}
\caption{The 3~mm continuum map of the protocluster observed as part of the ATOMS survey is shown in colour. The magenta contours are from the same map with contour levels of 3, 5, 10, 15, 25, and 50 times the noise $\rm \sigma$, where $\rm \sigma=0.0004~Jy/beam$. The yellow ellipses, along with labels, are the identified cores. A scale bar of 0.1~pc is shown on the top, and the beam size is in the bottom left corner. } 
\label{3mmcont_map}
\end{figure}

The emission from the cold dust at 870~$\rm \mu m$ appears as a single clump. However, the ALMA image at a higher angular resolution of $\sim2.5^{\prime\prime}$ can decipher and unfurl the clump's underlying core structures. In Fig. \ref{3mmcont_map}, we show the 3~mm continuum map with its contours overlaid. Above the $\rm 3\sigma$ level, the emission shows a central core with a diffuse, filamentary extension to the southeast harbouring two compact cores. With proper contrast, the central core also reveals the presence of substructures. Another separate core is also visible above the central core. These subsequent cores within the clump indicate the ongoing fragmentation within the massive clump. We match the morphology of the 3~mm continuum emission with the {\it Spitzer} image shown in Fig. \ref{mips_irac_map}. The central dense core of the 3~mm continuum map is associated with the less bright $\rm 24~\mu m$ blob along with the EGOs \citep{2008AJ....136.2391C}, and the 6~GHz $\rm CH_3OH$ maser \citep{1998MNRAS.297..215C}. This indicates that the region surrounding the central core is associated with high-mass star formation at an early stage of evolution. The other two brighter cores in the southeast direction are associated with the brighter $\rm 24~\mu m$ region. Thanks to ALMA's sensitivity and resolution, we can retrieve the cores within it, that appear as a single object at the MIR wavelengths. The massive core is associated with a 70~$\rm \mu m$ point source. Thus, the two cores are more evolved compared to the central core. We use this 3~mm dust continuum map to identify and derive the physical properties of the cores.

\subsubsection{Core identification}\label{3mm_core_ident}
The ATOMS 3~mm continuum map reveals cores and thus allows their identification. We follow a similar approach of core identification discussed in \citet{2020ApJ...896..110L} and applied by other works \citep{2021MNRAS.505.2801L,2022MNRAS.516.1983S}. We use {$\it astrodendro$} algorithm \citep{2008ApJ...679.1338R}, and the CASA-{$\it imfit$} task to identify the cores and derive their properties. The {$\it astrodendro$} algorithm decomposes the map into tree structures such as trunks, branches, and leaves. Leaves are the smallest structures, which are called cores. The {$\it astrodendro$} algorithm requires a threshold flux level, a minimum delta value (contour separation), and the number of pixels to decompose the map into different structures. We use the threshold flux as $\rm 1~mJy~beam^{-1}$ ($\rm 3\sigma$) and the minimum delta as $\rm \sigma$. To ensure proper detection of cores, we choose the number of pixels equivalent to the synthesized beam area in pixel units. These parameters are adopted after multiple trials for the robust detection of cores. As discussed in \cite{2021MNRAS.505.2801L}, we first use the {$\it astrodendro$} to identify the cores and get their parameters such as core position, major and minor axes sizes, peak and integrated flux, and position angle. In the next step, we use the parameters obtained from {$\it astrodendro$} as an initial guess to obtain a more accurate estimation using the CASA $\it imfit$. Fig. \ref{3mmcont_res_map} displays the residual map obtained from the task. Following the said procedure, we identified six cores towards IRAS 15394$-$5358 (see Fig. \ref{3mmcont_map}). Three cores are detected in the central part (C-1A, C-1B, C-1C), two (C-2, C-3) towards the southeast of the central core, and the other one (C-4) towards the north of the central core. We called the central region encompassed within the $\rm3\sigma$ contour level as core C-1. Cores C-1A, C-1B, and C-1C are the substructures of core C-1. A few dense clumpy structures are seen towards the edges of the map. However, we did not consider them because of the higher uncertainty at the edges of the map. The derived core properties are listed in Table \ref{core_prop}. In the residual map (Fig. \ref{3mmcont_res_map}), a bright region emerges to the south of the central core C-1A. Most likely, this is a low-mass core that is not visible in the original map due to the dominance of C-1A but becomes apparent in the residual map.

\subsubsection{Physical properties of cores}\label{phy_prop_core}
We derive several physical properties (such as mass, radius, and surface density) of the cores. We estimate the mass of each core using the equation B1 of \cite{2021MNRAS.505.2801L},

\begin{equation}
\rm M_{core} = \frac{F^{int}_{\nu}~R_{gd}~D^2}{B_{\nu}(T_d)~\kappa_{\nu}},
\label{core_mass}
\end{equation}

\noindent where, $\rm F^{int}_{\nu}$ is the integrated flux density obtained for each core from CASA-{\it imfit}, $\rm R_{gd}$ is the gas-to-dust ratio assumed to be 100, D is the distance to the target source, $\rm B_{\nu}$ is the Planck's function evaluated at dust temperature $\rm T_d$, and $\rm \kappa_{\nu}$ is the opacity assumed to be $\rm 0.18~cm^2~g^{-1}$ \citep{1994A&A...291..943O}. \citet{2018MNRAS.473.1059U} have mentioned the value of $\rm T_d = 20.7~K$ for the whole ATLASGAL clump. For a better estimate of each core's $\rm T_d$ value, we use the dust temperature map generated from {\it Herschel} images \citep{2017MNRAS.471.2730M}. All cores except core (C-4; $\rm T_d = 20~K$) have similar $\rm T_d$ values ($\rm \sim 23-25~K$). It is to be noted that the ALMA ATOMS image and the dust temperature map from {\it Herschel} maps have different resolutions and different pixel sizes. So, an exact comparison between both maps can not be incorporated.

The surface density ($\rm \Sigma_{core}$) is derived using the expression $\rm \Sigma_{core} = M_{core}/\pi R^2_{core}$. We present values of all the physical parameters in Table \ref{core_prop}. C-1A is the most massive and the largest among all the cores. This core holds a mass of $\rm \sim 270~M_{\odot}$ within a radius of 0.04~pc. We also derive the physical properties of core C-1. Since for C-1, we do not have information of deconvolved size from $\rm CASA-{\it imfit}$, we derive its physical radius assuming the core to be circular and using the expression $\rm r = (A/\pi)^{0.5}$, where A is the area of the core. The C-1 core's derived mass, radius, and surface density are 372.5$\pm$212~$\rm M_{\odot}$, 0.08~pc, and 3.9$\pm$2.2~$\rm g~cm^{-2}$, respectively. The mass of core C-1 is similar to the collective mass of cores C-1A, C-1B, and C-1C.

In a protocluster complex, the temperatures estimated for the cores from the dust temperature map using {\it Herschel} images might be an underestimation. To obtain more realistic temperature information, \cite{2021MNRAS.508.2964A} derived temperatures using the flux density obtained from the Hi-GAL 70~$\rm \mu m$ Compact Source Catalogue for cores associated with a 70~$\rm \mu m$ point source. This is because the 70~$\rm \mu m$ flux density is considered as a good tracer of the luminosity of embedded sources \citep{2008ApJS..179..249D,2012A&A...547A..49R}. In this estimation, the $\rm 70~\mu m$ flux density is converted into bolometric luminosities \citep{2017MNRAS.471..100E}, and the dust temperature is then derived under the assumption that dust emission from a protostellar core is optically thin and predominantly in the FIR.

Here, we follow the same approach, deriving the temperatures of the central core C-1 and core C-3, as they are associated with 70~$\rm \mu m$ point sources. The derived dust temperatures are 39.1$\pm$0.02 K and 46.8$\pm$0.1 K for C-1 and C-3, respectively. As C-1 contains three substructures, the same temperature is considered for the substructures. These derived temperatures of the cores are higher than the temperatures estimated from the dust temperature maps.

With the new temperatures, we derived the physical parameters of these cores, which are listed in Table \ref{core_prop} within parentheses, are used for the remainder of the analysis. The derived mass and surface density of C-1 are 214$\pm$121.6~$\rm M_{\odot}$ and 2.2$\pm$1.3~$\rm g~cm^{-2}$, respectively. The cores C-1 and C-3 in our work correspond to the cores MM1 and MM2 in the work of \cite{2021MNRAS.508.2964A}, whose derived masses are approximately twice as high as ours. In their study, they used a similar image at $\sim$3~mm. The other cores in our work are also relatively less massive compared to those derived by \cite{2021MNRAS.508.2964A}. The primary reasons for this difference are the varying distance values, opacity, and dust temperatures used in the respective studies. In our work, we adopt a distance of 1.82~kpc, whereas \cite{2021MNRAS.508.2964A} used a higher distance of 2.61~kpc. Nevertheless, given the high uncertainties associated with these derived physical parameters, the mass estimates in both the studies agree within their error limits.

\subsubsection{Uncertainties in the physical parameters} \label{uncer}
The parameters of distance and dust temperature can introduce uncertainties in the determination of the physical properties of the cores. In the case of our protocluster, we have a parallax distance estimation of 1.8~kpc \citep{2013ApJS..208...11L}. As there are no reported errors on the distance, we have not factored them into our calculations. The estimation of dust temperature is also subject to variability, influenced by factors such as the characteristics of dense cores and whether they host any hot cores and ultra-compact \hii\ (UC \hii\ ) regions. We have not detected any $\rm H_{40\alpha}$ emission towards the protocluster, and previous ATOMS studies have not reported any association of hot cores and UC \hii\ regions with the protocluster \citep{2021MNRAS.505.2801L,2023MNRAS.520.3245Z}. Consequently, the currently adopted dust temperature values for the cores are presumed to accurately represent their true nature. However, it's worth noting that considering higher temperatures, such as 100 K, would result in a reduction in core masses by a factor of four to five times.

Additionally, uncertainties in $\rm \kappa_{\nu}$ and $\rm R_{gd}$ significantly affect the uncertainty of mass and surface density estimates for the cores. In our current calculations, we have adopted a value of $\rm \kappa_{\nu}$ as $\rm 0.18~cm^2~g^{-1}$. This opacity value, as described in \citet{2021MNRAS.505.2801L}, is assumed at a frequency of approximately $\rm 94~GHz$, derived from the `OH5' dust model. The `OH5' model is a composite of dust properties from \citet{1994A&A...291..943O} and \citet{1994ApJ...421..615P}, extended towards longer wavelengths by \citet{2005ApJ...627..293Y}. While models from \citet{1994A&A...291..943O} have been predominantly favored in multi-wavelength studies of star-forming regions (e.g., \citealt{2011ApJ...728..143S}), it's important to note that $\rm \kappa_{\nu}$ remains one of the least precisely known parameters due to challenges in characterizing dust properties.

As discussed in other studies (e.g., \citealt{2017ApJ...841...97S,2019ApJ...886..102S}), we have also explored the literature for the potential range of $\rm \kappa_{\nu}$. Values for $\rm \kappa_{\nu}$ at ALMA band 3 range from $\rm 0.1~cm^2~g^{-1}$, derived using formulations from \citet{2015MNRAS.454.4282M,2017MNRAS.471.2730M}, to $\rm 0.5~cm^2~g^{-1}$ from the study by \citet{2021MNRAS.501.1316L}, assuming a standard gas-to-dust ratio of 100. As outlined in \citet{2017ApJ...841...97S}, if we assume a uniformly distributed range for $\rm \kappa_{\nu}$, then the standard deviation is $\sim 0.1$, corresponding to an error of $\sim 50\%$ on the adopted value in this work. A $50\%$ error on $\rm \kappa_{\nu}$ has also been reported in \citet{2014A&A...562A.138R,2021MNRAS.508.2964A}.

The next parameter that significantly contributes to the uncertainty in mass estimation is the gas-to-dust ratio, $\rm R_{gd}$. The value of Galactic $\rm R_{gd}$ varies between 70 to 162, depending on factors such as grain size, shape, and compositions \citep{1990ApJ...359...42D,2003A&A...408..581V,2011piim.book.....D,2017MNRAS.467.4322P}. Assuming $\rm R_{gd}$ follows a uniform distribution, the standard deviation is 27, implying an uncertainty of 27\% on the adopted value of 100.

Taken together, the uncertainties in both $\rm \kappa_{\nu}$ and $\rm R_{gd}$, along with the error in other parameters, result in an overall uncertainty of $\sim 60\%$ on the estimates of mass and surface density for the cores. These uncertainties in mass and surface densities are also presented in Table \ref{core_prop}.

\begin{landscape}
\begin{table}
\caption{Physical properties of the cores derived from the 3~mm continuum map. }
\label{core_prop}
\begin{tabular}{ccccccccccccc}
\\ \hline

Name & RA (2000) & DEC (2000) & Decon. FWHM & PA & Peak Flux & Int. Flux & $\rm V_{LSR}$ & $\rm T_d$ & $\rm R_{core}$ & $\rm M_{core}$ & $\rm \Sigma_{core}$ & $\rm \alpha_{vir}$ \\
     &           &            & ($\arcsec \times \arcsec$)    & (degree) & (mJy/beam) & (mJy) & (km/s) & (K) & (pc) & ($\rm M_{\odot}$) & ($\rm g\ cm^{-2} $) & \\ 
\hline
C-1A & 15:43:16.55 & -54:07:13.61 & 4.46$\times$3.41 & 153.0 & 48.3$\pm$3.0 & 201.0$\pm$15.0 & $\rm-40.0\pm3.0$ & 23.5$\pm$0.4 & 0.04 & 271$\pm$155 & 14.8$\pm$8.5 & 0.07 \\
     &             &              &                  &       &              &                &                  & (39.1$\pm$0.02) &  & (157$\pm$90) & (8.6$\pm$4.9) & (0.12) \\
C-1B & 15:43:17.21 & -54:07:13.72 & 2.75$\times$1.51 & 108.0 & 4.62$\pm$0.32 & 8.87$\pm$0.89 & $\rm-42.4\pm1.3$ & 23.5$\pm$0.4 & 0.02 & 12$\pm$7 & 2.4$\pm$1.4 & 0.54 \\
     &             &              &                  &       &               &               &                  & (39.1$\pm$0.02) &  & (7$\pm$4) & (1.5$\pm$0.8) & (0.93) \\ 
C-1C & 15:43:17.29 & -54:07:24.12 & 3.59$\times$1.39 & 128.0 & 1.91$\pm$0.31 & 4.32$\pm$0.98 & $\rm-42.6\pm0.9$ & 23.5$\pm$0.4 & 0.02 & 6$\pm$3.5 & 0.9$\pm$0.5 & 0.01 \\
     &             &              &                  &       &               &               &                  & (39.1$\pm$0.02) &  & (3.4$\pm$2.1) & (0.6$\pm$0.3) & (0.01) \\ 
C-2 & 15:43:18.00 & -54:07:32.76 & 2.52$\times$1.90 & 93.0 & 6.77$\pm$0.53 & 13.5$\pm$1.5 & $\rm-42.6\pm0.1$ & 25.1$\pm$0.5 & 0.02 & 17$\pm$10 & 2.9$\pm$1.7 & 0.35 \\ 
C-3 & 15:43:18.91 & -54:07:35.32 & 4.54$\times$2.62 & 108.0 & 7.69$\pm$0.47 & 27.1$\pm$2.1 & $\rm-42.0\pm7.0$ & 25.1$\pm$0.5 &  0.03 & 34$\pm$19.5 & 2.4$\pm$1.4 & 0.49 \\
    &             &              &                  &       &               &              &                  & (46.8$\pm$0.1) &    & (17.5$\pm$10) & (1.2$\pm$0.7) & (0.94) \\ 
C-4 & 15:43:16.95 & -54:06:58.52 & 3.14$\times$2.54 & 145.0 & 2.08$\pm$0.26 & 5.5$\pm$0.92 & $\rm-41.5\pm0.1$ & 19.5$\pm$0.8 & 0.03 &  9$\pm$5 & 0.9$\pm$0.5 & 0.30 \\ 
\hline
\end{tabular}
\\ 
Values in parentheses are obtained with the revised temperature estimation from the 70~$\rm \mu m$ flux density (see Section \ref{phy_prop_core}). These revised values are used for further analysis of these cores in this work.
\end{table}
\end{landscape}

\subsection{Molecular line emission}\label{Mol_emi}
\begin{figure}
\centering
\includegraphics[scale=0.3]{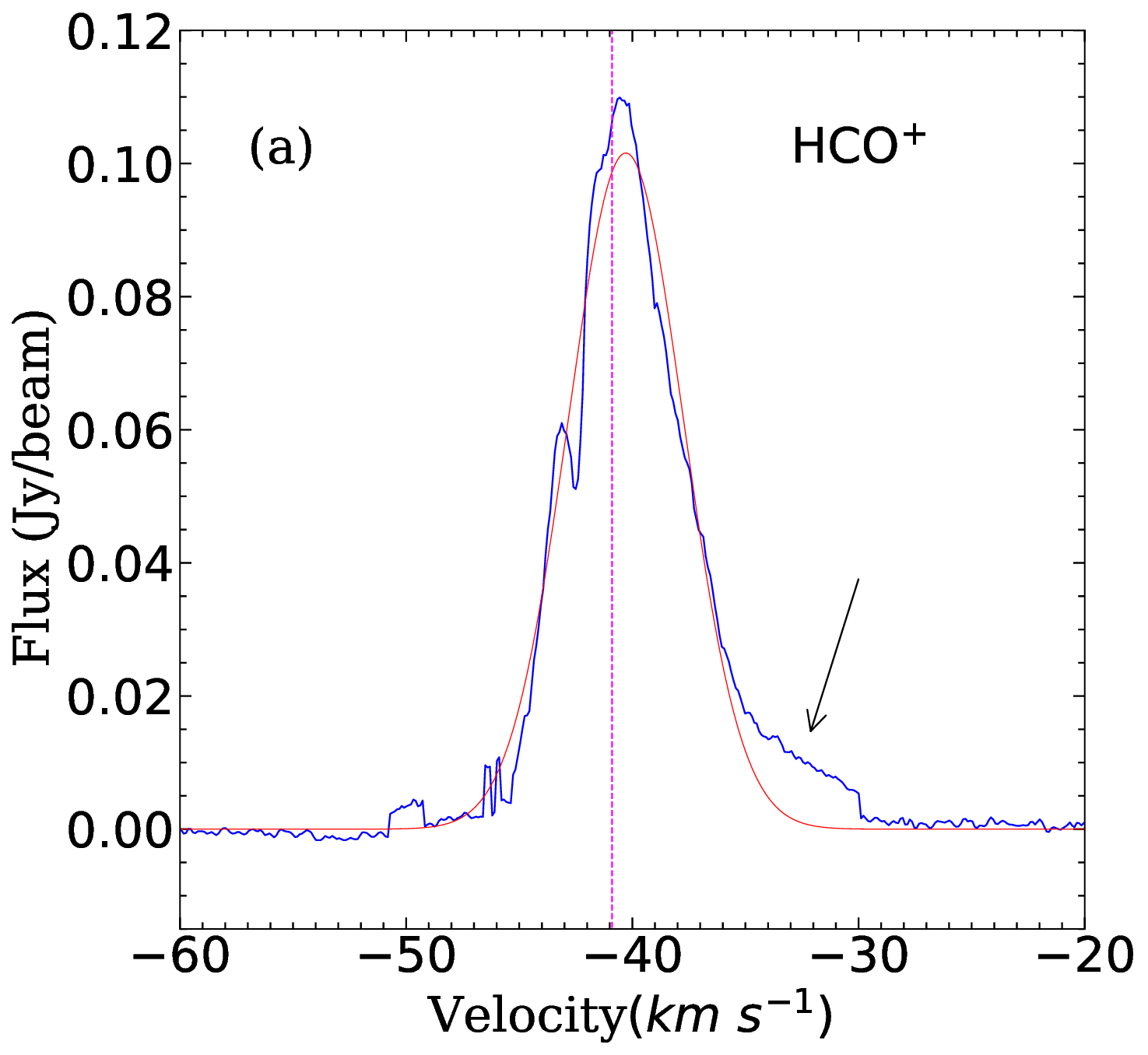}
\includegraphics[scale=0.3]{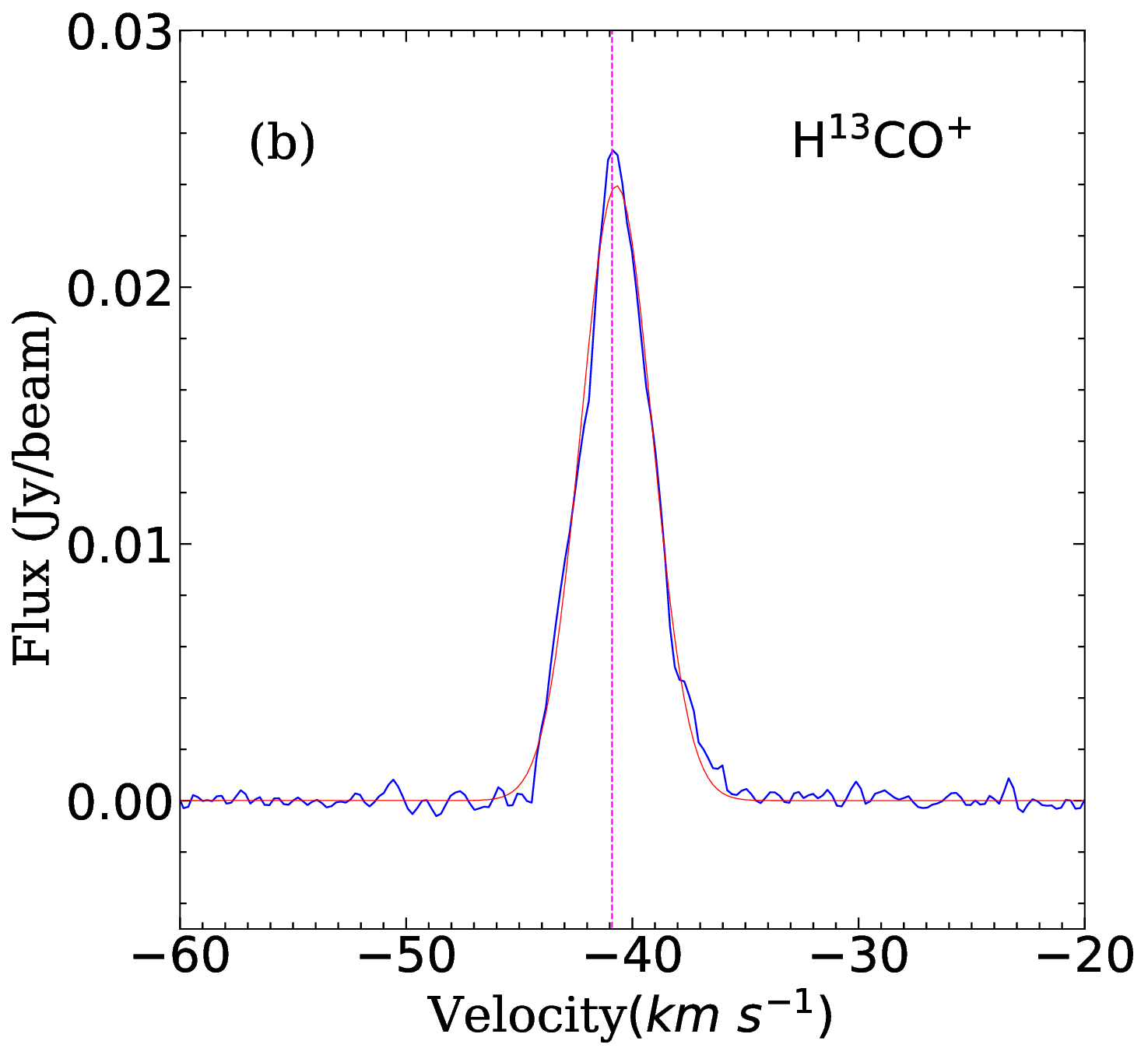}
\caption{Spectra of $\rm HCO^{+}$ and $\rm H^{13}CO^{+}$ lines averaged within $\rm 25\arcsec$ around the centre. The red curves are the Gaussian fitting to the spectra done with {\tt PySpecKit} package \citep{2022AJ....163..291G}. The vertical straight line in both panels marks the LSR velocity of $\rm -40.9~km~s^{-1}$ obtained from CS($\rm 2-1$) molecular line transition \citep{1996A&AS..115...81B}. The arrow in (a) points to the broad line wing in $\rm HCO^{+}$. } 
\label{spec_full}
\end{figure}

To get overall properties of the protocluster, we retrieve the average spectra of optically thick line $\rm HCO^{+}~(1-0)$ and optically thin line $\rm H^{13}CO^{+}~(1-0)$ within the $\rm 25\arcsec$ region around the central position. Due to their high critical density ($\rm \sim 6\times10^4~cm^{-3}$; \citealt{2015PASP..127..299S}), both lines are ideal for studying the dense gas properties. Also, the optically thick line, $\rm HCO^{+}$, can probe the infall and outflow signatures within the cloud. In Fig. \ref{spec_full}, we show the average spectra of both line transitions. Both lines show a single emission peak, where the intensity of optically thick $\rm HCO^{+}$ is approximately five times higher compared to the optically thin $\rm H^{13}CO^{+}$. The higher intensity of $\rm HCO^{+}$ compared to $\rm H^{13}CO^{+}$ can be attributed to both the abundance variation and the opacity effects of these isotopologues.

Gaussian fitting shows the LSR velocity to be $\rm -40.3~km~s^{-1}$ and $\rm -40.7~km~s^{-1}$ for $\rm HCO^{+}$ and $\rm H^{13}CO^{+}$, respectively. This is similar to $\rm -40.9~km s^{-1}$ obtained with CS($\rm 2-1$) line \citep{1996A&AS..115...81B}. Velocity widths are ~$\rm 2.25~km~s^{-1}$ and $\rm 1.47~km~s^{-1}$ for $\rm HCO^{+}$ and $\rm H^{13}CO^{+}$, respectively. The $\rm HCO^{+}$ spectra show a small dip at $\rm \sim -43~km~s^{-1}$ and broad line wings, mainly on the redshifted side. These features indicate a possible outflow in the protocluster region regulated by massive star formation activity. The central hub (C-1) hosts the most massive core and preferentially forms massive star(s), which is discussed later. Hence, we discuss the kinematics of the central hub of this protocluster complex in a later section. The different molecular lines can reveal the central hub's infall/outflow characteristics.

\subsubsection{Spatial distribution}\label{mol_spat_dis}
\begin{figure*}
\centering
\includegraphics[scale=0.52]{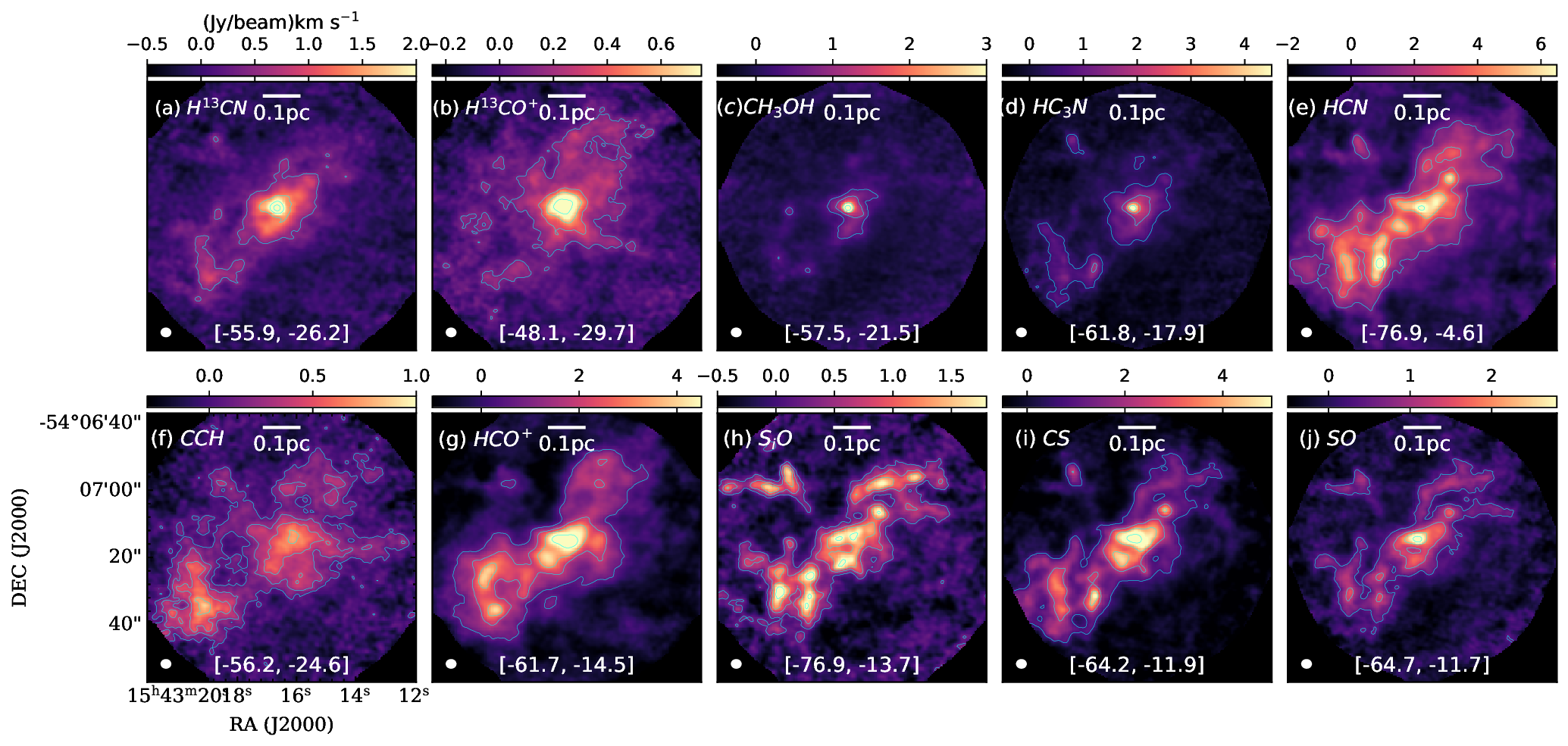}
\caption{Integrated intensity (moment0) maps of the molecules are displayed. The contours on each panel are from the same map. The contour levels are 10, 30, 50, and 80$\%$ of the peak flux. The peak values are 4.60, 1.13, 4.70, 5.30, 9.22, 0.90, 6.30, 2.33, 7.84, and 4.23~$\rm (Jy/beam)km~s^{-1}$ for $\rm H^{13}CN,~ H^{13}CO^{+},~ CH_3OH,~ HC_3N,~ HCN,~ CCH,~ HCO^{+},~ SiO,~ CS,~ and~ SO$, respectively. A scale bar of 0.1~pc and the beam size are shown on each panel. The numbers within square brackets in all panels gives the velocity range over which the lines are intigrated.}
\label{mom0_map}
\end{figure*}

Fig. \ref{mom0_map} presents the integrated intensity (moment0) maps for ten molecules, showing the morphological features of the line emissions in the protocluster region IRAS 15394$-$5358. We do not present $\rm H_{40\alpha}$ because there is no detected emission in that line transition, consistent with previous studies \citep{2006ApJ...651..914G}. $\rm H^{13}CO^{+}$ traces emission from the dense central region with little extended emission in the surroundings. A very faint emission is seen towards the cores C-2 and C-3. The morphology of $\rm H^{13}CN$ and $\rm HC_3N$ is similar. Both these lines show enhanced emission towards the massive dense core C-1, with little diffuse emission in the direction of cores C-2 and C-3. $\rm CH_3OH$ shows emission from the dense core, with no diffuse emission in the surrounding area. $\rm HCN$, $\rm HCO^{+}$, $\rm CCH$, $\rm SiO$, CS, and SO reveal much more extended emissions. Emission of $\rm CCH$, which is a good tracer of photo dissociation region (PDR), shows much more uniform extended emission. Faint emission is seen in the central massive dense core region and towards the other southern cores. This could be due to the presence of \hii\ regions generating PDRs around them. The morphology of $\rm SiO$ is extended with multiple elongated compact structures. These could be the outflows observed towards IRAS 15394$-$5358. A few of such features are also seen in the emissions of $\rm HCN$ and $\rm HCO^{+}$.

\subsubsection{Similarity and differences between molecules} \label{pca}
The spatial distribution of the emission in the molecular line transitions (Fig. \ref{mom0_map}) shows similarities and differences. Since the different transitions trace different physical conditions, it is worth analyzing their similarities and differences, allowing us to find the best tracers for different structures within the clouds. We use the principal component analysis (PCA) to conduct such study. Because of the robustness in finding the similarities and differences between line transitions, the PCA has been widely used in studying molecular clouds \citep{2009MNRAS.395.1021L,2012MNRAS.419.2961J,2013MNRAS.433..221J,2020MNRAS.496.2790L}. PCA is part of an unsupervised machine learning algorithm, which describes the multidimensional data set through a linear combination of uncorrelated variables called the principal components (PCs). These PCs represent the variances that capture the most significant common features in the data. The PC with the largest variance identifies the most significant feature. Mathematically, PCs are derived by performing eigenvalue decomposition on the covariance matrix (or correlation matrix) of the input variables, resulting in eigenvectors (PCs) and their corresponding eigenvalues, which quantify the variance explained by each PC.
We use the {\it scikit-learn}\footnote{https://scikit-learn.org/stable/modules/generated/ \\ sklearn.decomposition.PCA.html} package of Python to derive the correlation matrix and the variances of the PCs. We analyze this using the integrated intensity maps shown in Fig. \ref{mom0_map}. As described in \citet{2020MNRAS.496.2790L}, all the maps are smoothed and re-gridded, so all maps have the same resolution and pixel sizes. Before running the analysis, data standardisation was performed, resulting in the data having a mean value of zero and a standard deviation of one. This process mitigates the effect of brighter pixels. In this analysis, we used all pixels from the integrated intensity maps of all the molecules. However, we tested the impact of missing zero-spacing data or negative pixels in the analysis by considering the pixels above 3$\sigma$ for individual molecules. We verified that the presence of missing zero-spacing data or negative pixels in the integrated intensity maps does not alter the final outcome of the analysis.

Table \ref{res_pca} presents the correlation matrix, eigenvalues, and eigenvectors (PCs). In the correlation matrix, a correlation coefficient exceeding $80\%$ indicates a strong correlation between components, while a coefficient surpassing $90\%$ signifies a very strong correlation between the components. $\rm H^{13}CN,~ and~ HC_3N$, and $\rm HCN,~ HCO^{+}$, and CS are strongly correlated, with a correlation coefficient of above 0.93. SiO and SO are highly correlated with a coefficient of 0.99, suggesting both lines trace similar physical conditions such as outflows and shocks in the protocluster region.

\begin{table*}
\small
\centering
\caption{Results of PCA analysis.}
\label{res_pca}
\begin{tabular}{ccccccccccc}
\\ \hline

 & $\rm H^{13}CN$ & $\rm H^{13}CO^{+}$ & CCH & $\rm SiO$ & HCN & $\rm HCO^{+}$ & $\rm CH_{3}OH$ & $\rm HC_{3}N$ & $\rm CS$ & $\rm SO$\\ 
\hline
\multicolumn{9}{c}{Correlation matrix values of the data set} \\
\hline 
$\rm H^{13}CN$     & 1.00 & {\bf 0.88} & 0.76 & 0.79 & {\bf 0.84} & {\bf 0.86} & {\bf 0.80} & {\bf 0.94} & {\bf 0.92} & {\bf 0.90} \\ 
$\rm H^{13}CO^{+}$ & {\bf 0.88} & 1.00       & 0.70       & 0.62 & 0.70       & 0.76       & 0.68 & {\bf 0.81} & 0.77 & 0.77 \\ 
CCH                & 0.76       & 0.70       & 1.00       & 0.64 & {\bf 0.80} & {\bf 0.84} & 0.44 & 0.76 & {\bf 0.81} & 0.66 \\ 
$\rm SiO$          & 0.79       & 0.62       & 0.64       & 1.00 & {\bf 0.87} & 0.78       & 0.59 & 0.77 & {\bf 0.84} & {\bf 0.91} \\ 
HCN                & {\bf 0.84} & 0.70 & {\bf 0.80} & 0.87 & 1.00       & {\bf 0.94} & 0.54 & {\bf 0.80} & {\bf 0.93} & {\bf 0.85} \\ 
$\rm HCO^{+}$      & {\bf 0.86} & 0.76 & {\bf 0.84} & 0.78 & {\bf 0.94} & 1.00       & 0.58 & {\bf 0.82} & {\bf 0.93} & {\bf 0.83} \\ 
$\rm CH_{3}OH$     & {\bf 0.80} & 0.68       & 0.44       & 0.59 & 0.54       & 0.58       & 1.00 & {\bf 0.81} & 0.67 & 0.77 \\
$\rm HC_{3}N$      & {\bf 0.94} & {\bf 0.81} & 0.76 & 0.77 & {\bf 0.80} & {\bf 0.82} & {\bf 0.81} & 1.00 & {\bf 0.91} & {\bf 0.88} \\  
$\rm CS$           & {\bf 0.92} & 0.77 & {\bf 0.81} & {\bf 0.84} & {\bf 0.93} & {\bf 0.93} & 0.67 & {\bf 0.91} & 1.00 & {\bf 0.90} \\
$\rm SO$           & {\bf 0.90} & 0.77 & 0.66 & {\bf 0.91} & {\bf 0.85} & {\bf 0.83} & 0.77 & {\bf 0.88} & {\bf 0.90} & 1.0 \\ 
\hline
\multicolumn{9}{c}{Eigen values and eigen vectors or contributions of each line in the directions of prinicipal components} \\
\hline
Component & 1 & 2 & 3 & 4 & 5 & 6 & 7 & 8 & 9 & 10 \\
\hline
Variance (\%) & 81.15 &  7.47 & 4.88 & 2.33 & 1.50 & 1.06 & 0.56 & 0.41 & 0.33 & 0.27 \\
$\rm H^{13}CN$     & 0.33 & 0.15  & -0.12 & -0.07 & -0.04 & 0.18  & 0.27  & -0.84 & 0.08  & 0.10 \\
$\rm H^{13}CO^{+}$ & 0.29 & 0.19  & -0.46 & -0.74 & 0.07  & -0.07 & 0.04  & 0.26  & -0.02 & -0.10 \\ 
CCH                & 0.28 & -0.41 & -0.47 & 0.37  & 0.49  & -0.31 & -0.09 & -0.03 & 0.16  & -0.04 \\ 
$\rm SiO$          & 0.30 & -0.12 & 0.60  & -0.17 & 0.47  & -0.15 & 0.26  & -0.01 & -0.34 & -0.24 \\ 
HCN                & 0.32 & -0.33 & 0.18  & -0.03 & -0.31 & -0.03 & 0.46  & 0.29  & 0.49  & 0.29 \\ 
$\rm HCO^{+}$      & 0.32 & -0.28 & -0.06 & 0.04  & -0.55 & -0.30 & -0.19 & -0.06 & -0.59 & 0.09 \\ 
$\rm CH_{3}OH$     & 0.26 & 0.69  & 0.02  & 0.38  & -0.13 & -0.44 & 0.14  & 0.14  & 0.08  & -0.17 \\ 
$\rm HC_{3}N$      & 0.33 & 0.19  & -0.11 & 0.28  & 0.19  & 0.58  & 0.06  & 0.31  & -0.34 & 0.39 \\ 
CS                 & 0.33 & -0.12 & 0.04  & 0.12  & -0.22 & 0.45  & -0.22 & 0.05  & 0.17  & -0.71 \\ 
SO                 & 0.33 & 0.12  & 0.33  & -0.14 & 0.10  & -0.06 & -0.71 & -0.05 & 0.30  & 0.34 \\
\hline
\end{tabular}
\\Correlation coefficient values above 0.8 are highlighted in boldface. 
\end{table*}

\begin{figure}
\centering
\includegraphics[scale=0.65]{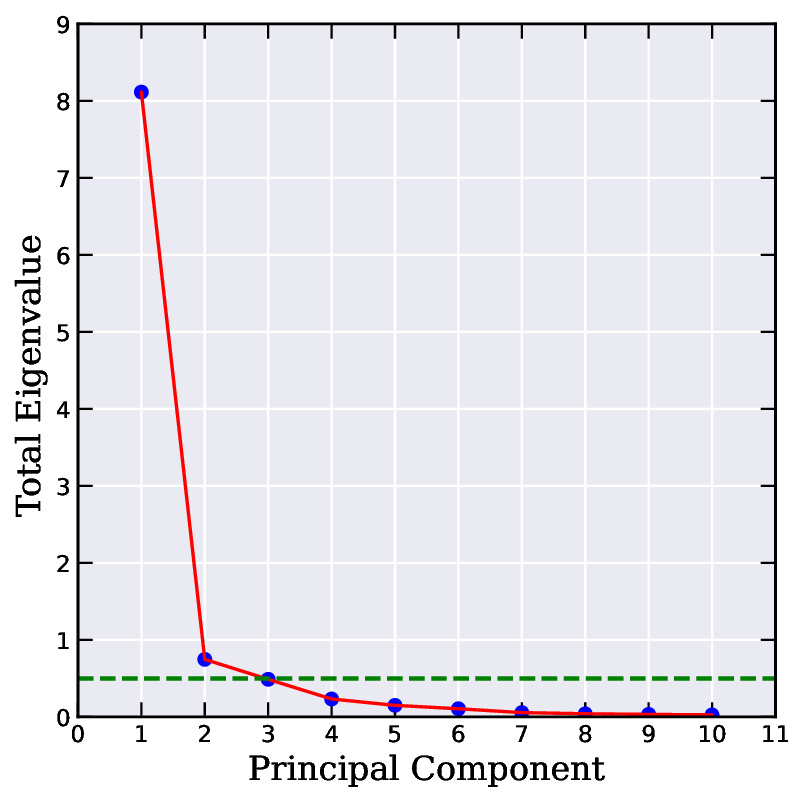}
\caption{The plot shows the derived eigenvalues of PCs. A threshold eigenvalue of 0.5 is marked with the dashed line.} 
\label{pca_variance}
\end{figure}

\begin{figure*}
\includegraphics[scale=0.45]{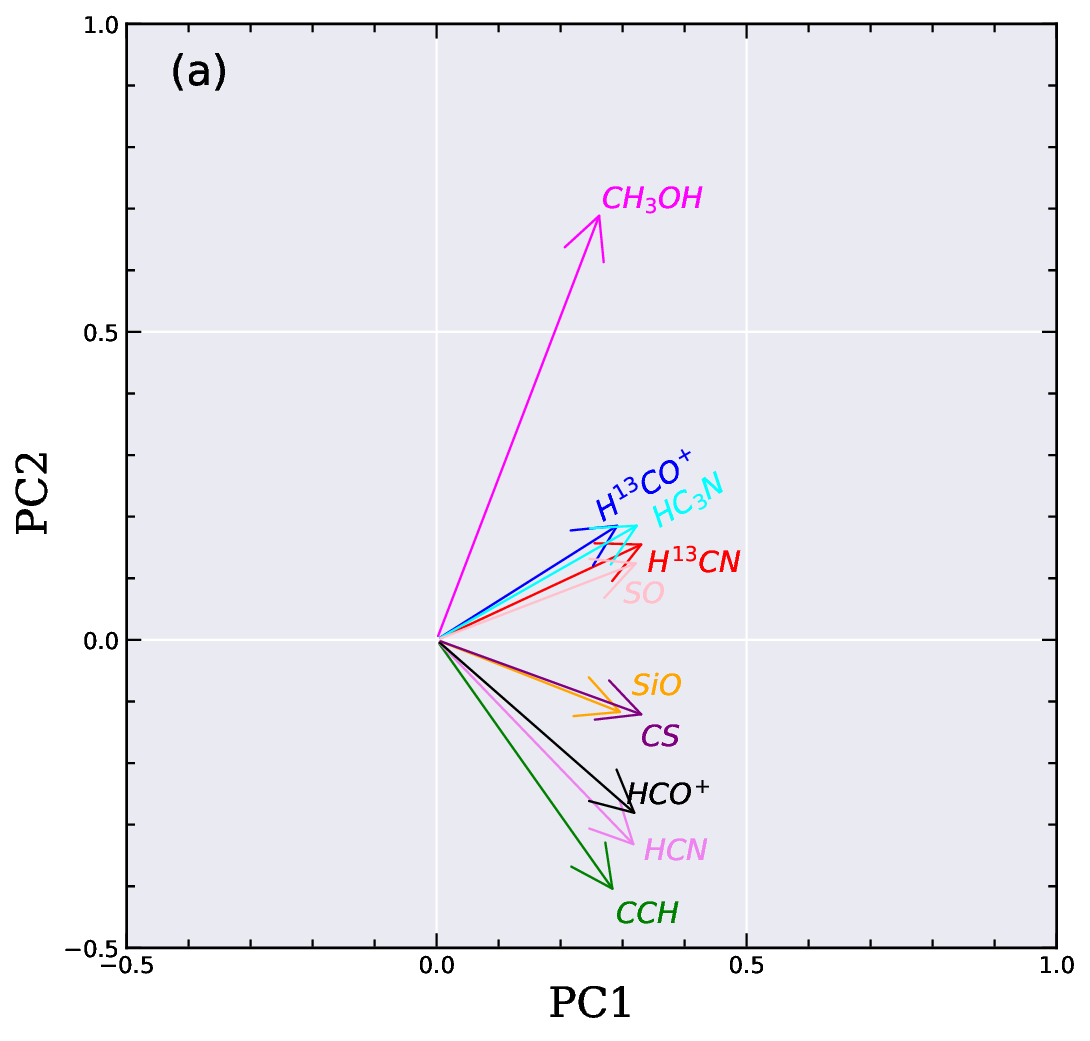}
\includegraphics[scale=0.45]{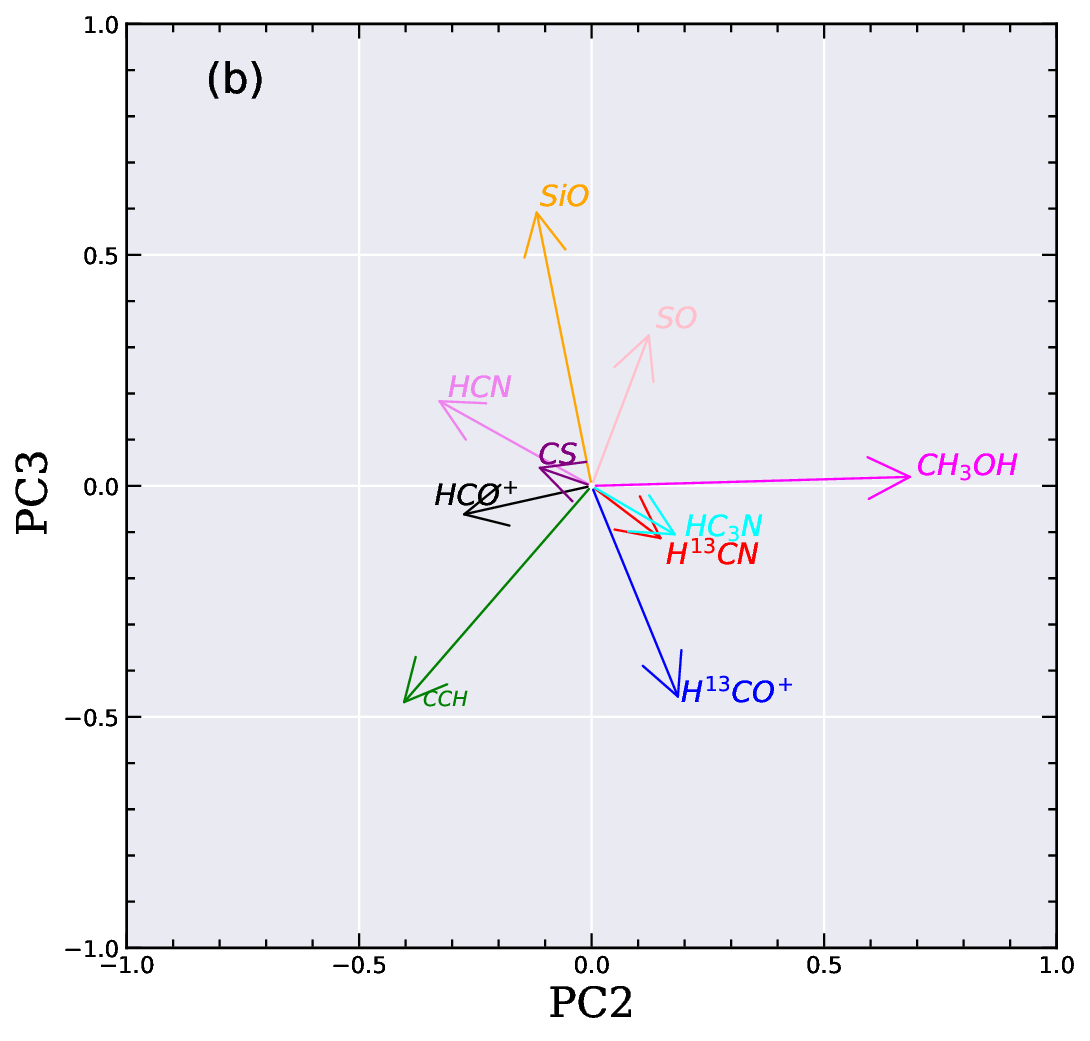}
\caption{The plots show variations between PC1 and PC2 (a) and between PC2 and PC3 (b). Contributions from each molecule to the PCs are represented as arrows. } 
\label{pca1_2_3}
\end{figure*}

\begin{figure*}
\centering
\includegraphics[scale=0.52]{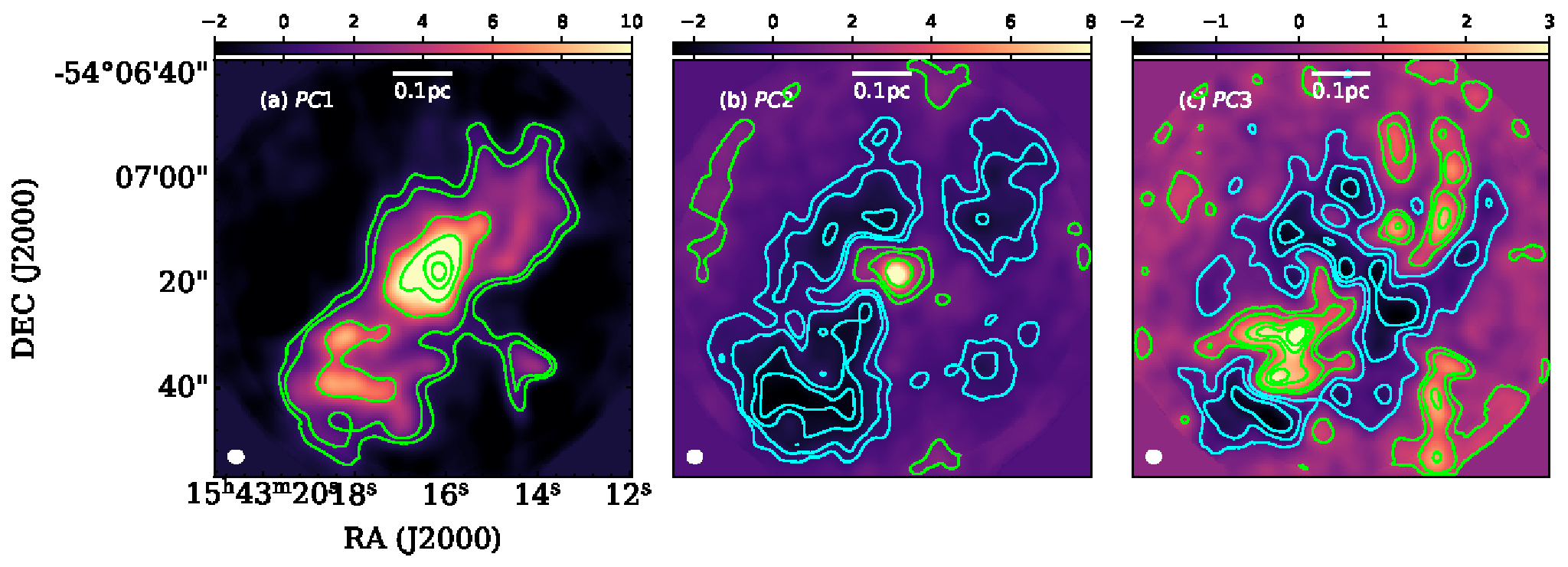}
\caption{Maps of the generated first three PCs. (a) PC1 is overlaid with its contours at levels of 0, 1, 5, 10, 15, and 20. (b) PC2 has contour levels of -2.0, -1.5, -0.7, -0.3, 1, 2, and 5. (c) PC3 has contour levels of -1.5, -1.0, -0.5, 0.5, 1, 1.5, 2, 2.5, and 3. The maps show contours with negative and positive values in cyan and green colours, respectively. The different components, especially PC2 and PC3 demarcates the distinctions among the different molecules. } 
\label{pca_maps}
\end{figure*}

In Fig. \ref{pca_variance}, we show the distribution of eigenvalues with respect to the PCs. The plot shows a steep decline in eigenvalues. In our analysis, we identify two PCs with eigenvalues greater than 0.5. These two PCs contain the maximum information of the data sets, with a total contribution of about $90\%$ of the total variance. PC1 and PC2 have eigenvalues of 81.15 and 7.47\%. We also present the PC3, whose eigenvalue is 4.88$\%$, which is very close to 5\%. Fig. \ref{pca1_2_3} shows the variation between pairs of PCs and the contributions of different molecules to these PCs. Fig. \ref{pca1_2_3}(a) is between PC1 and PC2, and Fig. \ref{pca1_2_3}(b) between PC2 and PC3. We can broadly segregate the molecules into three groups by looking at Fig. \ref{pca1_2_3}(a). To get a better insight into their nature, we present the constructed PC maps (PC1, PC2, and PC3) in Fig. \ref{pca_maps}.

In PCA, PC1 represents the direction of maximum variance in the dataset, indicating that it captures the most significant variability among the molecules. As a result, delineating differences among molecules solely based on PC1 can be challenging. However, by examining the contributions of molecules along PC2 and PC3, which represent orthogonal directions capturing additional variance, it becomes feasible to discern distinctions among the molecules more effectively. On PC2, the cyan contours indicate $\rm HCN,~ HCO^{+},~ CCH,~ SiO, ~ and ~ CS$. The lines CCH, HCN, and $\rm HCO^{+}$ are highly anticorrelated with PC2. These molecules are more likely to trace gas from the surrounding region of the central massive dense core. The green contours indicate $\rm CH_3OH$, which is strongly correlated to PC2. This is primarily tracing the dense gas from the central core. $\rm H^{13}CO^{+}$, $\rm H^{13}CN$, $\rm HC_3N$, and SO are positively correlated to PC2 because they trace same emission around the central massive core. Similarly, SiO and CS are negatively correlated to PC2, suggesting these might trace some outflow emission in the protocluster. On PC3 (Fig. \ref{pca_maps}(c)) map, the cyan contours represent CCH and $\rm H^{13}CO^{+}$. These molecules appear to be highly anticorrelated to $\rm SiO$ and SO, which are shown in green contours. This anticorrelation could be because CCH and $\rm H^{13}CO^{+}$ trace some emission around the central core, whereas SiO and SO trace the shock and outflows in the protocluster complex. SO appears to be better correlated to $\rm H^{13}CO^{+}$, $\rm H^{13}CN$, $\rm HC_3N$ in PC2, since this molecule can also trace dense gas, and the spatial distribution (see Fig. \ref{mom0_map}) supports the similarity. HCN and CS are slightly correlated with PC3, which means these lines also trace some outflow emission.

Considering the PC1 vs PC2 plot (Fig. \ref{pca1_2_3}(a)), we broadly segregate the molecules into three groups. First group contains $\rm H^{13}CN,~ H^{13}CO^{+},~ HC_3N,~ and ~SO$, and the second group contains $\rm HCN,~ HCO^{+},~ CCH,~ SiO, ~ and ~ CS$. The third group is the $\rm CH_3OH$. Note that this classification is not strict since the molecules might trace multiple physical conditions. This is visible on the PC2 vs PC3 plot (Fig. \ref{pca1_2_3}(b)), where the correlation between molecules appears slightly different. This analysis provides an opportunity to segregate the molecules into different groups, which trace emissions of different physical conditions in the protocluster complex IRAS 15394$-$5358. In a similar analysis towards G9.62+019, \cite{2020MNRAS.496.2790L} see a slightly different grouping of molecules, suggesting differences in the nature of the surrounding environment in different star-forming regions. More studies of the ATOMS survey will help get a deeper insight into the molecular line emissions by characterizing the gas distributions.

\subsubsection{Hub-Filament system} \label{filament}
Within the massive clump of 15394$-$5358, we can see the presence of a hub-filament system. The growing observational evidences indicates these hub-filament systems are the preferable sites of massive star formation, where matter from the surroundings gets funnelled to the central core or hub through the filaments. We can detect the filamentary structures in the star-forming clump using the $\rm H^{13}CO^{+}$ line transition. The reason for using this transition in filament identification is the following. 
The high critical density ($\rm 6.2\times10^4~cm^{-3}$ at 10~K) and low optical depth make the $\rm H^{13}CO^{+}~J=1-0$ transition a good tracer of dense gas. Effective excitation density ($\rm 3.9\times10^4~cm^{-3}$; \citealt{2015PASP..127..299S}) of the molecule is comparable to its critical density. Also, the $\rm H^{13}CO^{+}~J=1-0$ line emission correlates well with the dense gas column density traced by {\it Herschel} data \citep{2017A&A...604A..74S,2022ApJ...926..165L}. Due to this property, \citet{2022MNRAS.514.6038Z} used this line to identify the filaments toward the targets of the ATOMS survey. Applying the FILFINDER algorithm \citep{2015MNRAS.452.3435K} over the moment0 maps of $\rm H^{13}CO^{+}~J=1-0$, these authors have identified the filaments associated with all the targets.

\begin{figure}
\centering
\includegraphics[scale=0.15]{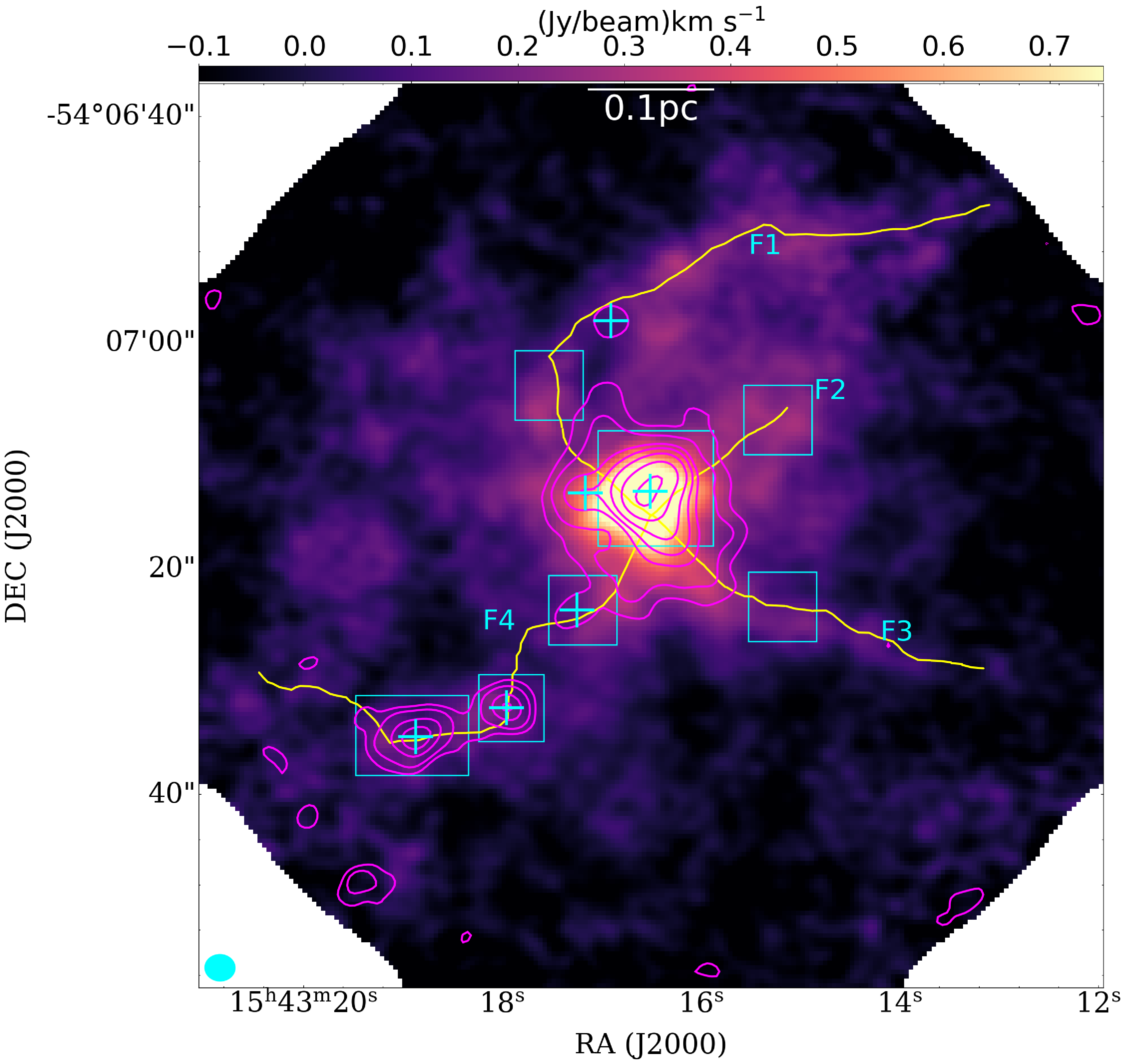}
\caption{Integrated intensity map of $\rm H^{13}CO^{+}$. The filaments are shown as yellow lines and labelled from F1 to F4. The 3~mm continuum emission is shown as magenta contours, with the same contour levels as of Fig. \ref{3mmcont_map}. The `+' marks display the locations of the identified dense cores in the region. The cyan boxes on the central core and the filaments are the regions where spectra are extracted to analyze the infall signature in that region. } 
\label{fila_box}
\end{figure}

\citet{2022MNRAS.514.6038Z} identified four major filamentary structures extending from the central hub (C-1), creating the hub-filament system in the protocluster complex 15394$-$5358. In this work, we use the same filaments identified by \citet{2022MNRAS.514.6038Z}. Fig. \ref{fila_box} presents the identified hub-filament system. The filaments (F1 to F4) are the yellow lines. In a next section, we discuss the nature of infall onto the central core through the filaments.

\subsubsection{Gas kinematics}\label{gas_kin}
\begin{figure*}
\centering
\includegraphics[width=4.2cm,height=4.2cm]{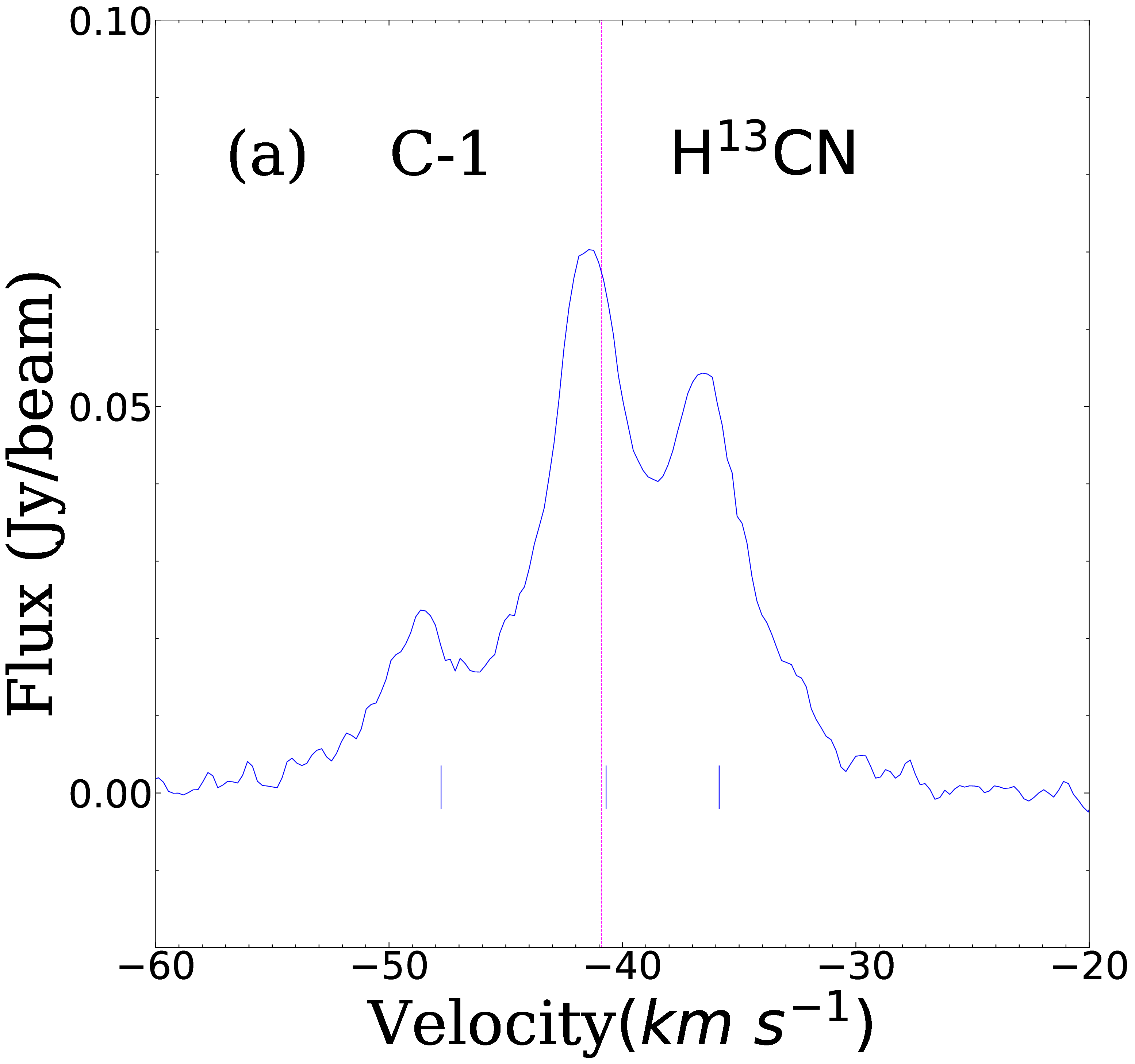}
\includegraphics[width=4.2cm,height=4.2cm]{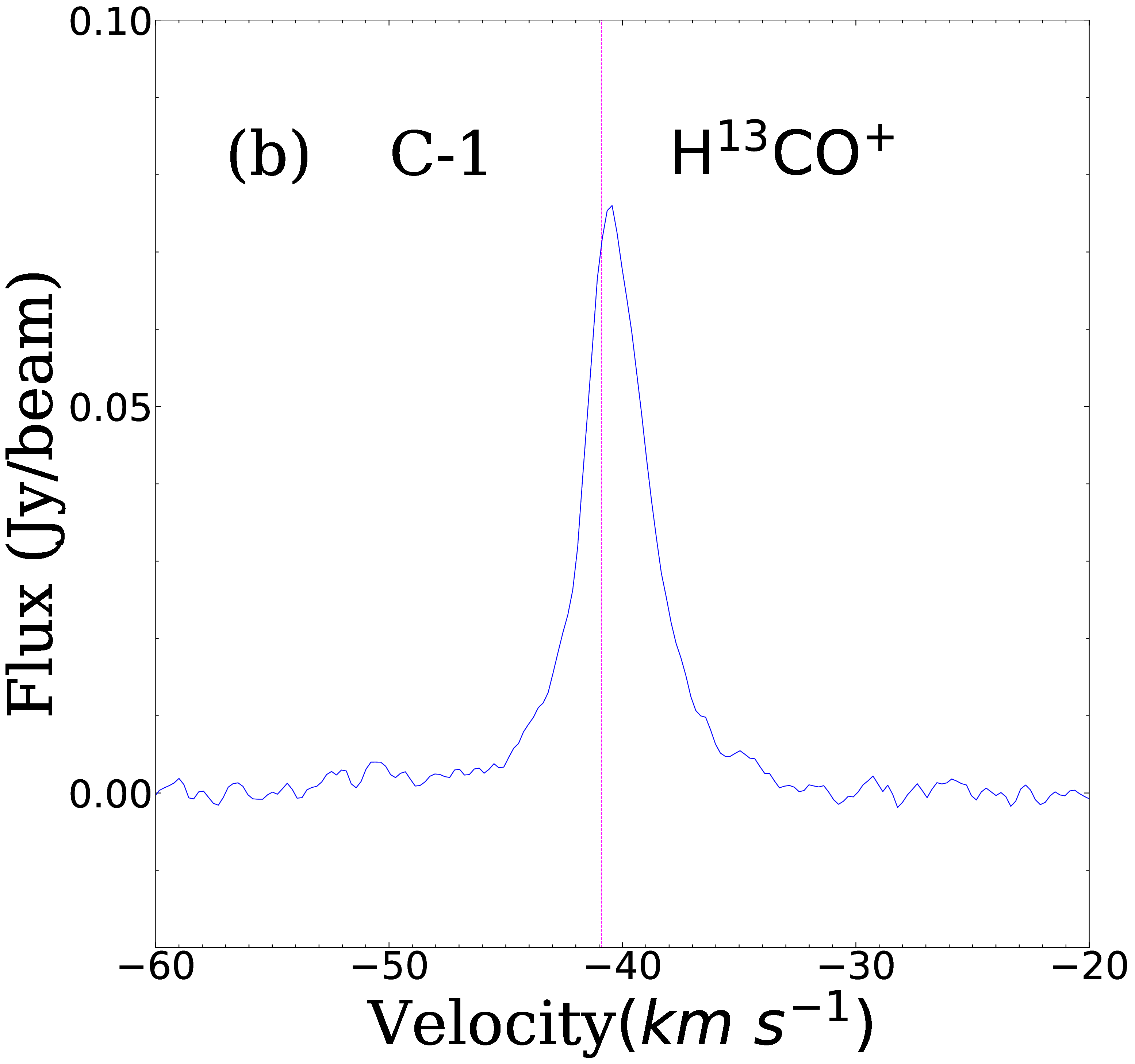}
\includegraphics[width=4.2cm,height=4.2cm]{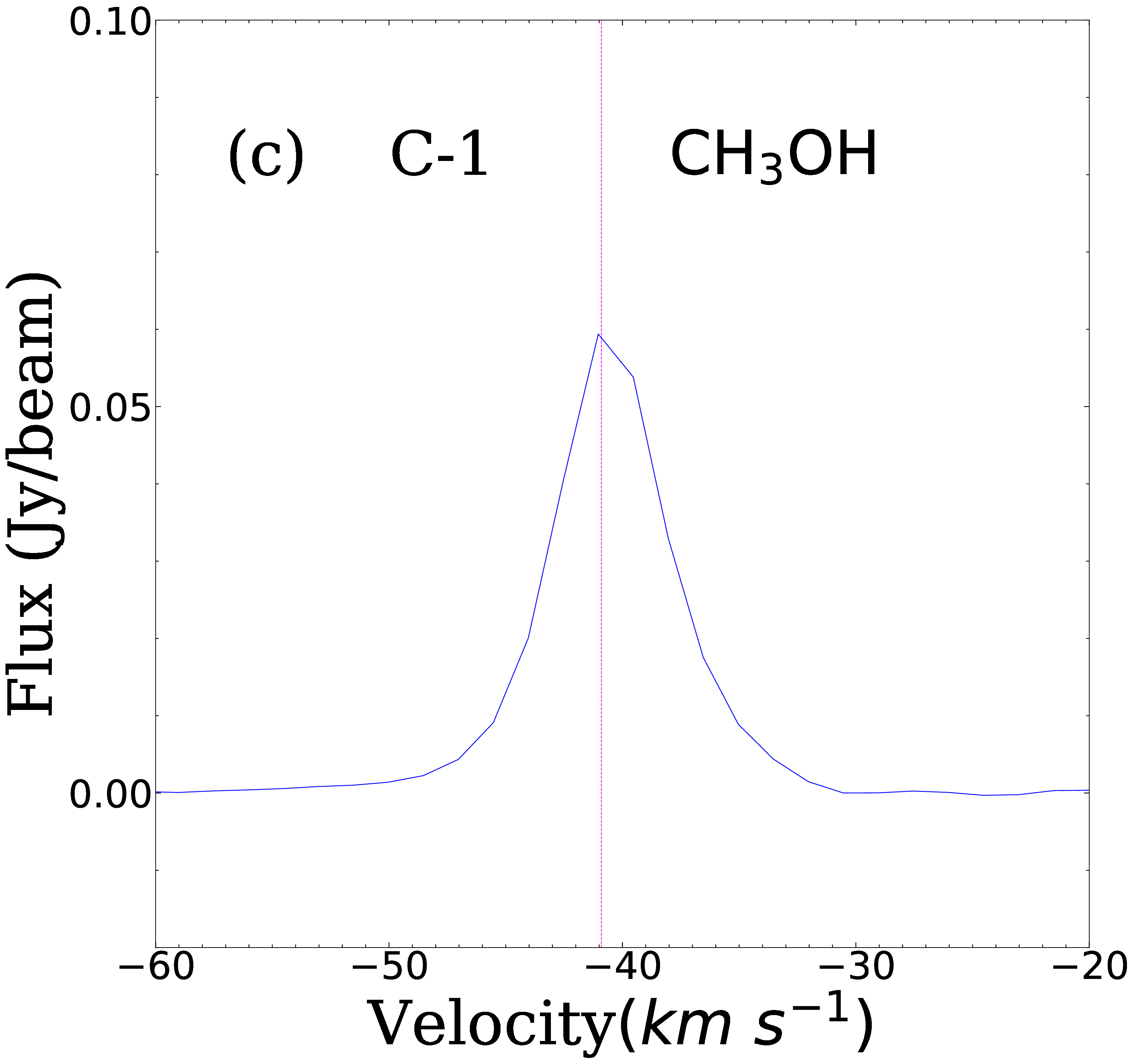}
\includegraphics[width=4.2cm,height=4.2cm]{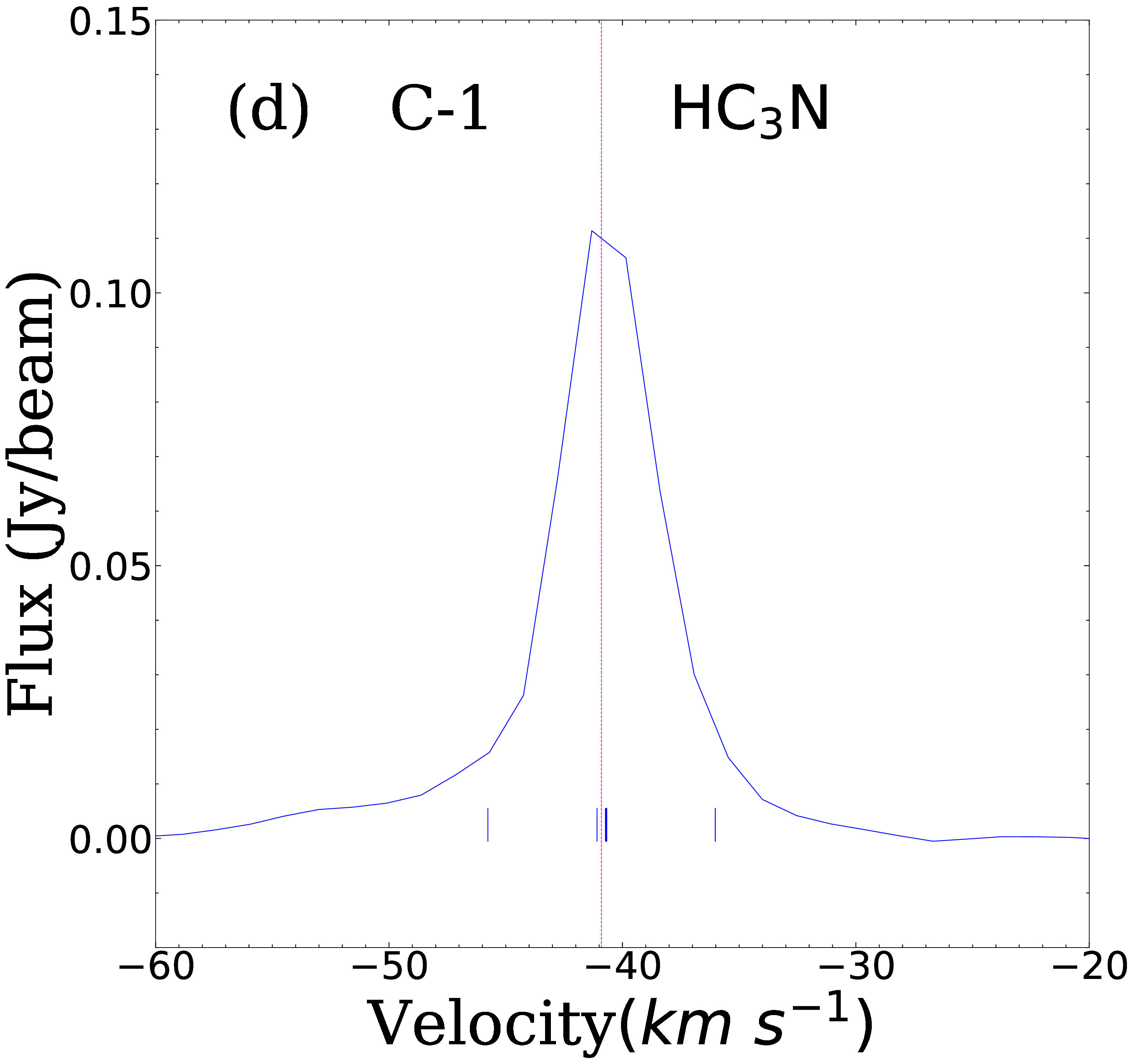}
\includegraphics[width=4.2cm,height=4.2cm]{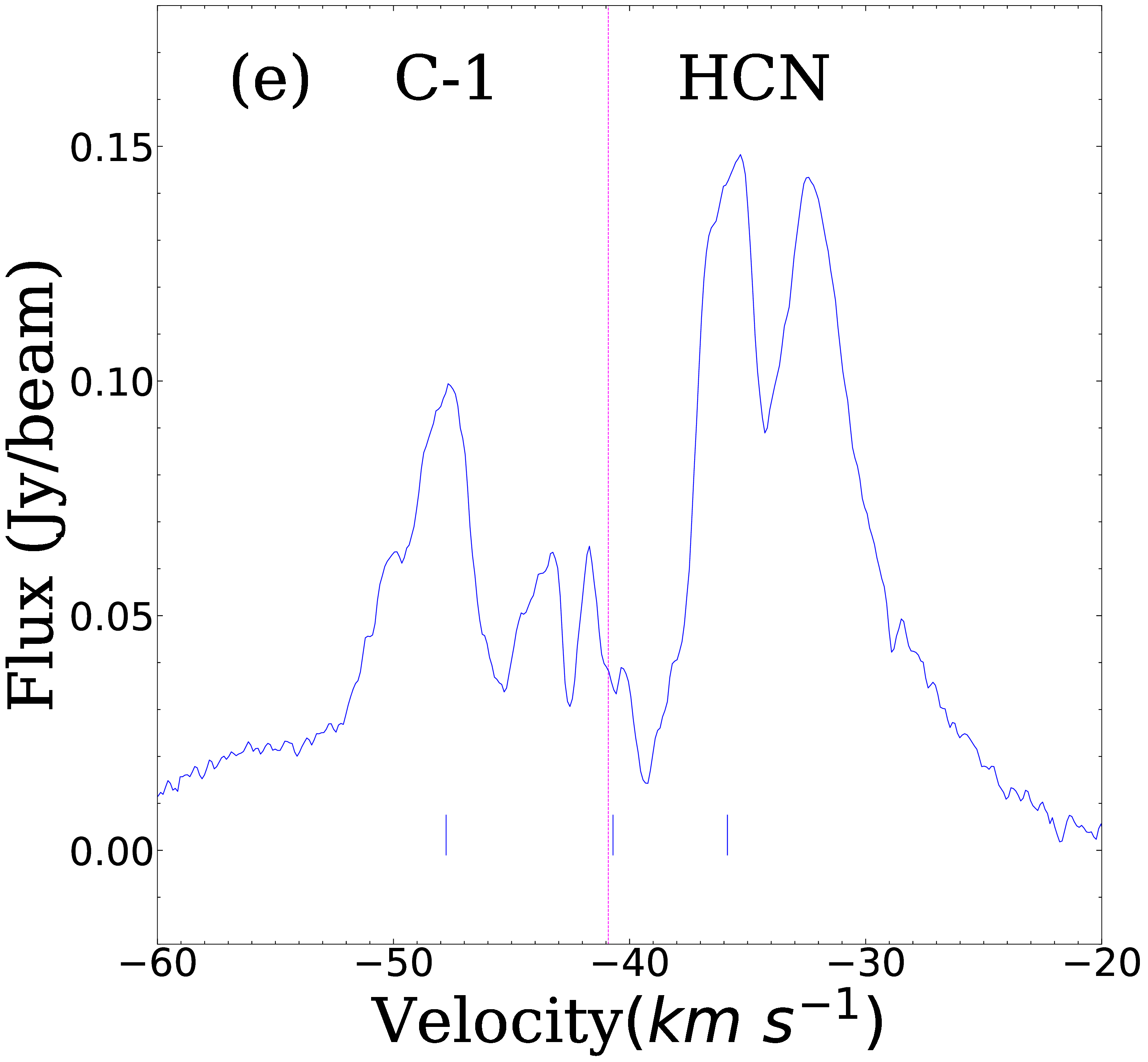}
\includegraphics[width=4.2cm,height=4.2cm]{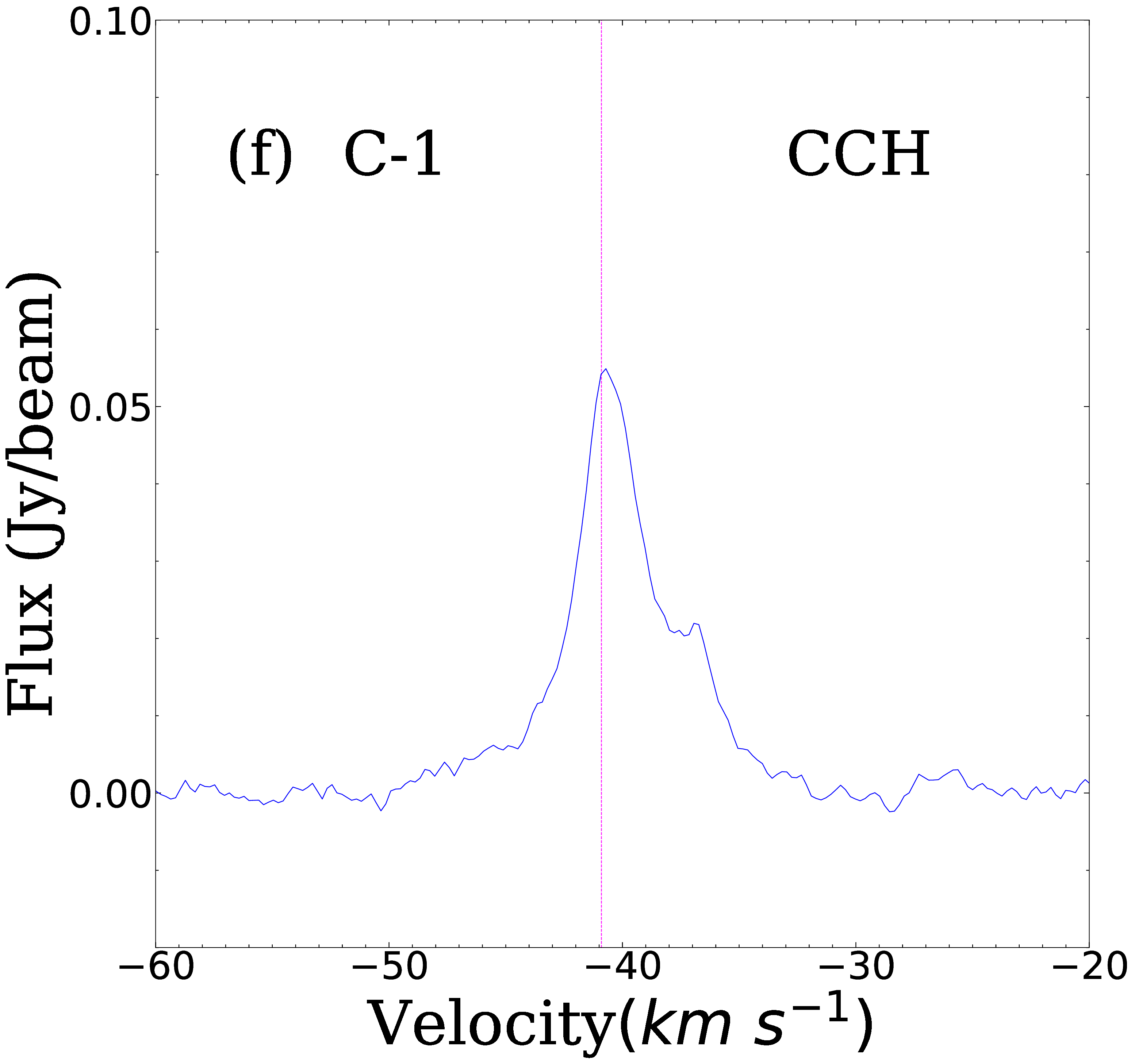}
\includegraphics[width=4.2cm,height=4.2cm]{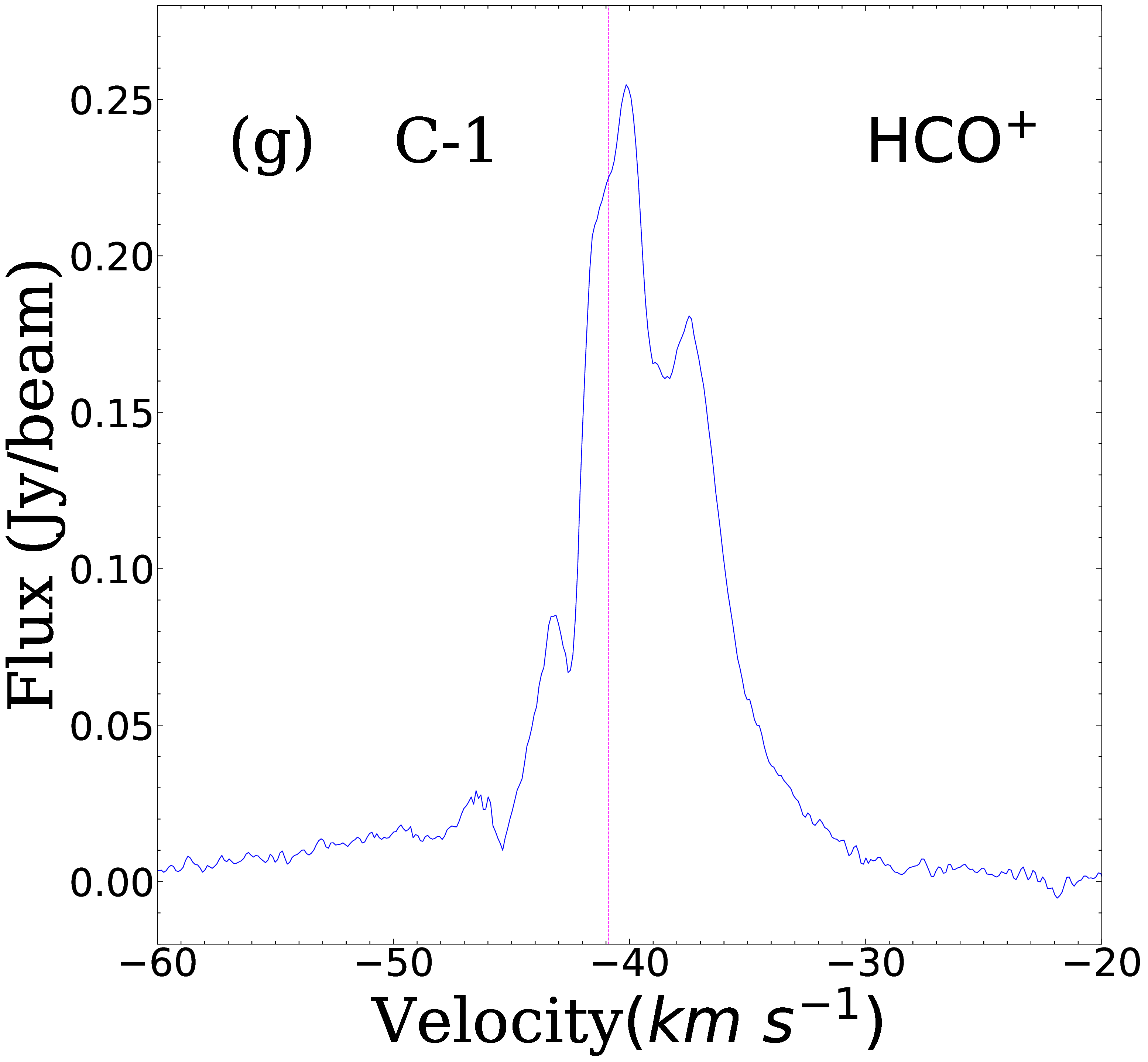}
\includegraphics[width=4.2cm,height=4.2cm]{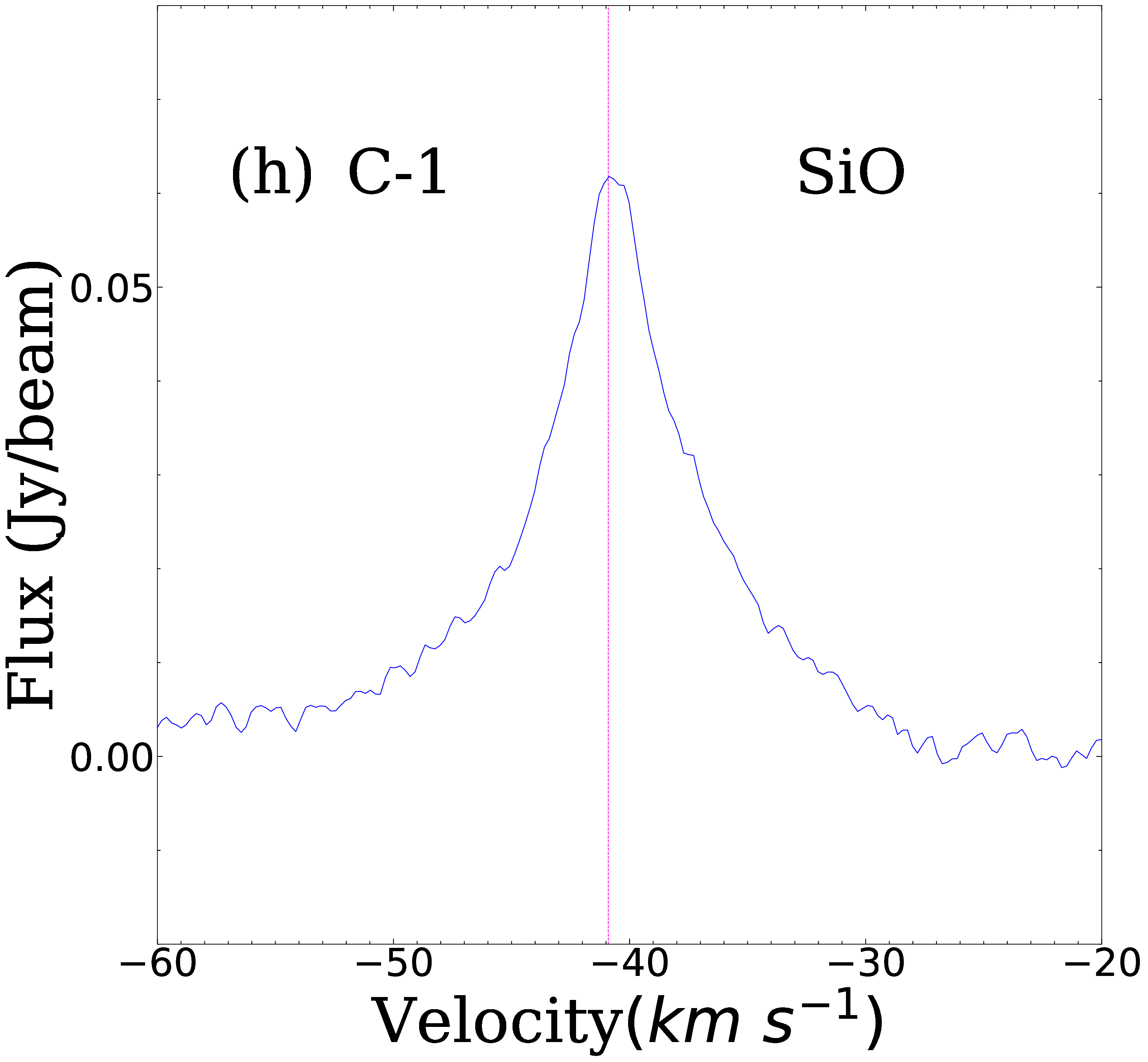}
\includegraphics[width=4.2cm,height=4.2cm]{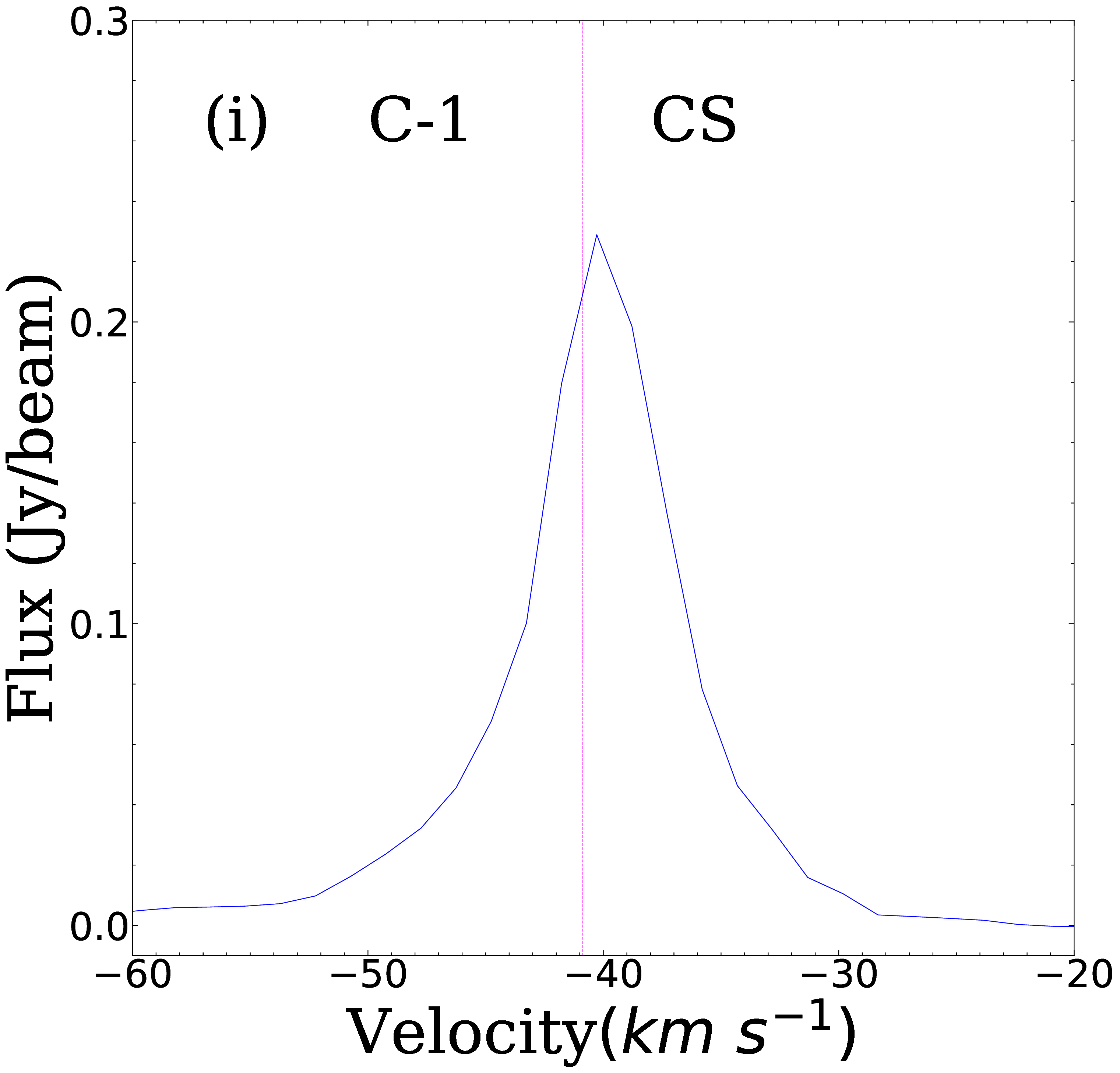}
\includegraphics[width=4.2cm,height=4.2cm]{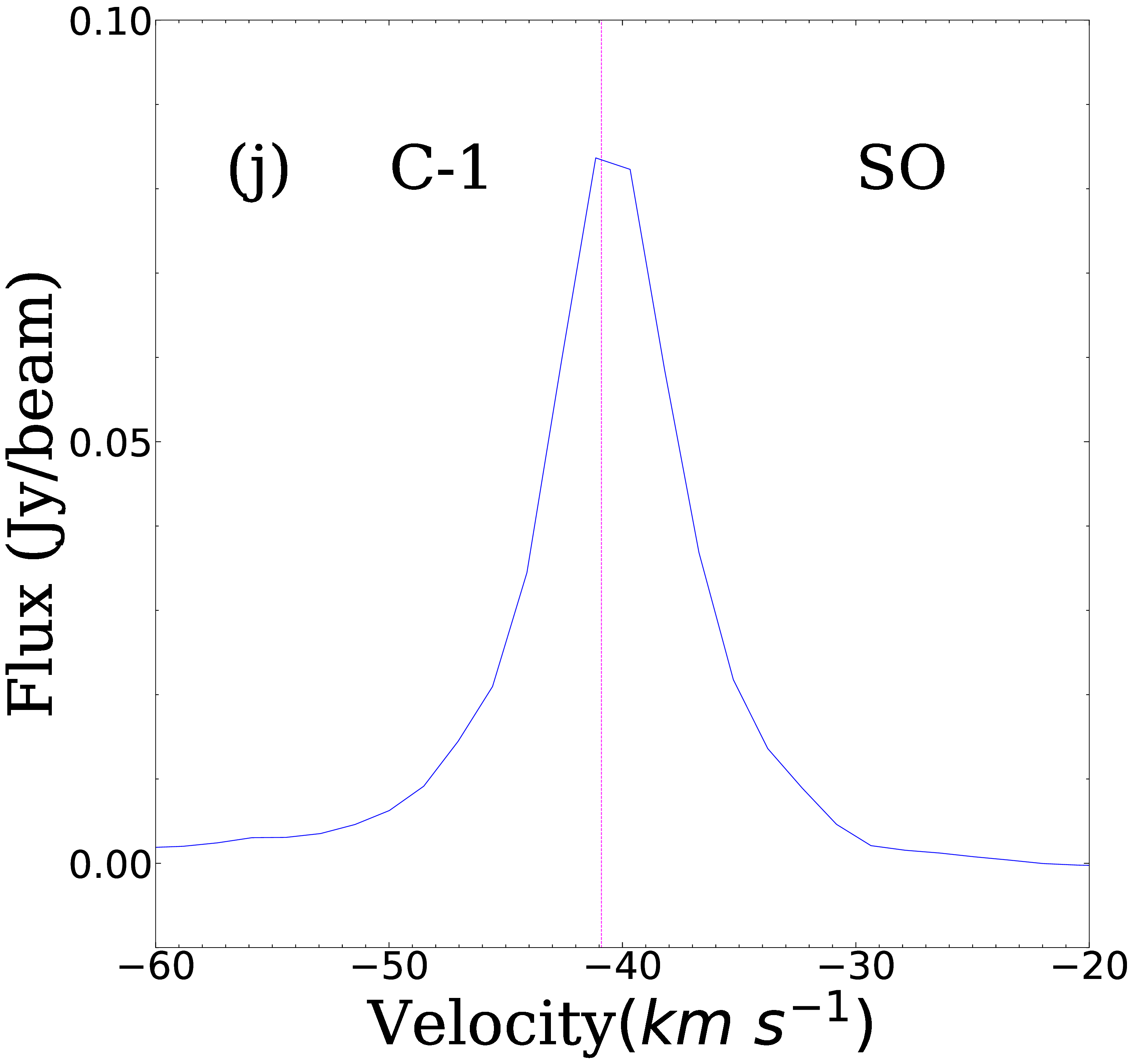}
\caption{Average spectra of all the molecular lines extracted over the central core C-1. The vertical straight line in each panel marks the LSR velocity of $\rm -40.9~km~s^{-1}$ obtained from CS($\rm 2-1$) molecular line transition \citep{1996A&AS..115...81B}. The small vertical blue lines on plots of $\rm H^{13}CN$, $\rm HC_3N$, and HCN mark the positions of hyperfine components in those line transitions. } 
\label{spec_C1}
\end{figure*}

In this section, we discuss the kinematics of the central hub of the protocluster complex. To obtain the kinematics of the hub, we extract the spectra over the core C-1. This will provide a global kinematic property of the central part of the protocluster region. In Fig. \ref{spec_C1}, we show the average spectra of the ten molecular line transitions. Spectra of $\rm H^{13}CO^{+},~CH_3OH,~HC_3N,~CCH,~SiO,~CS,~and~SO$ show single profiles. Extended line wings are seen especially in SiO, CS, and SO is a signature of outflows. The multiple peaks in $\rm H^{13}CN$ and HCN are due to their hyperfine nature.

\subsubsection{Molecular outflows}\label{outflow}
\begin{figure*}
\centering
\includegraphics[scale=0.5]{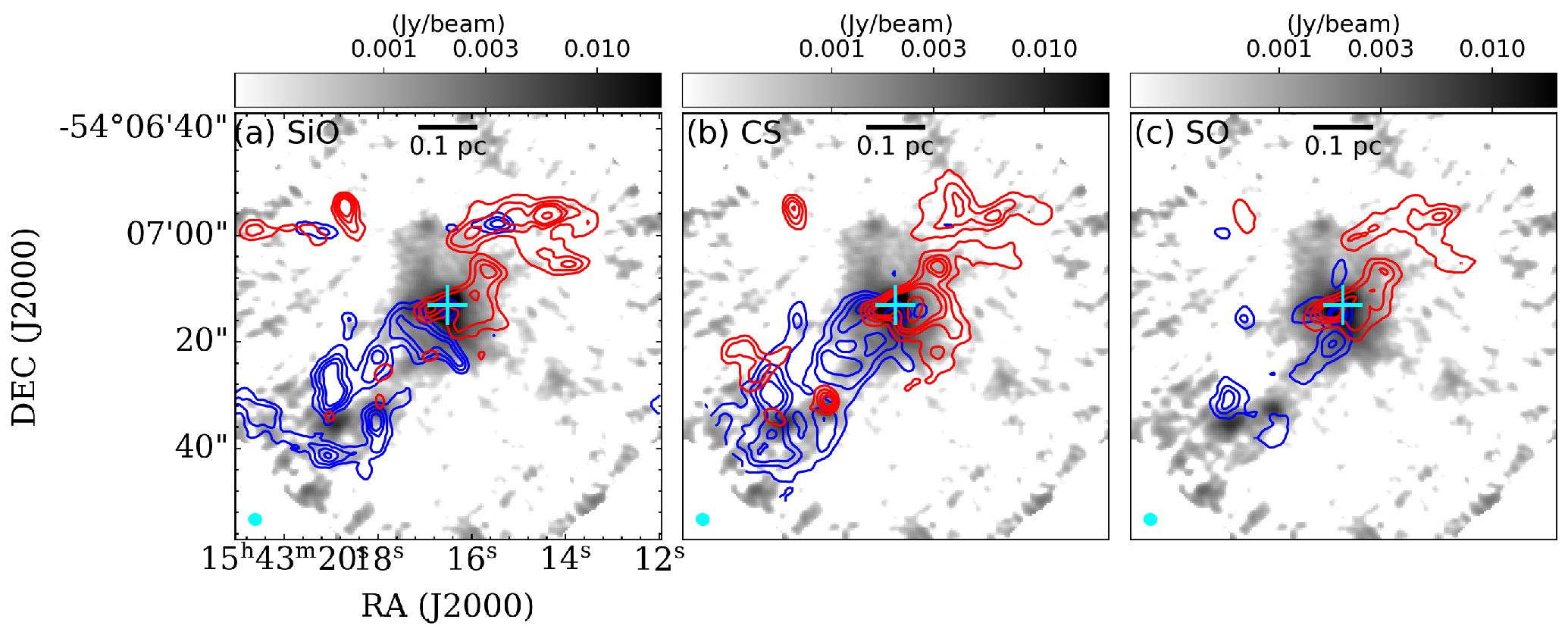}
\caption{The background grayscale represents the 3~mm dust continuum map, on top of which outflows from SiO (a), CS (b), and SO (c) are overlaid. The contour levels are set at 3, 5, 7, 9, 15, and 20 times $\sigma$, where $\rm \sigma = 0.07, 0.17, ~and~ 0.1~(Jy/beam)~km~s^{-1}$ for SiO, CS, and SO, respectively. The `+' mark indicates the location of the central core C-1A.} 
\label{outf_plot}
\end{figure*}

Towards the protocluster, we conducted a search for outflows in the SiO, CS, and SO lines. The blue and red-shifted gas in these lines were integrated within the velocity ranges of [-56.25, -44.87] and [-37.92, -27.17]~$\rm km~s^{-1}$ for SiO, [-49.24, -43.27] and [-37.29, -28.33]~$\rm km~s^{-1}$ for CS, and [-45.57, -42.62] and [-38.20, -33.77]~$\rm km~s^{-1}$ for SO. The outflow lobes identified from these lines are illustrated in Fig. \ref{outf_plot}.

The outflow emissions appear extended and exhibit similar morphology for the SiO and CS lines, whereas the outflow is more compact in SO. It is plausible that the low-level extended emission of SO lies below the sensitivity level of the observations. The outflows extend diagonally in the northwest and southeast directions emanating from the central massive core, with the blue-shifted gas extending in the southeast direction and the red-shifted gas extending in the northwest direction. The detection of outflows from the central core indicates underlying star formation activity.

\subsubsection{Velocity gradient and activity within central hub} \label{feedback}

\begin{figure*}
\includegraphics[scale=0.38]{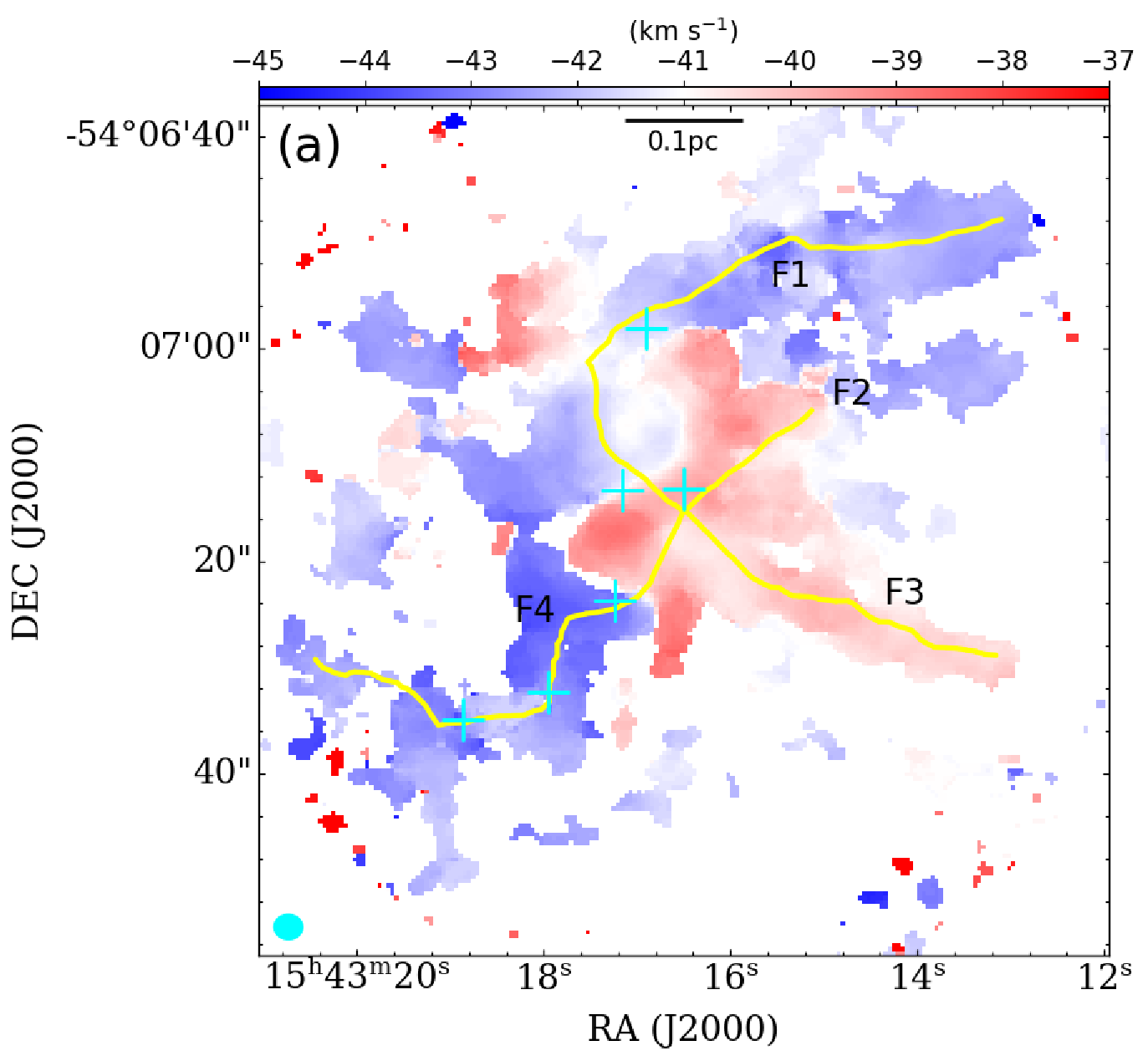}
\includegraphics[scale=0.38]{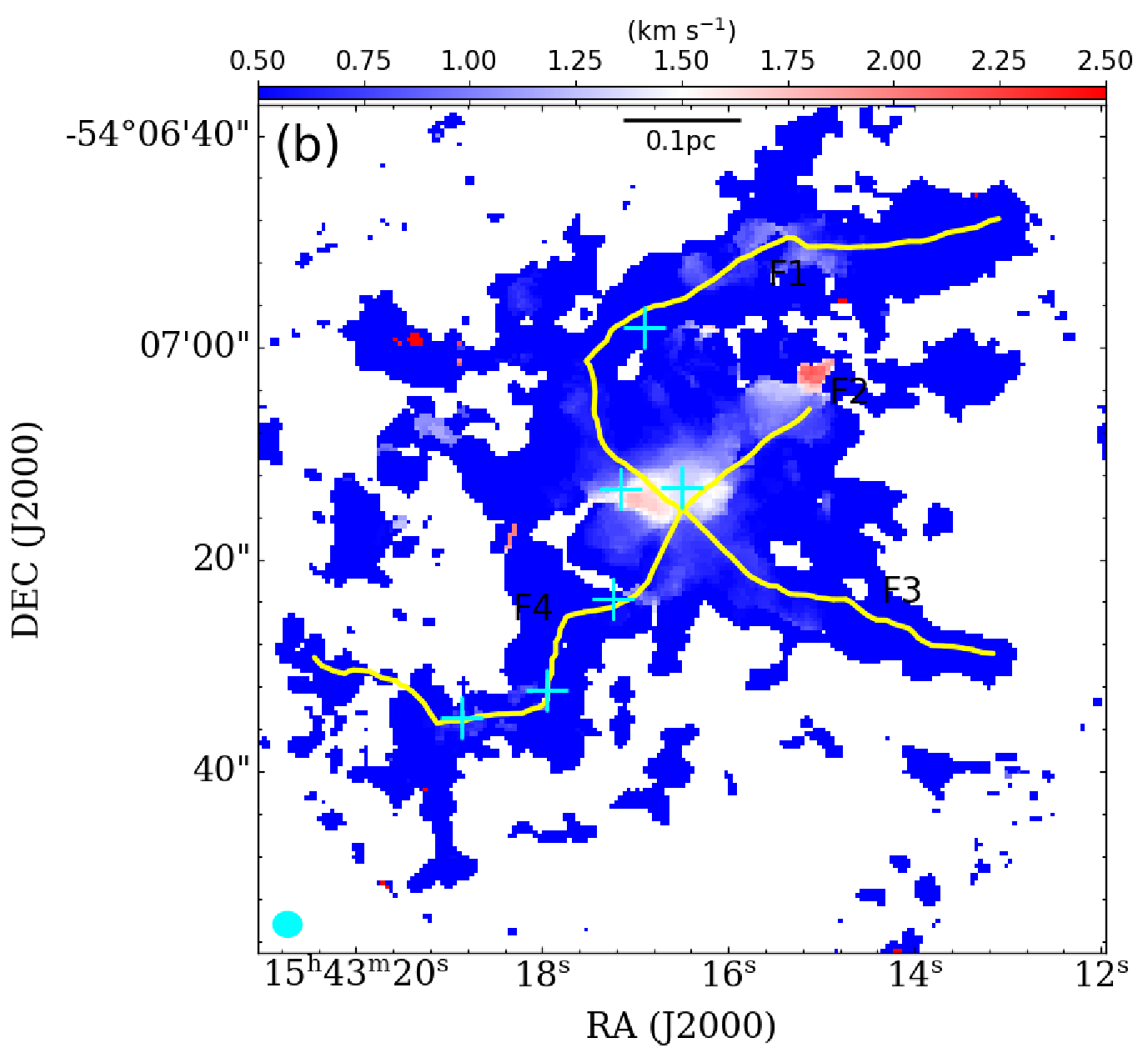}
\caption{Moment1 (a) and moment2 (b) maps of $\rm H^{13}CO^{+}$ line transition. Filaments, along with their labels, are overplotted on the maps. The `+' marks are the locations of the dense cores. Both the moment maps are generated within velocity range of [-48.1, -29.7]~$\rm km~s^{-1}$ and considering pixels above $5\sigma$, where $\rm \sigma=0.009~Jy/beam$. Scale bar and beam size are also shown on the maps.} 
\label{h13cop_mom1_mom2}
\end{figure*}

Fig. \ref{h13cop_mom1_mom2} presents the $\rm H^{13}CO^{+}$ moment1 and moment2 images of IRAS 15394$-$5358. The moment1 map is displayed in Fig. \ref{h13cop_mom1_mom2}(a). From this map, we can see a gradual gradient in velocity along the filamentary structures. The gradient is blue-shifted on the hub's eastern side and red-shifted on its western side. The filaments F1 and F4 are located along the blue-shifted gas, whereas the filaments F2 and F3 lie on the red-shifted gas. So, essentially, the gas is transferred to the hub from its surroundings. We discuss the nature of the velocity gradients along the filaments in section \ref{discuss}.

Fig. \ref{h13cop_mom1_mom2}(b) illustrates the moment2 map. This map exhibits a higher value ($\rm \sim 1.8~km~s^{-1}$) in the hub compared to its surrounding region. The other prominent cores, C-2 and C-3, do not show significantly enhanced linewidths. The increased linewidth of the central core reflects underlying feedback-driven activity. It could also indicate core rotation, not resolved by the current observations. The primary reason behind the enhanced linewidth is likely the energetic outflows and stellar winds from the massive protostars. Additionally, inflows dominated by gravity or turbulence towards the hub could contribute to the higher linewidth values.

The $\rm H_{40\alpha}$ line does not exhibit any emission towards the protocluster. We also searched for radio emission towards the region from surveys such as the 843~MHz SUMSS (Sydney University Molonglo Sky Survey; \citealt{2003MNRAS.342.1117M}) and 21~cm continuum SGPS (The Southern Galactic Plane Survey; \citealt{2005ApJS..158..178M}) and found no emission towards the protocluster. Previous studies also reported non-detection of radio emission towards IRAS 15394$-$5358 \citep{2006ApJ...651..914G}. Also, no luminous young stellar object (YSO) or UC \hii\ region has been detected from these surveys or previous studies towards the region \citep{2021MNRAS.505.2801L,2023MNRAS.520.3245Z}.

The core C-1 is associated with massive star formation and is at an early stage of evolution, as evidenced by the presence of a 6~GHz $\rm CH_3OH$ maser \citep{1998MNRAS.297..215C} and EGOs \citep{2008AJ....136.2391C}. It is likely that the dynamical motion of the inflows driven by gravity/turbulence, along with outflows and stellar winds, are playing a significant role in the enhanced linewidth. The map displays a few pixels with higher values (near filament F2). These higher values appear as red spots in the linewidth map, which might be due to local gas distributions but could also be artifacts.

\subsubsection{Velocity-integrated channel and position-velocity maps} \label{chh_pv}
\begin{figure*}
\includegraphics[scale=1.5]{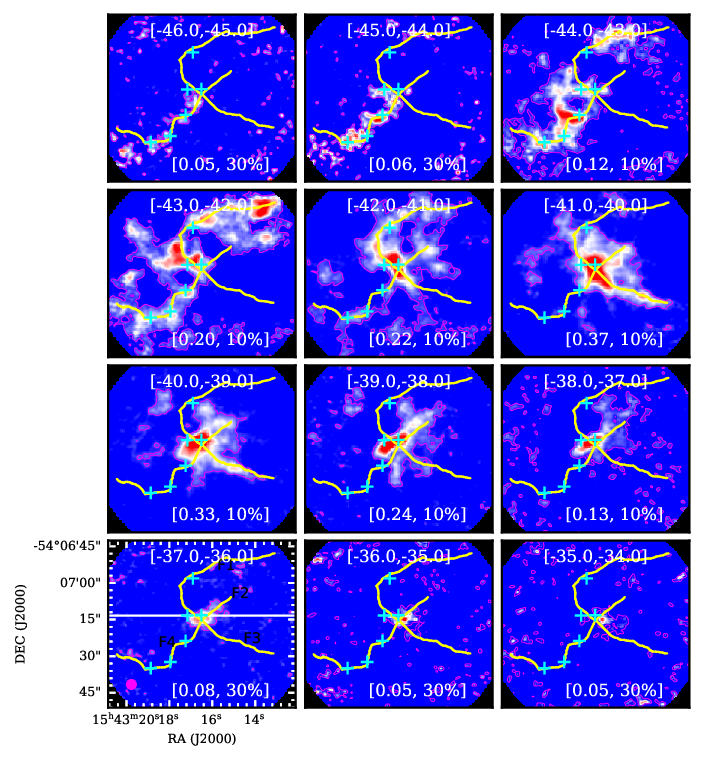}
\caption{Channel maps of $\rm H^{13}CO^{+}$ are presented. Each panel displays the velocity-integrated intensity within a velocity range of $\rm 1~km~s^{-1}$. The numbers in square brackets at the top of each panel indicate the velocity range, while the numbers in square brackets at the bottom right of each panel represent the peak value of the map and the percentage of the peak value shown as contour. The locations of the dense cores and filaments are marked in all panels, as in Fig. \ref{h13cop_mom1_mom2}. The beam size of the map is shown in the bottom left corner of the bottom left panel. The PV-map, generated along the white straight line shown in the bottom left panel, compressed over the full declination range and is presented in Fig. \ref{pv_map}.}
\label{chh_map}
\end{figure*}

For a deeper understanding of the kinematics of the hub-filament system, we generated velocity-integrated channel and position-velocity (PV) maps using the $\rm H^{13}CO^{+}$ line. In Fig. \ref{chh_map}, we present the channel maps generated by integrating the line in steps of $\rm 1~km~s^{-1}$, starting from $\rm -46~km~s^{-1}$ to $\rm -34~km~s^{-1}$. The channel maps provide a detailed view of the kinematic properties of the hub-filament system. Each panel displays the emission within a specific velocity range, capturing the dynamic motions of gas within the system. As the velocity increases from $\rm -46~km~s^{-1}$ to $\rm -34~km~s^{-1}$, we observe distinct shifts in the intensity and spatial distribution of the emission.

At higher velocities (e.g., $\rm -46~kms^{-1}$ to $\rm -44~kms^{-1}$), the intensity is predominantly associated with filament F4, which may imply active gas flows along this structure. Between velocities $\rm -44~km~s^{-1}$ and $\rm -42~km~s^{-1}$, the emission becomes more prominent around filament F1 as well as F4. The emission is dominated by filament F1 within the velocity range of $\rm -42~km~s^{-1}$ to $\rm -41~km~s^{-1}$. The velocity range of $\rm -41~km~s^{-1}$ to $\rm -39~km~s^{-1}$ shows significant emission from filaments F2 and F3. In the lower velocity ranges (e.g., $\rm -39~km~s^{-1}$ to $\rm -34~km~s^{-1}$), the intensity around filaments F2 and F3 starts to fade, and the emission appears as a red spot towards the centre. The significant red regions on the maps highlight areas of higher gas density, many of which are associated with the identified dense cores of the protocluster. These channel maps suggest that the filaments in the protocluster are crucial conduits for channelling gas into the central region and the dense cores, especially filaments F1 and F4. The dense cores, marked by plus signs, are consistently located along these filaments, indicating that they are focal points for gas accumulation and potential star formation.

This kinematic structure illustrates how gas is funnelled along the filaments towards the dense cores, supporting the ongoing star formation processes in the hub-filament system. The observed velocity gradients further corroborate the information regarding velocity gradients seen in the $\rm H^{13}CO^{+}$ moment1 map shown in Fig. \ref{h13cop_mom1_mom2}(a), where filaments F1 and F4 appear blue-shifted, and filaments F2 and F3 are red-shifted with respect to the central hub. This scenario is depicted in Fig. \ref{spec_fila} and discussed in Section \ref{mass_transf}.

\begin{figure}
\centering
\includegraphics[scale=0.7]{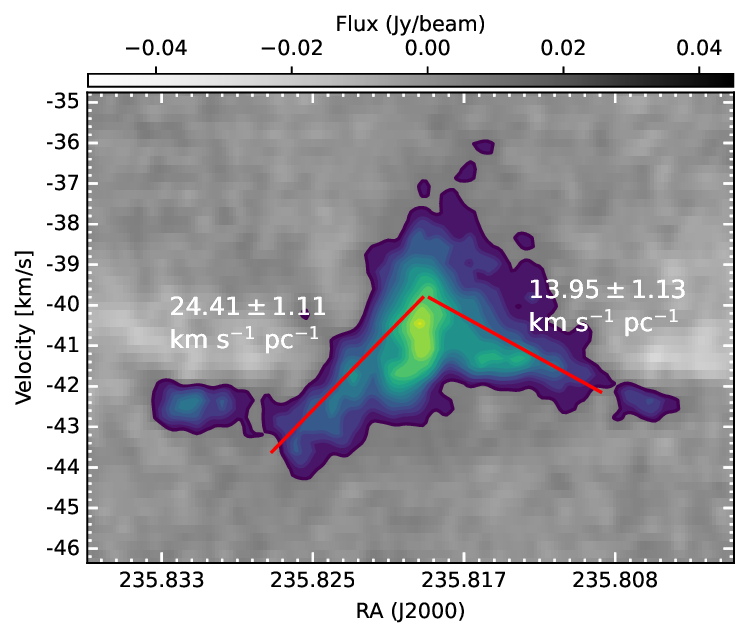}
\caption{The position-velocity diagram of $\rm H^{13}CO^{+}$ is extracted along the white horizontal line shown in Fig. \ref{chh_map} (at $\rm DEC=-54:07:13.80$) and compressed over the entire declination range. The coloured filled contours range from 10\% to 90\% of the peak value (0.045~Jy/beam) in steps of 10\%. The red lines represent the fitted velocity gradients from both the eastern and western sides of the central hub.}
\label{pv_map}
\end{figure}

To further explore the velocity gradient and the funneling of gas towards the central hub, we generated the PV-map along the white horizontal line shown in Fig. \ref{chh_map}, compressing it over the entire declination range. The PV-map, shown in Fig. \ref{pv_map}, provides complete velocity information across the protocluster, spanning the full right ascension range. The PV-map reveals a typical V-shaped structure, usually attributed to strong velocity streams converging towards the central location \citep{2019ApJ...886...15T,2021ApJ...908L..43N,2022A&A...660A..56A}. This V-shaped structure indicates a velocity gradient from both the eastern and western sides towards the central hub. The velocity gradients on the eastern and western sides predominantly correspond to the velocity structures of filaments F4 and F1, respectively. Although filaments F2 and F3, located on the western side of the hub, also show velocity gradients, their gradients are shallower compared to filament F1. The presence of this gradient feature in the protocluster is further discussed in Section \ref{mass_transf} and illustrated in Figures \ref{mom0_vel_cir} and \ref{spec_fila}.

We fitted the two velocity streams and found that the eastern side has a higher gradient of $\rm 24.41\pm1.11~km~s^{-1}pc^{-1}$, while the western side has a lower gradient of $\rm 13.95\pm1.13~km~s^{-1}pc^{-1}$. This is nearly half the gradient observed on the eastern side of the hub. This may imply that the eastern side is more active in funnelling mass to the hub, due to filament F4. This observation is also supported by the discussions in Section \ref{mass_transf} and Fig. \ref{spec_fila}, where it is evident that filament F4 is relatively more blue-shifted and may be more active in channelling gas to the hub compared to the other filaments. The mean velocity gradient towards the hub is $\rm \sim 20~km~s^{-1}pc^{-1}$, facilitating the channelling of gas through the filaments towards the hub, thereby increasing its mass and aiding in the formation of massive star(s) within the hub.

Given the significant mean velocity gradient observed, it can be interpreted as indicative of a globally collapsing scenario in this massive star-forming clump. The velocity gradient observed towards IRAS 15394$-$5358 is comparable to those observed on small scales in massive star-forming regions \citep{2022MNRAS.514.6038Z,2023MNRAS.520.3259X,2023ASPC..534..153H,2024ApJ...961L..35M} and in simulation studies \citep{2020MNRAS.492.1594S}. At such small scales ($\rm <1~pc$), the velocity gradient could primarily be gravity-driven \citep{2022MNRAS.514.6038Z}, leading to mass inflows and variations in density structures due to local conditions. This velocity gradient in the protocluster clearly indicates a globally collapsing scenario, where mass inflow through the filaments is evident (e.g., \citealt{2013A&A...555A.112P}). In such a scenario, the gravity-driven inflows within clumps might become more complex and chaotic, with evolution leading to the mass growth of the associated cores \citep{2024MNRAS.528.1172R}. However, on a larger scale ($\rm >1~pc$), the velocity gradient could be much smaller ($\rm <5~km~s^{-1}pc^{-1}$), showing signatures of global collapse \citep{2022MNRAS.514.6038Z,2023A&A...676A..69Z,2024MNRAS.527.4244S}.

\section{Star formation properties of the protocluster region} \label{star_form_reg}
\subsection{Virial analysis of cores} \label{virial}
We carry out the virial analysis to asses the gravitational stability and kinematics of the cores. We calculated the virial parameter using the expression \citep{1992ApJ...395..140B}:

\begin{equation}
\rm \alpha_{vir} = \frac{5 \sigma^2_{v} R_{core}}{G M_{core}},
\end{equation} 

\noindent where G is the gravitational constant, $\rm \sigma_v$ is the velocity dispersion, $\rm R_{core}$, and $\rm M_{core}$ are the radius and mass of the core, respectively. The above equation neglects the effect of magnetic field. The velocity dispersion is expressed as $\rm \sigma_v = \Delta V/\sqrt{8 ln2}$, where $\rm \Delta V$ is the full width half maximum (FWHM) velocity width. To estimate this parameter, we use the $\rm H^{13}CO^{+}$ (1$-$0) line, which can be optically thin in nature. The FWHM velocity width of the cores is obtained by fitting their $\rm H^{13}CO^{+}$ spectra. The derived value of $\rm \alpha_{vir}$ for all the cores are listed in the last column of Table \ref{core_prop}.

For a core to undergo gravitational collapse, it needs to be subvirial, for which $\rm \alpha_{vir} \leqslant 2$. In contrast, if the core is supported against the gravitational collapse, it must be supervirial. Such cores must expand or might stay confined under the effect of external pressure or magnetic force \citep{2013ApJ...779..185K}. For all cores, the value of $\rm \alpha_{vir}$ lies in the range of $0.01 - 0.94$. This indicates that all the cores are subvirial and gravitationally bound. These cores are expected to collapse leading to the formation of stars.

\subsection{Fragmentation} \label{fragm}
The ALMA 3~mm continuum image reveals the presence of multiple cores within the single ATLASGAL clump. If the fragmentation is happening by the criteria of thermal Jeans instabilities, then the homogenous self-gravitating clump will break into smaller objects defined by the Jeans length ($\rm \lambda_J$) and Jeans mass ($\rm M_J$). These are given by (e.g., \citealt{2014MNRAS.439.3275W,2015MNRAS.453.3785P,2017ApJ...841...97S,2019ApJ...886..102S}).

\begin{equation}
\rm \lambda_J = \sigma_{th} \left(\frac{\pi}{G \rho_{}} \right)^{1/2}~ and ~ M_J = \frac{\pi^{5/2}}{6}~\frac{\sigma^3_{th}}{\sqrt{G^3\rho_{}}},
\end{equation}   

\noindent where $\rm \rho$ is the volume density of the clump derived using the mass and radius values from \cite{2018MNRAS.473.1059U} ,and $\rm \sigma_{th}$ is the thermal velocity dispersion given by

\begin{equation}
\rm \sigma_{th} = \left(\frac{k_B T_{kin}}{\mu m_H}\right)^{1/2},
\end{equation} 

\noindent where $\rm k_B$, $\rm T_{kin}$, $\mu$ and $\rm m_H$ are Boltzmann's constant, kinetic temperature, mean molecular weight, and the mass of hydrogen atom, respectively. We assume the clump's kinetic temperature is the same as the dust temperature of 20~K and $\mu$=2.37 \citep{2008A&A...487..993K,2022MNRAS.510..658P}. Importing other parameters into the above equations, we obtain the thermal Jeans length and Jeans mass to be 0.16~pc and 4.4~$\rm M_{\odot}$, respectively.

It is also essential to consider the non-thermal contribution arising from turbulence in the calculations. To take into account the non-thermal contribution, the thermal velocity dispersion needs to be replaced with total velocity dispersion as follows \citep{2014MNRAS.439.3275W,2019MNRAS.487.1259L,2023ApJ...949..109L}

\begin{equation}
\rm \sigma_{tot} = \left(\sigma^2_{nt} + \sigma^2_{th} \right)^{1/2}, 
\end{equation}

\noindent where the non-thermal velocity dispersion can be derived from the observed lines using the following expression

\begin{equation}
\rm \sigma_{nt} = \sqrt{\sigma^2_{line} - \frac{k_B T_{kin}}{m_{line}}},
\end{equation}

\noindent where, $\rm \sigma_{line}$ is the velocity dispersions observed from the molecular line transition, $\rm m_{line}$ is the molecular mass of the molecule. We use the optically thin $\rm H^{13}CO^{+}$ line to obtain the velocity dispersion $\rm \sigma_{line}$. We determine the velocity dispersion from spectra obtained from a region away from the cores. This is to ensure that the measured velocity dispersion is not affected by the current star formation activity of the complex. We derive the turbulent Jeans length and mass as 0.30~pc and 28~$\rm M_{\odot}$, respectively. Table \ref{jeans_prop} gives all the derived Jeans parameters.

\begin{table}
\small
\centering
\caption{Density of clump and derived Jeans parameters. }
\label{jeans_prop}
\begin{tabular}{ccccc}
\\ \hline

$\rm \rho$ & $\rm \lambda^{Th}_{Jeans}$ & $\rm M^{Th}_{Jeans}$ & $\rm \lambda^{Turb}_{Jeans}$ & $\rm M^{Turb}_{Jeans}$ \\
($\rm g~cm^{-3}$) & (pc) & ($\rm M_{\odot}$) & (pc) & ($\rm M_{\odot}$) \\ 
\hline
$1.4\times10^{-19}$ & 0.16 & 4.4 & 0.30 & 28.0 \\
\hline
\end{tabular}
\\
\end{table}

\subsection{Thermal versus turbulent Jeans fragmentation} \label{ther_non}

\begin{figure}
\centering
\includegraphics[scale=0.47]{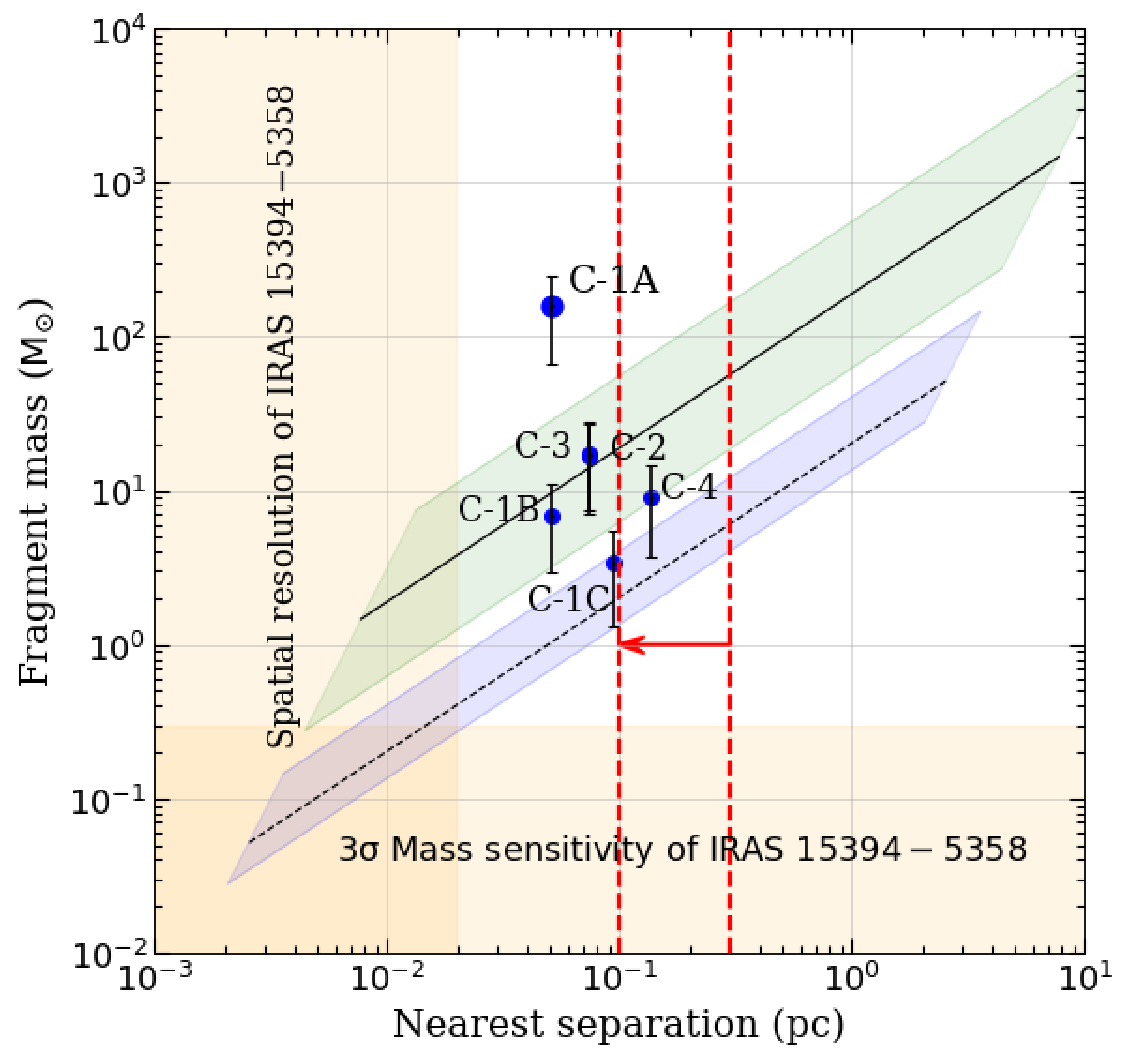}
\caption{Fragment mass vs. the nearest separation. Here, the fragment masses are the core masses. The blue-filled circles represent the associated cores of IRAS 15394$-$5358. The dotted, solid black line and the shaded regions are plotted following the outlines described in \citet{2014MNRAS.439.3275W,2023MNRAS.520.3259X}. Spatial and $\rm 3\sigma$ mass sensitivity of IRAS 15394$-$5358 with ALMA Band-3 observations is shown as an orange-shaded region. The $\rm 3\sigma$ mass sensitivity estimated using the $3\sigma$ flux density and dust temperature of 100~K. The black dotted line is for thermal Jeans fragmentation with T = 15~K and density n = [$\rm 10^2,~10^8~cm^{-3}$]. The blue-shaded region corresponds to the same density range, but for temperatures, T = [10, 30]~K. The solid black line is for the turbulent Jeans fragmentation corresponding to velocity dispersion $\rm \sigma~ =~ 0.7~km~s^{-1}$ and density n = [$\rm 10^2,~10^8~cm^{-3}$]. The green shaded region is for the same density but with velocity dispersion $\rm \sigma~ =~ [0.4,1.2]~km~s^{-1}$. All cores are labelled on the plot. Due to their similar masses, cores C-2 and C-3 overlap with each other. The red vertical dashed lines mark the location of 0.3~pc ($\rm \lambda^{Turb}_{Jeans}$ or Jeans length assuming the cloud fragmented when its density was $\rm 10^4~cm^{-3}$) and 0.1~pc, the median value of the observed core separation, respectively. The red arrow shows the length of 0.2~pc, highlighting the difference.}
\label{mass_mst}
\end{figure}

In this section, we try to understand the dominant mechanism responsible for the fragmentation of the clump into several cores. For this purpose, it is necessary to compare the core separation and masses. Here we apply the minimum spanning tree (MST) method to derive the core separation. Using MST, we can generate a set of straight lines connecting a set of points or nodes. This algorithm generates the straight lines so that the sum of lengths of all straight lines is minimum. In our analysis, the nodes are the locations of cores, and the straight lines connecting the nodes provide spatial separation between the cores. Due to the projection effect, the derived separation between the cores is smaller than the actual separation by a factor of ($\rm 2/\pi$) \citep{2019ApJ...886..102S,2022MNRAS.516.1983S}. We derive the core separations in the range of 0.03$-$0.14~pc with a median value of 0.07~pc. For cores, the mass lies in the range of 3.4$-$157~$\rm M_{\odot}$ with the median value of 13~$\rm M_{\odot}$.

In our case $\rm \lambda_{obs}$ $<$ $\rm \lambda^{Th}_{Jeans}$ $<$ $\rm \lambda^{Turb}_{Jeans}$ and $\rm M^{Th}_{Jeans}$ $<$ $\rm M_{obs}$ $<$ $\rm M^{Turb}_{Jean}$. To compare the thermal and turbulent Jeans fragmentation, \citet{2014MNRAS.439.3275W} have analyzed the core mass distribution with core separation. The same procedure has also been used by \citet{2023MNRAS.520.3259X,Morii}. In this work, we follow the same procedure to compare the thermal and turbulent Jeans fragmentation mechanism. In Fig. \ref{mass_mst}, we show the mass separation distribution for all six cores in this work. The straight black lines (dotted and solid) and the shaded region are plotted following discussions of \citet{2014MNRAS.439.3275W,2023MNRAS.520.3259X}. The caption of the figure has the details of the lines and shaded regions.

The observed core separation matches the thermal Jeans length, whereas the thermal Jeans mass is significantly less than the observed median value. However, it is close to the lowest mass among all cores. 
In Fig. \ref{mass_mst}, C-1A lies above the green shaded region, whereas cores C-2 and C-3 lie within it. Among the other three cores, C-1C is within the blue shaded region, C-4 is very close, and C-1B is nearer when considering its error. However, caution should be taken when interpreting the mechanism responsible for fragmentation. The recent work by \citet{2024arXiv240810406V} suggests that turbulence might not be an effective mechanism to counteract gravity, and to understand the fragmentation mechanism, it is essential to compare the separation between the cores rather than the mass of the cores.

In the protocluster, the observed median separation between the cores is approximately 0.1~pc, whereas it is 0.3~pc as derived from turbulent Jeans fragmentation (see Fig. \ref{mass_mst}). Assuming that fragmentation took place when the cloud was $\rm 10^4~cm^{-3}$ and at 20~K, this yields a Jeans length of approximately 0.3~pc. It is possible that fragmentation occurred at an earlier stage of evolution. To understand this delay, we refer to the discussions in \citet{2024ApJS..270....9X}, which explain that the core separation would tighten over time by a length ($\Delta l$), which is expressed as follows:

\begin{equation} 
\Delta l = \rm 0.088 \times \left(\frac{R_{cl}}{1~pc}\right) \left(\frac{n_{H_2}}{10^4~cm^{-3}}\right)^{0.5} \left(\frac{t_{life}}{5\times10^4~yr}\right)~pc 
\end{equation}

\noindent where, $\rm R_{cl}$ is the clump radius \citep{2018MNRAS.473.1059U}, and $\rm n_{H_2} = \rho/ \mu_{H_2}m_H$ is the number density of the clump, estimated to be $\rm 3\times10^4~cm^{-3}$. Using these values, we derive the value of time, $\rm t_{life}$, over which the cores will tighten by 0.2~pc, to be $\rm 1.3\times10^5~yr$. This is in agreement with the typical statistical lifetime of massive starless clumps \citep{2018ARA&A..56...41M}.

This suggests that in the protocluster, the fragmentation occurred at an earlier evolutionary stage dominated by the thermal Jeans fragmentation mechanism. At a later stage, due to collapse, the cores grew in mass through accretion and became closer, as observed by \citet{2024ApJS..270....9X}. The growth in mass of cores C-2 and C-3 is due to filament F4, which is the major contributor of gas inflow with a higher velocity gradient. However, the extremely high mass of C-1A compared to its neighbours might have resulted from mass accumulation due to the inflow of gas through the filaments, whereas the mass of the other cores may be primarily due to fragmentation alone \citep{2021ApJ...916...13L}. Hence, the discrepancies observed between the observed and the thermal Jeans parameters can be attributed to the delay between the time of fragmentation, when the density was lower, and the time of observation.

In the SDC335 star-forming region, it was mentioned that turbulence dominates the fragmentation of clumps into cores and cores into condensations \citep{2023MNRAS.520.3259X}. Some earlier studies have also observed similar results, where turbulence pressure seems to dominate over thermal pressure for hierarchical fragmentation in protocluster regions \citep{2014MNRAS.439.3275W,2020ApJ...891..113R,2023MNRAS.520.3259X}. Contrary to this, fragmentation driven by thermal pressure in protoclusters has also been observed in several studies \citep{2015MNRAS.453.3785P,2022MNRAS.510..658P,2022MNRAS.516.1983S}.

In a systematic study of a series of protoclusters, \citet{2024ApJS..270....9X} have found that massive protoclusters are in a totally dynamic state, where the cores will keep accreting mass. This will make the observed core mass much larger than expected from the static Jeans fragmentation \citep{2024MNRAS.528.7333L}. So another possibility is that the core is small in mass and keeps accreting mass until it reaches a mass comparable to the turbulent Jeans fragmentation.
Therefore, caution must be taken when interpreting super-Jeans masses as being due to turbulent support, as they may correspond to thermal fragmentation within a globally collapsing clump.

In Fig. \ref{mass_mst}, core C-1A stands out from the rest of the cores due to its higher mass compared to all other cores. Core C-1B is closest to C-1A, separated by 0.05~pc, just above the spatial resolution of IRAS 15394$-$5358. It is expected that the central core C-1A might hold further sub-structures/condensations within itself, similar to SDC335 MM1 \citep{2023MNRAS.520.3259X}, G335-MM1 \citep{2021ApJ...909..199O,2022ApJ...929...68O}. Further high-resolution observations with a spatial resolution of $\sim$0.01~pc would enable to probe of the fragmentation within the core C-1A.

\subsection{Nature of high-mass star formation in the protocluster region} \label{OBstar}
In previous sections, we derived the physical parameters and analyzed the virial state of the cores. All the cores are gravitationally bound and prone to form stars. The core C-1A is the most massive with a mass of {\bf $\rm 157\pm90~M_{\odot}$}, with a radius of 0.04~pc and a surface density of {\bf $\rm 8.6\pm4.9~g~cm^{-2}$.} In this section, we discuss the star formation properties of the cores.

\begin{figure}
\includegraphics[scale=0.47]{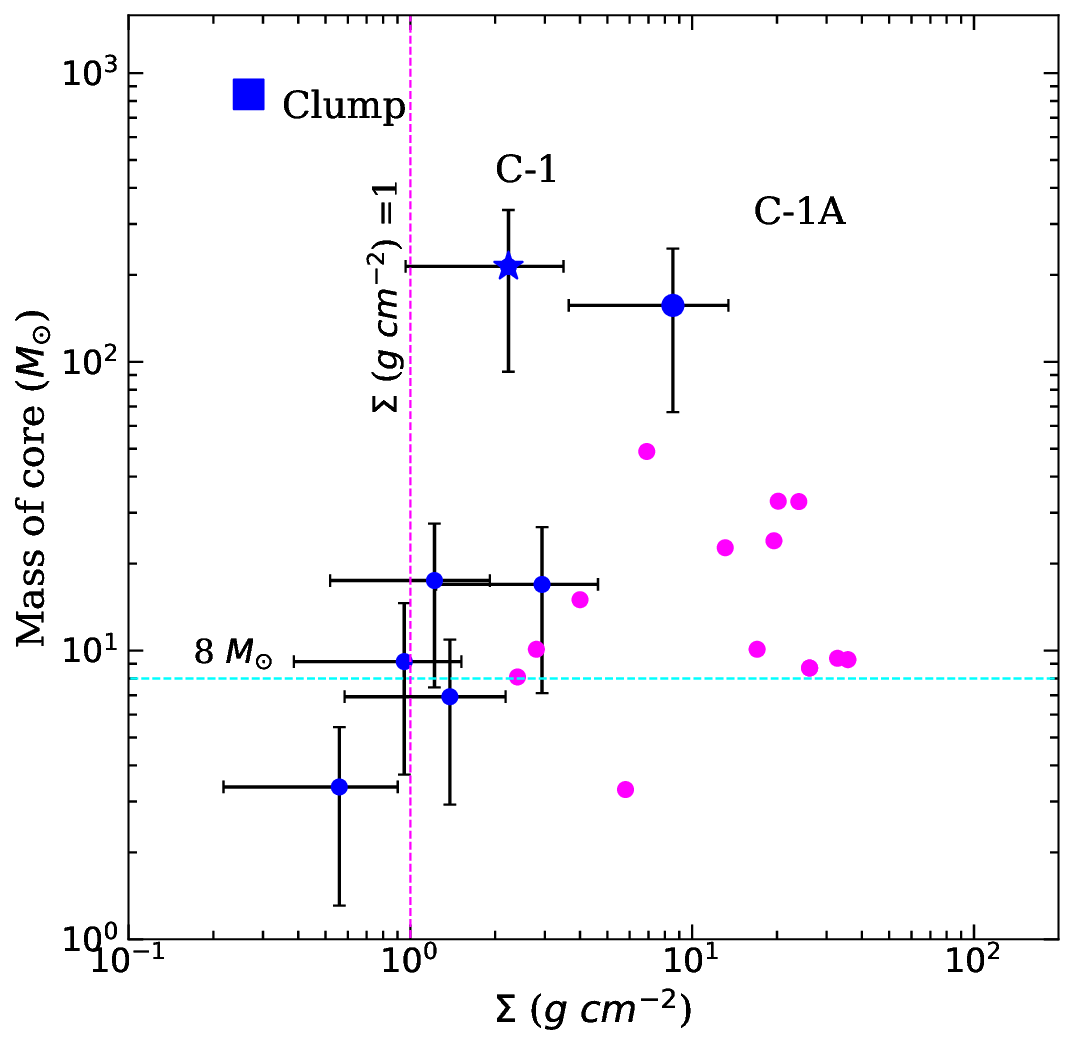}
\caption{Surface density vs. core mass. The blue-filled circles are the cores associated with IRAS 15394$-$5358, and the magenta-filled circles are from \citet{2022MNRAS.516.1983S}. The blue-filled star and square represent the hub C-1 and the whole clump, respectively. The vertical and the horizontal lines are for surface density of $\rm 1~g~cm^{-2}$, and core mass of 8~$\rm M_{\odot}$, respectively.}
\label{mass_sur_den}
\end{figure}

In Fig. \ref{mass_sur_den}, we plot the core masses with respect to their surface density. For comparison, we also plotted the central hub C-1 and the full ATLASGAL clump. We derive the clumps' surface density using the mass and radius values from \citet{2018MNRAS.473.1059U} (see Section \ref{intro}). The clump's surface density (0.27~$\rm g~cm^{-2}$) is less than 1~$\rm g~cm^{-2}$. However, the surface density of the central hub C-1 is similar to the massive core C-1A. Along with the cores associated with IRAS 15394$-$5358 (blue-filled circles), we also plot the cores associated with protoclusters G12.42+0.50 and G19.88$-$0.53 \citep{2022MNRAS.516.1983S} from previous ATOMS studies. Among all the cores, C-1A stands out as the most massive. Out of six cores, two cores (C-1C, ad C-4) have {\bf $\rm \Sigma<1~g~cm^{-2}$.} C-1A is one of the densest and most compact protostellar cores. To analyze the star formation nature of the cores, we compare the mass of cores with their radius.

\begin{figure}
\includegraphics[scale=0.45]{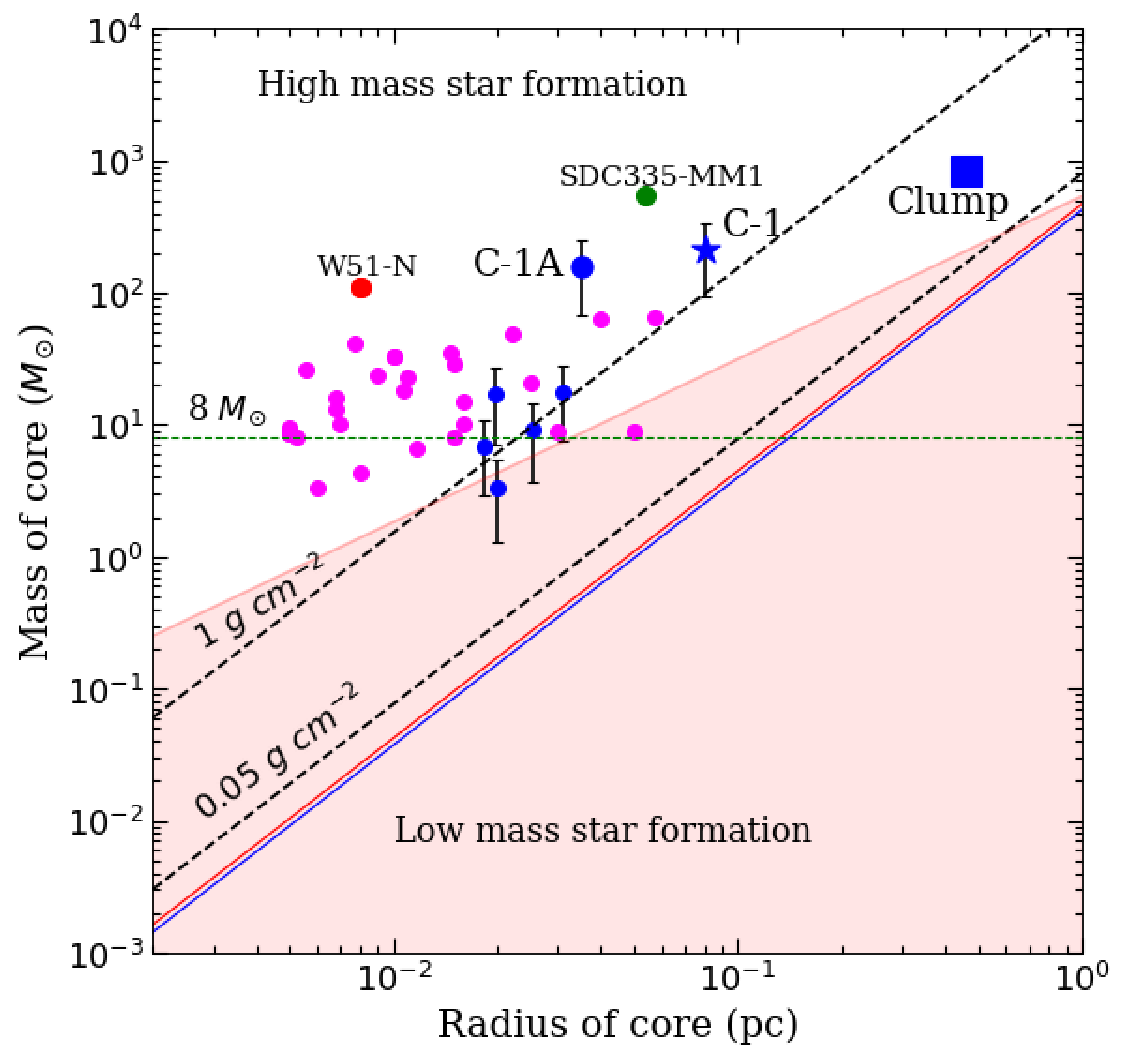}
\caption{Radius vs. mass. The blue-filled circles mark the positions of the cores associated with IRAS 15394$-$5358. The blue-filled star and square represent the hub C-1 and the whole clump. Cores from literature \citep{2007A&A...464..983P,2009ApJ...698.1422Z,2011ApJ...741..120R,2011ApJ...735...64W,2013A&A...555A.112P,2022MNRAS.516.1983S} are also shown in the plot as magenta-filled circles. SDC335-MM1 \citep{2013A&A...555A.112P}, and W51-N \citep{2009ApJ...698.1422Z} are displayed as green and red-filled circles, respectively. The blue and red lines represent the surface density thresholds of 116~$\rm M_{\odot}~pc^{-2}$ ($\rm \sim 0.024~g~cm^{-2}$) and 129~$\rm M_{\odot}~pc^{-2}$ ($\rm \sim 0.027~g~cm^{-2}$) from \citet{2010ApJ...724..687L} and \citet{2010ApJ...723.1019H}, respectively. The shaded region is for the low-mass star formation, which does not satisfy the criteria of $\rm m(r)>870~M_{\odot}~(r/pc)^{1.33}$ \citep{2010ApJ...716..433K}. Surface density thresholds of 0.05~$\rm g~cm^{-2}$ \citep{2008Natur.451.1082K}, and 1~$\rm g~cm^{-2}$ \citep{2014MNRAS.443.1555U} are shown as dashed black lines. The green dashed line is for a core mass of 8~$\rm M_{\odot}$.}
\label{mass_radius}
\end{figure}

In Fig. \ref{mass_radius}, we present the distribution of core mass relative to their radius, accompanied by the central hub C-1, the entire massive clump, and individual cores, offering insights into their star formation characteristics. Notably, the high-mass star formation properties of both C-1 and the entire clump are discernible from the plot. Additionally, for comparison purposes, cores from the literature are included in this figure. Among cores of similar radius, core C-1A stands out by its significant mass. This core ranks among the most massive protostellar cores in the Galaxy. Its mass is comparable to that of SDC335-MM1 \citep{2013A&A...555A.112P} on a larger scale, and on a smaller scale, to the massive cores detected in Cygnus-X \citep{2010A&A...524A..18B}.

We compare the massive star formation properties of the core C-1A with already known massive star-forming cores SDC335-MM1 \citep{2013A&A...555A.112P} and W51-N \citep{2009ApJ...698.1422Z}. SDC335-MM1 has a mass of $\rm \sim 550~M_{\odot}$ within a radius of $\sim 0.05$~pc. Along with other low-mass objects, this core has the potential of forming a massive star of $\rm \sim 50 - 100~M_{\odot}$. The other massive protostellar core W51-N has a mass of 110~$\rm M_{\odot}$. According to \citet{2009ApJ...698.1422Z}, this core is hosting a massive star of $\rm \sim65~M_{\odot}$ with an accreting disc of $\rm \sim40~M_{\odot}$. Compared with these protostellar cores, it is expected that the core C-1A can potentially host a massive O/B-type star(s). Theoretical studies suggest that in the case of compact cores, the amount of matter accreted onto the star is $\sim50\%$ of the core mass \citep{2003ApJ...585..850M}. We have to take into account the uncertainty (see Section \ref{uncer}) involved in mass estimations of the cores. With the uncertainty and also taking into consideration the unresolved components within, it is possible that core C-1A can potentially form a cluster with at least one single massive star with mass greater than $\rm 8-10~M_{\odot}$.

Except for C-1B and C-1C, all other cores associated with IRAS 15394$-$5358 are more massive than 8 $\rm M_{\odot}$. As depicted in Fig. \ref{mass_radius}, all the cores except C-1C lie outside the low-mass star-forming zone and meet the criterion of $\rm m(r)>870M_{\odot}(r/pc)^{1.33}$ \citep{2010ApJ...716..433K}, suggesting that these cores might foster the formation of massive stars within themselves. As discussed earlier, compact cores typically utilise a significant fraction of their mass to form the most massive object. Cores C-2 and C-3 have masses of $\sim$17~$\rm M_{\odot}$. Taking into account the uncertainty in mass along with unresolved components into consideration, it is likely that the cores can potentially form a cluster of low or intermediate mass stars. Similarly, the remaining cores, C-1B, C-1C, and C-4, have masses of 7, 3.4, and 9~$\rm M_{\odot}$, respectively, and it is highly likely that low-mass stars will form within these cores. Consequently, the region of IRAS 15394$-$5358 is likely evolving into a stellar cluster, where the most massive member could potentially be a massive O/B-type star.

\section{Discussion} \label{discuss}
\subsection{Mass transfer to the central hub} \label{mass_transf}

\begin{figure}
\centering
\includegraphics[scale=0.36]{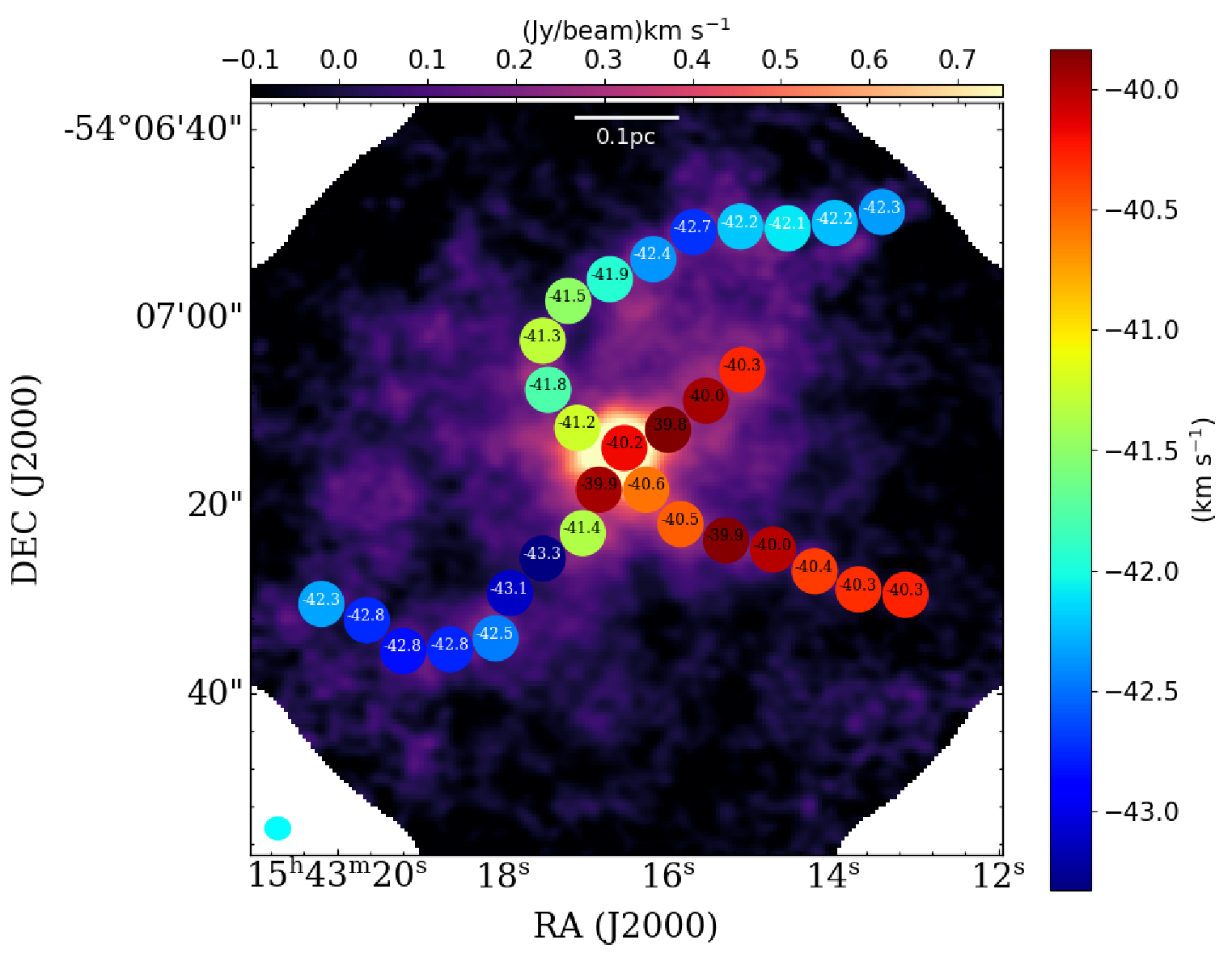}
\caption{The background image is the moment0 map of the $\rm H^{13}CO^{+}$ molecular line transitions. LSR velocities obtained within circular regions ($2.5^{\prime\prime}$), equivalent to the beam size, are overplotted on the image. The colour coding of the circles represents the variation in LSR velocities. The colour bar of the variation is also shown in the plot. The scale bar is shown on the top, and the beam size is shown in the bottom left corner. }
\label{mom0_vel_cir}
\end{figure}

We detect a network of four filaments extending from the massive central core, thus creating a hub-filament system. As discussed earlier, in such a system, the hub serves as a progenitor for the formation of massive stars, where the filaments help in funnelling matter into the hub \citep{2013A&A...555A.112P,2019ApJ...875...24C,2019ApJ...877..114C,2021ApJ...919....3C,2023MNRAS.522.3719L}. Here, we attempt to study any such mass infall into the central hub through the filaments.

\begin{figure}
\centering
\includegraphics[scale=0.6]{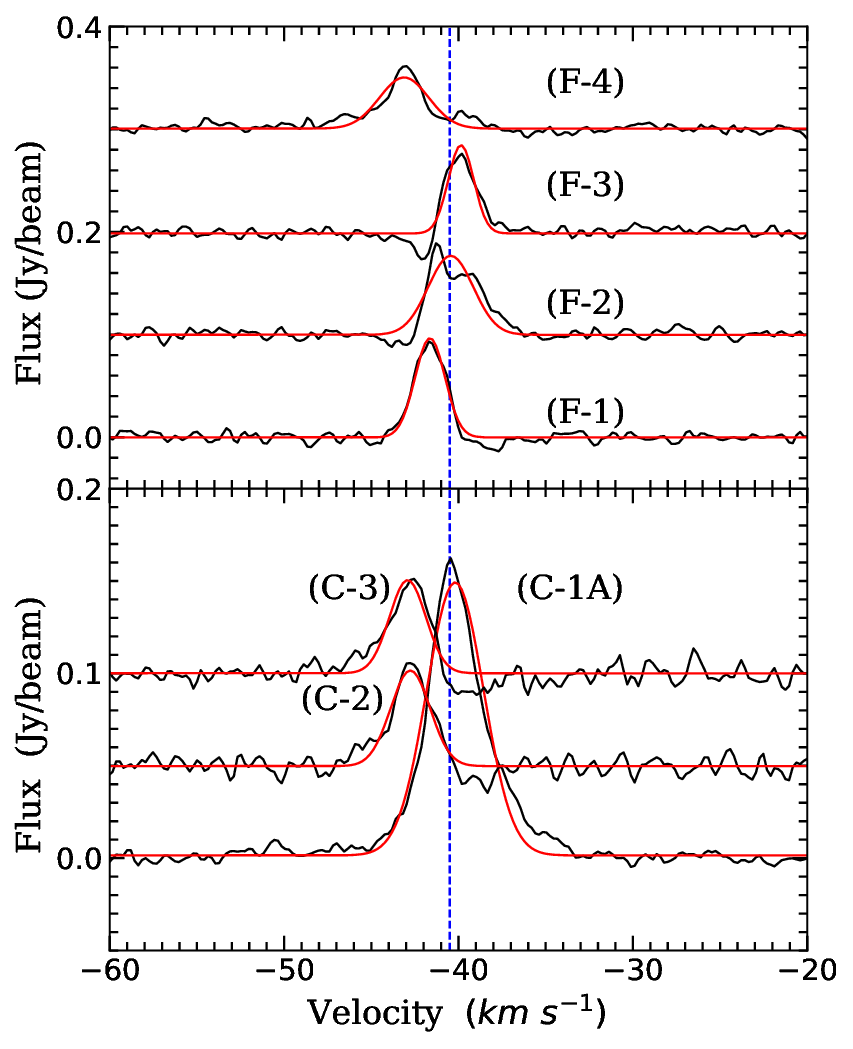}
\caption{$\rm H^{13}CO^{+}$ average spectra for the cores C-1A, C-2, and C-3 and over all filaments. The spectra are taken within the boxes shown in Fig. \ref{fila_box}. Gaussian fits to all spectra are shown as red curves. Spectras of C-2, C-3, and filaments F2 to F4 are scaled up for a better representation. The vertical blue dashed line shows the LSR velocity of the central core. }
\label{spec_fila}
\end{figure}

We observe a velocity gradient in the protocluster region (Section \ref{feedback} and \ref{chh_pv}). To further explore the velocity gradient along the filaments, we extract $\rm H^{13}CO^{+}$ average spectra within circular regions (2.5$^{\prime\prime}$) equivalent to the beam size. In Fig. \ref{mom0_vel_cir}, we provide the LSR velocity values obtained from the $\rm H^{13}CO^{+}$ average spectra within the circular regions. The LSR velocities' distribution also reveals the velocity gradient along the filaments.    
Also, we extract spectra within the rectangular boxes shown in Fig. \ref{fila_box} to analyze such infall signatures within the region. In Fig. \ref{spec_fila} we show the $\rm H^{13}CO^{+}$ spectral profile of cores C-1A, C-2, and C-3. Also, we plot the average spectra taken over rectangular regions on all the filaments. It is interesting to notice that the velocity of filaments F1 and F4 are shifted to the lower side of C-1A, or they are blue-shifted, and the filaments F2 and F3 are shited to the higher side of C-1A or they are red-shifted. The shift ($\rm \sim 3~km~s^{-1}$) of filament F4 is highest compared to all other filaments. The LSR velocities of the filaments F1 and F4 differ by $\rm \sim 2-3~km~s^{-1}$ from the systemic velocity, whereas it is $\rm \sim1~km~s^{-1}$ for filaments F2 and F3. This suggests that more matter might be funnelled to the hub by F1 and F4 compared to the filaments F2 and F3. This further supports the fact that the observed velocity gradient is due to gas inflow to the central hub, as indicated by the PV-map (Fig. \ref{pv_map}). In the protocluster region SDC 335, \cite{2013A&A...555A.112P} have observed a similar mass inflow to the hub from its surroundings and with high-resolution data \cite{2023MNRAS.520.3259X} observed the presence of condensations within the massive core MM1 of the source SDC 335. All condensations can grow due to the mass infall into the central hub. In our case, further deep, higher-resolution data would enable us to observe the presence of condensations within the central hub C-1/core C-1A and their mass accretion features.

It is clear that an active mass transfer is happening to the central hub C-1 from its surroundings. This is also evident in moment2 map (Fig. \ref{h13cop_mom1_mom2}(b)), where the central hub region has higher velocity dispersion compared to its surroundings. The value is $\rm \sim2~km~s^{-1}$, and we assume it as the infall rate ($\rm V_{in}$) to make an empirical estimate of mass accretion rate in the protocluster region of IRAS 15394$-$5358. 
Using the expression $\rm \dot{M}_{in} = 3V^3_{in}/2G$ \citep{2012A&A...544L...7P}, we obtain the mass infall rate to be $\rm 3\times10^{-3}~M_{\odot}~yr^{-1}$. Assuming a typical free-fall time scale of $\rm 5\times10^4~yr$ \citep{2019ApJ...887..108R}, the mass infall rate can collect matter of $\rm 150~M_{\odot}$ in the central hub. This accretion is enough to gather matter similar to the mass of hub C-1 over a few free-fall times. This suggests that most of the central hub's mass is collected through a large-scale molecular cloud collapse over a few free-fall times. A simulation study by \citet{2010A&A...520A..49S} also shows that such mass accumulation to the central hub is possible through a large parsec-scale molecular cloud collapse. Note that a better estimation of infall velocity is essential to derive the mass accretion rate in the complex.

\subsection{Driving mechanism for massive star formation in IRAS 15394$-$5358} \label{drive_model}
In the introduction, we discuss the several models proposed to understand the massive star formation process. Our kinematic analysis using the ATOMS data helps to shed light on the possible driving mechanism at work for massive star formation in the protocluster complex IRAS 15394$-$5358. A hub-filament system is seen in the protocluster complex, where four filaments are linked with the hub. A velocity gradient is seen from the surroundings towards the centre, indicating a collective mass inflow towards the hub. In this section, we discuss the possible mechanisms at work for massive star formation in the protocluster.

All the dense cores of the protocluster complex have low virial parameters ($\rm \alpha_{vir}<1$), suggesting they are gravitationally bound in nature. Competitive accretion would be possible for a case of $\rm \alpha_{vir}<1$ \citep{2005Natur.438..332K}. Also, this model requires global hierarchical collapse \citep{2011MNRAS.411...65B,2017MNRAS.467.1313V} assuming initial fragmentation of a molecular cloud into Jeans core masses. The cores near the gravitational potential minimum are favoured with higher mass accretion. In the protocluster, the most massive core (C-1A) is present within the hub, and its location could be the primary reason for its higher mass compared to other dense cores of the complex. Also, it is observed that the other two dense cores, C-2 and C-3, with masses of $\sim$17~$\rm M_\odot$, are present on the filament F4, which has the maximum velocity gradient among all the filaments. This higher velocity gradient might have helped to form these cores by supplying enough material. This morphology also supports the role of the GHC model, where multiscale infall motions are possible, with varying velocity gradients along filaments at different scales. The IF model suggests large-scale, converging, inertial flows occurring due to turbulence. Both the GHC and IF models predict very small velocity gradients {\bf ($\rm <~5~km~s^{-1}~pc^{-1}$)} at large scales ($>$ 1~pc). A recent study of 26 IRDC clouds reports that the majority of these clouds are self-gravitating, ranging from tenths of a parsec to several tens of parsecs \citep{2023MNRAS.525.2935P}, thereby questioning the importance of IF model in star formation. However, further studies are needed to deepen our understanding of the role of these models in massive star formation.

At scales smaller than $\rm \sim 1~pc$ in the hub-filament systems, the longitudinal flow along the filaments can be attributed to GHC model. At such small scales, it is more likely that the infall will be primarily dominated by the gravity of the hub \citep{2022MNRAS.514.6038Z}. Similar results were also seen in a few previous studies \citep{2022MNRAS.510.5009L,2022MNRAS.511.4480L,2022MNRAS.516.1983S,2024ApJ...960...76P}. In IRAS 15394$-$5358, the velocity gradient observed is $\rm \sim 20~km~s^{-1}~pc^{-1}$, comparable with the values observed in massive star-forming regions at smaller scales ($<$ 1~pc). This suggests competitive accretion, along with the GHC model, could be playing significant roles in the star-formation activity within the protocluster. Due to the large statistically significant sample size, the ATOMS survey can help to gain more knowledge regarding inflow properties and the role of gravity/turbulence in massive star formation by characterizing the gas environment of hub-filament systems.

\section{Summary} \label{summ}
Using the dust continuum and multiple molecular line transitions data from the ATOMS survey, we conducted a detailed analysis of the massive protocluster IRAS 15394$-$5358, where clump fragmentation and indications of massive star formation through a large-scale collapse have been observed. Significant findings of this work are the following.

\begin{enumerate}
\item The 3~mm dust continuum map shows the fragmentation of the massive clump, revealing six cores within it. The dense central part, called core C-1, with a derived mass of $\rm 214~M_{\odot}$, has three substructures called cores C-1A, C1-B, and C-1C. There are three other cores labelled as C-2, C-3, and C-4. Among all the cores, C-1A is the largest (radius = 0.04~pc), massive ($\rm 157~M_{\odot}$) and dense ($\rm \Sigma=8.6~g~cm^{-2}$).

\item Two $\rm 24~\mu m$ blobs are seen towards the massive clump. The dense central part is associated with the less bright $\rm 24~\mu m$ blob, EGOs, and a methanol maser. The brighter $\rm 24~\mu m$ blob is associated with cores C-2 and C-3. This suggests that the dense central region hosts massive star formation and is likely at an early evolutionary stage compared to the surrounding cores.

\item Virial analysis reveals that all the cores are gravitationally bound. Our analysis shows that the protocluster region's fragmentation might have occured at an earlier time following a thermal Jeans fragmentation process. Except for cores C-1B, C-1C and C-4, all other cores might form a massive star. The massive core C-1A is prone to form a single massive O/B-type star.

\item The multiple line transitions of the protocluster region display different morphological features. Our PCA analysis shows that the molecular transitions can be broadly segregated into three groups. The first group contains $\rm H^{13}CN,~ H^{13}CO^{+},~ HC_3N,~ and ~SO$, the second group contains $\rm HCN,~ HCO^{+},~ CCH,~ SiO, ~ and ~ CS$, and the third group includes $\rm CH_3OH$.

\item Using the $\rm H^{13}CO^{+}$ line, a hub-filament system was identified in IRAS 15394$-$5358, where the central hub is linked with four filaments labelled as F1 to F4. The $\rm H^{13}CO^+$ moment1 and moment2 reveal the velocity gradient from the surrounding toward the hub. An average velocity gradient of $\rm \sim 20~km~s^{-1}~pc^{-1}$ is observed towards the hub from the surrounding regions. This supports a global collapsing scenario, where the filaments, primarily F1 and F4, are contributing to the hub through gas inflow. $\rm H^{13}CO^+$ moment2 map reveals an enhanced line-width value ($\rm \sim 1.8~km~s^{-1}$) towards the hub, suggesting the ongoing star-forming activity in the dense central region.

\item Assuming the infall velocity to be $\rm 2~km~s^{-1}$ same as the velocity dispersion in the central hub, we derive an empirical mass accretion rate of $\rm 3\times10^{-3}~M_{\odot}~yr^{-1}$. This accretion rate would be enough to gather mass equivalent to the hub (C-1) in a few free-fall times. Considering the morphology of the dense cores distribution and velocity gradient, we speculate that competitive accretion along with GHC model might be at work for star formation activity in IRAS 15394$-$5358.

\end{enumerate}

\addcontentsline{toc}{section}{Acknowledgements}
\section*{Acknowledgements}
We thank the anonymous referee for a constructive review of the manuscript, which helped in improving the quality of the paper. SRD acknowledges support from Fondecyt Postdoctoral fellowship (project code 3220162). SRD, MM, LB, GG, DM, and AS gratefully acknowledges support by the ANID BASAL project FB210003. SRD acknowledges membership in the BRICS-SO. AS gratefully acknowledges support by the Fondecyt Regular (project code 1220610). PS was partially supported by a Grant-in-Aid for Scientific Research (KAKENHI Number JP22H01271 and JP23H01221) of JSPS. H.-L. Liu is supported by Yunnan Fundamental Research Project (grant No. 202301AT070118, 202401AS070121). EVS, GCG and AP acknowledge financial support from the UNAM-PAPIIT IG100223 grant. AP acknowledges support from the Sistema Nacional de Investigadores of CONAHCyT, and from the CONAHCyT project number 86372 of the `Ciencia de Frontera 2019’ program, entitled `Citlalc\'oatl: A multiscale study at the new frontier of the formation and early evolution of stars and planetary systems’, M\'exico. JH and LZ acknowledge that the work is sponsored (in part) by the Chinese Academy of Sciences (CAS), through a grant to the CAS South America Center for Astronomy (CASSACA) in Santiago, Chile. MJ acknowledges the support of the Research Council of Finland Grant No. 348342. This paper makes use of the following ALMA data: ADS/JAO.ALMA\#2019.1.00685.S. ALMA is a partnership of ESO (representing its member states), NSF (USA), and NINS (Japan), together with NRC (Canada), MOST and ASIAA (Taiwan), and KASI (Republic of Korea), in cooperation with the Republic of Chile. The Joint ALMA Observatory is operated by ESO, AUI/NRAO, and NAOJ. This research has made use of the SIMBAD database, operated at CDS, Strasbourg, France. This work made use of various packages of Python programming language.
\section*{Data Availability}
The data underlying this article can be shared on reasonable request to the PI of the ATOMS survey, Tie Liu (liutiepku@gmail.com).


\bibliographystyle{mn2e}
\bibliography{refer} 

\begin{thebibliography}{133}
\expandafter\ifx\csname natexlab\endcsname\relax\def\natexlab#1{#1}\fi

\bibitem[{{Anderson} {et~al}\mbox{.}(2021){Anderson}, {Peretto}, {Ragan},
  {Rigby}, {Avison}, {Duarte-Cabral}, {Fuller}, {Shirley}, {Traficante}, \&
  {Williams}}]{2021MNRAS.508.2964A}
{Anderson} M. {et~al.}, 2021, \mnras, 508, 2964

\bibitem[{{Arzoumanian} {et~al}\mbox{.}(2022){Arzoumanian}, {Russeil},
  {Zavagno}, {Chun-Yuan Chen}, {Andr{\'e}}, {Inutsuka}, {Misugi},
  {S{\'a}nchez-Monge}, {Schilke}, {Men'shchikov}, \&
  {Kohno}}]{2022A&A...660A..56A}
{Arzoumanian} D. {et~al.}, 2022, \aap, 660, A56

\bibitem[{{Ballesteros-Paredes} {et~al}\mbox{.}(2011){Ballesteros-Paredes},
  {Hartmann}, {V{\'a}zquez-Semadeni}, {Heitsch}, \&
  {Zamora-Avil{\'e}s}}]{2011MNRAS.411...65B}
{Ballesteros-Paredes} J., {Hartmann} L.~W., {V{\'a}zquez-Semadeni} E.,
  {Heitsch} F., {Zamora-Avil{\'e}s} M.~A., 2011, \mnras, 411, 65

\bibitem[{{Baug} {et~al}\mbox{.}(2018){Baug}, {Dewangan}, {Ojha}, {Tachihara},
  {Pandey}, {Sharma}, {Tamura}, {Ninan}, \& {Ghosh}}]{2018ApJ...852..119B}
{Baug} T. {et~al.}, 2018, \apj, 852, 119

\bibitem[{{Benjamin} {et~al}\mbox{.}(2003){Benjamin}, {Churchwell}, {Babler},
  {Bania}, {Clemens}, {Cohen}, {Dickey}, {Indebetouw}, {Jackson}, {Kobulnicky},
  {Lazarian}, {Marston}, {Mathis}, {Meade}, {Seager}, {Stolovy}, {Watson},
  {Whitney}, {Wolff}, \& {Wolfire}}]{2003PASP..115..953B}
{Benjamin} R.~A. {et~al.}, 2003, \pasp, 115, 953

\bibitem[{{Bertoldi} \& {McKee}(1992)}]{1992ApJ...395..140B}
{Bertoldi} F., {McKee} C.~F., 1992, \apj, 395, 140

\bibitem[{{Bonnell} {et~al}\mbox{.}(2001){Bonnell}, {Bate}, {Clarke}, \&
  {Pringle}}]{2001MNRAS.323..785B}
{Bonnell} I.~A., {Bate} M.~R., {Clarke} C.~J., {Pringle} J.~E., 2001, \mnras,
  323, 785

\bibitem[{{Bonnell}, {Vine} \& {Bate}(2004){Bonnell}, {Vine}, \&
  {Bate}}]{2004MNRAS.349..735B}
{Bonnell} I.~A., {Vine} S.~G., {Bate} M.~R., 2004, \mnras, 349, 735

\bibitem[{{Bontemps} {et~al}\mbox{.}(2010){Bontemps}, {Motte}, {Csengeri}, \&
  {Schneider}}]{2010A&A...524A..18B}
{Bontemps} S., {Motte} F., {Csengeri} T., {Schneider} N., 2010, \aap, 524, A18

\bibitem[{{Bronfman}, {Nyman} \& {May}(1996){Bronfman}, {Nyman}, \&
  {May}}]{1996A&AS..115...81B}
{Bronfman} L., {Nyman} L.~A., {May} J., 1996, \aaps, 115, 81

\bibitem[{{Caratti o Garatti} {et~al}\mbox{.}(2015){Caratti o Garatti},
  {Stecklum}, {Linz}, {Garcia Lopez}, \& {Sanna}}]{2015A&A...573A..82C}
{Caratti o Garatti} A., {Stecklum} B., {Linz} H., {Garcia Lopez} R., {Sanna}
  A., 2015, \aap, 573, A82

\bibitem[{{Caswell}(1998)}]{1998MNRAS.297..215C}
{Caswell} J.~L., 1998, \mnras, 297, 215

\bibitem[{{Chen} {et~al}\mbox{.}(2019){Chen}, {Zhang}, {Wright}, {Busquet},
  {Lin}, {Liu}, {Olguin}, {Sanhueza}, {Nakamura}, {Palau}, {Ohashi},
  {Tatematsu}, \& {Liao}}]{2019ApJ...875...24C}
{Chen} H.-R.~V. {et~al.}, 2019, \apj, 875, 24

\bibitem[{{Chung} {et~al}\mbox{.}(2021){Chung}, {Lee}, {Kim}, {Gopinathan},
  {Tafalla}, {Caselli}, {Myers}, {Liu}, {Yoo}, {Kim}, {Kim}, {Soam}, {Cho},
  {Kwon}, {Lee}, \& {Kang}}]{2021ApJ...919....3C}
{Chung} E.~J. {et~al.}, 2021, \apj, 919, 3

\bibitem[{{Chung} {et~al}\mbox{.}(2019){Chung}, {Lee}, {Kim}, {Kim}, {Caselli},
  {Tafalla}, {Myers}, {Soam}, {Liu}, {Gopinathan}, {Kim}, {Kim}, {Kwon},
  {Kang}, \& {Lee}}]{2019ApJ...877..114C}
{Chung} E.~J. {et~al.}, 2019, \apj, 877, 114

\bibitem[{{Contreras} {et~al}\mbox{.}(2018){Contreras}, {Sanhueza}, {Jackson},
  {Guzm{\'a}n}, {Longmore}, {Garay}, {Zhang}, {Nguyễn-Lu'o'ng}, {Tatematsu},
  {Nakamura}, {Sakai}, {Ohashi}, {Liu}, {Saito}, {Gomez}, {Rathborne}, \&
  {Whitaker}}]{2018ApJ...861...14C}
{Contreras} Y. {et~al.}, 2018, \apj, 861, 14

\bibitem[{{Cyganowski} {et~al}\mbox{.}(2008){Cyganowski}, {Whitney}, {Holden},
  {Braden}, {Brogan}, {Churchwell}, {Indebetouw}, {Watson}, {Babler},
  {Benjamin}, {Gomez}, {Meade}, {Povich}, {Robitaille}, \&
  {Watson}}]{2008AJ....136.2391C}
{Cyganowski} C.~J. {et~al.}, 2008, \aj, 136, 2391

\bibitem[{{Devereux} \& {Young}(1990)}]{1990ApJ...359...42D}
{Devereux} N.~A., {Young} J.~S., 1990, \apj, 359, 42

\bibitem[{{Dewangan} {et~al}\mbox{.}(2020){Dewangan}, {Ojha}, {Sharma},
  {Palacio}, {Bhadari}, \& {Das}}]{2020ApJ...903...13D}
{Dewangan} L.~K., {Ojha} D.~K., {Sharma} S., {Palacio} S.~d., {Bhadari} N.~K.,
  {Das} A., 2020, \apj, 903, 13

\bibitem[{{Draine}(2011)}]{2011piim.book.....D}
{Draine} B.~T., 2011, {Physics of the Interstellar and Intergalactic Medium}

\bibitem[{{Dunham} {et~al}\mbox{.}(2008){Dunham}, {Crapsi}, {Evans}, {Bourke},
  {Huard}, {Myers}, \& {Kauffmann}}]{2008ApJS..179..249D}
{Dunham} M.~M., {Crapsi} A., {Evans}, Neal~J. I., {Bourke} T.~L., {Huard}
  T.~L., {Myers} P.~C., {Kauffmann} J., 2008, \apjs, 179, 249

\bibitem[{{Elia} {et~al}\mbox{.}(2017){Elia}, {Molinari}, {Schisano},
  {Pestalozzi}, {Pezzuto}, {Merello}, {Noriega-Crespo}, {Moore}, {Russeil},
  {Mottram}, {Paladini}, {Strafella}, {Benedettini}, {Bernard}, {Di Giorgio},
  {Eden}, {Fukui}, {Plume}, {Bally}, {Martin}, {Ragan}, {Jaffa}, {Motte},
  {Olmi}, {Schneider}, {Testi}, {Wyrowski}, {Zavagno}, {Calzoletti},
  {Faustini}, {Natoli}, {Palmeirim}, {Piacentini}, {Piazzo}, {Pilbratt},
  {Polychroni}, {Baldeschi}, {Beltr{\'a}n}, {Billot}, {Cambr{\'e}sy},
  {Cesaroni}, {Garc{\'\i}a-Lario}, {Hoare}, {Huang}, {Joncas}, {Liu}, {Maiolo},
  {Marsh}, {Maruccia}, {M{\`e}ge}, {Peretto}, {Rygl}, {Schilke}, {Thompson},
  {Traficante}, {Umana}, {Veneziani}, {Ward-Thompson}, {Whitworth}, {Arab},
  {Bandieramonte}, {Becciani}, {Brescia}, {Buemi}, {Bufano}, {Butora},
  {Cavuoti}, {Costa}, {Fiorellino}, {Hajnal}, {Hayakawa}, {Kacsuk}, {Leto}, {Li
  Causi}, {Marchili}, {Martinavarro-Armengol}, {Mercurio}, {Molinaro},
  {Riccio}, {Sano}, {Sciacca}, {Tachihara}, {Torii}, {Trigilio}, {Vitello}, \&
  {Yamamoto}}]{2017MNRAS.471..100E}
{Elia} D. {et~al.}, 2017, \mnras, 471, 100

\bibitem[{{Fa{\'u}ndez} {et~al}\mbox{.}(2004){Fa{\'u}ndez}, {Bronfman},
  {Garay}, {Chini}, {Nyman}, \& {May}}]{2004A&A...426...97F}
{Fa{\'u}ndez} S., {Bronfman} L., {Garay} G., {Chini} R., {Nyman} L.~{\r{A}}.,
  {May} J., 2004, \aap, 426, 97

\bibitem[{{Garay} {et~al}\mbox{.}(2006){Garay}, {Brooks}, {Mardones}, \&
  {Norris}}]{2006ApJ...651..914G}
{Garay} G., {Brooks} K.~J., {Mardones} D., {Norris} R.~P., 2006, \apj, 651, 914

\bibitem[{{Garay} {et~al}\mbox{.}(2007){Garay}, {Mardones}, {Brooks}, {Videla},
  \& {Contreras}}]{2007ApJ...666..309G}
{Garay} G., {Mardones} D., {Brooks} K.~J., {Videla} L., {Contreras} Y., 2007,
  \apj, 666, 309

\bibitem[{{Ginsburg} {et~al}\mbox{.}(2022){Ginsburg}, {Sokolov}, {de
  Val-Borro}, {Rosolowsky}, {Pineda}, {Sip{\H{o}}cz}, \&
  {Henshaw}}]{2022AJ....163..291G}
{Ginsburg} A., {Sokolov} V., {de Val-Borro} M., {Rosolowsky} E., {Pineda}
  J.~E., {Sip{\H{o}}cz} B.~M., {Henshaw} J.~D., 2022, \aj, 163, 291

\bibitem[{{G{\'o}mez} \& {V{\'a}zquez-Semadeni}(2014)}]{2014ApJ...791..124G}
{G{\'o}mez} G.~C., {V{\'a}zquez-Semadeni} E., 2014, \apj, 791, 124

\bibitem[{{Hacar} {et~al}\mbox{.}(2023){Hacar}, {Clark}, {Heitsch},
  {Kainulainen}, {Panopoulou}, {Seifried}, \& {Smith}}]{2023ASPC..534..153H}
{Hacar} A., {Clark} S.~E., {Heitsch} F., {Kainulainen} J., {Panopoulou} G.~V.,
  {Seifried} D., {Smith} R., 2023, in Astronomical Society of the Pacific
  Conference Series, Vol. 534, Protostars and Planets VII, {Inutsuka} S.,
  {Aikawa} Y., {Muto} T., {Tomida} K., {Tamura} M., eds., p. 153

\bibitem[{{He} {et~al}\mbox{.}(2023){He}, {Liu}, {Tang}, {Qin}, {Zhou},
  {Esimbek}, {Pan}, {Li}, {Zhao}, {Ji}, \& {Komesh}}]{2023ApJ...957...61H}
{He} Y.-X. {et~al.}, 2023, \apj, 957, 61

\bibitem[{{Heiderman} {et~al}\mbox{.}(2010){Heiderman}, {Evans}, {Allen},
  {Huard}, \& {Heyer}}]{2010ApJ...723.1019H}
{Heiderman} A., {Evans}, Neal~J. I., {Allen} L.~E., {Huard} T., {Heyer} M.,
  2010, \apj, 723, 1019

\bibitem[{{Henshaw} {et~al}\mbox{.}(2014){Henshaw}, {Caselli}, {Fontani},
  {Jim{\'e}nez-Serra}, \& {Tan}}]{2014MNRAS.440.2860H}
{Henshaw} J.~D., {Caselli} P., {Fontani} F., {Jim{\'e}nez-Serra} I., {Tan}
  J.~C., 2014, \mnras, 440, 2860

\bibitem[{{Hwang} {et~al}\mbox{.}(2022){Hwang}, {Kim}, {Pattle}, {Lee}, {Koch},
  {Johnstone}, {Tomisaka}, {Whitworth}, {Furuya}, {Kang}, {Lyo}, {Chung},
  {Arzoumanian}, {Park}, {Kwon}, {Kim}, {Tamura}, {Kwon}, {Soam}, {Han},
  {Hoang}, {Kim}, {Onaka}, {Eswaraiah}, {Ward-Thompson}, {Liu}, {Tang}, {Chen},
  {Matsumura}, {Hoang}, {Chen}, {Le Gouellec}, {Kirchschlager}, {Poidevin},
  {Bastien}, {Qiu}, {Hasegawa}, {Lai}, {Byun}, {Cho}, {Choi}, {Choi}, {Choi},
  {Jeong}, {Kang}, {Kim}, {Kim}, {Lee}, {Lee}, {Lee}, {Lee}, {Kim}, {Yoo},
  {Yun}, {Chen}, {Di Francesco}, {Fiege}, {Fissel}, {Franzmann}, {Houde},
  {Lacaille}, {Matthews}, {Sadavoy}, {Moriarty-Schieven}, {Tahani}, {Ching},
  {Dai}, {Duan}, {Gu}, {Law}, {Li}, {Li}, {Li}, {Li}, {Liu}, {Lu}, {Qian},
  {Wang}, {Wu}, {Xie}, {Yuan}, {Zhang}, {Zhang}, {Zhang}, {Zhou}, {Zhu},
  {Berry}, {Friberg}, {Graves}, {Liu}, {Mairs}, {Parsons}, {Rawlings}, {Doi},
  {Hayashi}, {Hull}, {Inoue}, {Inutsuka}, {Iwasaki}, {Kataoka}, {Kawabata},
  {Kim}, {Kobayashi}, {Nagata}, {Nakamura}, {Nakanishi}, {Pyo}, {Saito},
  {Seta}, {Shimajiri}, {Shinnaga}, {Tsukamoto}, {Zenko}, {Chen}, {Duan},
  {Fanciullo}, {Kemper}, {Lee}, {Lin}, {Liu}, {Ohashi}, {Rao}, {Tang}, {Wang},
  {Yang}, {Yen}, {Bourke}, {Chrysostomou}, {Debattista}, {Eden}, {Eyres},
  {Falle}, {Fuller}, {Gledhill}, {Greaves}, {Griffin}, {Hatchell}, {Karoly},
  {Kirk}, {K{\"o}nyves}, {Longmore}, {van Loo}, {de Looze}, {Peretto},
  {Priestley}, {Rawlings}, {Retter}, {Richer}, {Rigby}, {Savini}, {Scaife},
  {Viti}, {Diep}, {Ngoc}, {Tram}, {Andr{\'e}}, {Coud{\'e}}, {Dowell},
  {Friesen}, \& {Robitaille}}]{2022ApJ...941...51H}
{Hwang} J. {et~al.}, 2022, \apj, 941, 51

\bibitem[{{Issac} {et~al}\mbox{.}(2019){Issac}, {Tej}, {Liu}, {Varricatt},
  {Vig}, {Ishwara Chandra}, \& {Schultheis}}]{2019MNRAS.485.1775I}
{Issac} N., {Tej} A., {Liu} T., {Varricatt} W., {Vig} S., {Ishwara Chandra}
  C.~H., {Schultheis} M., 2019, \mnras, 485, 1775

\bibitem[{{Jones} {et~al}\mbox{.}(2012){Jones}, {Burton}, {Cunningham},
  {Requena-Torres}, {Menten}, {Schilke}, {Belloche}, {Leurini},
  {Mart{\'\i}n-Pintado}, {Ott}, \& {Walsh}}]{2012MNRAS.419.2961J}
{Jones} P.~A. {et~al.}, 2012, \mnras, 419, 2961

\bibitem[{{Jones} {et~al}\mbox{.}(2013){Jones}, {Burton}, {Cunningham},
  {Tothill}, \& {Walsh}}]{2013MNRAS.433..221J}
{Jones} P.~A., {Burton} M.~G., {Cunningham} M.~R., {Tothill} N.~F.~H., {Walsh}
  A.~J., 2013, \mnras, 433, 221

\bibitem[{{Kauffmann} {et~al}\mbox{.}(2008){Kauffmann}, {Bertoldi}, {Bourke},
  {Evans}, \& {Lee}}]{2008A&A...487..993K}
{Kauffmann} J., {Bertoldi} F., {Bourke} T.~L., {Evans}, N.~J. I., {Lee} C.~W.,
  2008, \aap, 487, 993

\bibitem[{{Kauffmann}, {Pillai} \& {Goldsmith}(2013){Kauffmann}, {Pillai}, \&
  {Goldsmith}}]{2013ApJ...779..185K}
{Kauffmann} J., {Pillai} T., {Goldsmith} P.~F., 2013, \apj, 779, 185

\bibitem[{{Kauffmann} {et~al}\mbox{.}(2010){Kauffmann}, {Pillai}, {Shetty},
  {Myers}, \& {Goodman}}]{2010ApJ...716..433K}
{Kauffmann} J., {Pillai} T., {Shetty} R., {Myers} P.~C., {Goodman} A.~A., 2010,
  \apj, 716, 433

\bibitem[{{Kirk} {et~al}\mbox{.}(2013){Kirk}, {Myers}, {Bourke}, {Gutermuth},
  {Hedden}, \& {Wilson}}]{2013ApJ...766..115K}
{Kirk} H., {Myers} P.~C., {Bourke} T.~L., {Gutermuth} R.~A., {Hedden} A.,
  {Wilson} G.~W., 2013, \apj, 766, 115

\bibitem[{{Koch} \& {Rosolowsky}(2015)}]{2015MNRAS.452.3435K}
{Koch} E.~W., {Rosolowsky} E.~W., 2015, \mnras, 452, 3435

\bibitem[{{Krumholz} \& {McKee}(2008)}]{2008Natur.451.1082K}
{Krumholz} M.~R., {McKee} C.~F., 2008, \nat, 451, 1082

\bibitem[{{Krumholz}, {McKee} \& {Klein}(2005){Krumholz}, {McKee}, \&
  {Klein}}]{2005Natur.438..332K}
{Krumholz} M.~R., {McKee} C.~F., {Klein} R.~I., 2005, \nat, 438, 332

\bibitem[{{Kumar} {et~al}\mbox{.}(2020){Kumar}, {Palmeirim}, {Arzoumanian}, \&
  {Inutsuka}}]{2020A&A...642A..87K}
{Kumar} M.~S.~N., {Palmeirim} P., {Arzoumanian} D., {Inutsuka} S.~I., 2020,
  \aap, 642, A87

\bibitem[{{Lada}, {Lombardi} \& {Alves}(2010){Lada}, {Lombardi}, \&
  {Alves}}]{2010ApJ...724..687L}
{Lada} C.~J., {Lombardi} M., {Alves} J.~F., 2010, \apj, 724, 687

\bibitem[{{Li}(2024)}]{2024MNRAS.528.7333L}
{Li} G.-X., 2024, \mnras, 528, 7333

\bibitem[{{Li}, {Cao} \& {Qiu}(2021){Li}, {Cao}, \&
  {Qiu}}]{2021ApJ...916...13L}
{Li} G.-X., {Cao} Y., {Qiu} K., 2021, \apj, 916, 13

\bibitem[{{Li} {et~al}\mbox{.}(2022){Li}, {Sanhueza}, {Lee}, {Zhang},
  {Beuther}, {Palau}, {Liu}, {Smith}, {Liu}, {Jim{\'e}nez-Serra}, {Kim},
  {Feng}, {Liu}, {Wang}, {Li}, {Qiu}, {Lu}, {Girart}, {Wang}, {Li}, {Li},
  {Cao}, {Kim}, \& {Strom}}]{2022ApJ...926..165L}
{Li} S. {et~al.}, 2022, \apj, 926, 165

\bibitem[{{Li} {et~al}\mbox{.}(2023){Li}, {Sanhueza}, {Zhang}, {Guido},
  {Sabatini}, {Morii}, {Lu}, {Tafoya}, {Nakamura}, {Izumi}, {Tatematsu}, \&
  {Li}}]{2023ApJ...949..109L}
{Li} S. {et~al.}, 2023, \apj, 949, 109

\bibitem[{{Li} {et~al}\mbox{.}(2020){Li}, {Zhang}, {Liu}, {Beuther}, {Palau},
  {Girart}, {Smith}, {Hora}, {Lin}, {Qiu}, {Strom}, {Wang}, {Li}, \&
  {Yue}}]{2020ApJ...896..110L}
{Li} S. {et~al.}, 2020, \apj, 896, 110

\bibitem[{{Lin} {et~al}\mbox{.}(2021){Lin}, {Lee}, {Li}, {Tobin}, \&
  {Turner}}]{2021MNRAS.501.1316L}
{Lin} Z.-Y.~D., {Lee} C.-F., {Li} Z.-Y., {Tobin} J.~J., {Turner} N.~J., 2021,
  \mnras, 501, 1316

\bibitem[{{Liu} {et~al}\mbox{.}(2012){Liu}, {Quintana-Lacaci}, {Wang}, {Ho},
  {Li}, {Zhang}, \& {Zhang}}]{2012ApJ...745...61L}
{Liu} H.~B., {Quintana-Lacaci} G., {Wang} K., {Ho} P. T.~P., {Li} Z.-Y.,
  {Zhang} Q., {Zhang} Z.-Y., 2012, \apj, 745, 61

\bibitem[{{Liu} {et~al}\mbox{.}(2021){Liu}, {Liu}, {Evans}, {Wang}, {Garay},
  {Qin}, {Li}, {Stutz}, {Goldsmith}, {Liu}, {Tej}, {Zhang}, {Juvela}, {Li},
  {Wang}, {Bronfman}, {Ren}, {Wu}, {Kim}, {Lee}, {Tatematsu}, {Cunningham},
  {Liu}, {Wu}, {Hirota}, {Lee}, {Li}, {Kang}, {Mardones}, {Ristorcelli},
  {Zhang}, {Luo}, {Toth}, {Yi}, {Yun}, {Peng}, {Li}, {Zhu}, {Shen}, {Baug},
  {Dewangan}, {Chakali}, {Liu}, {Xu}, {Wang}, {Zhang}, {Li}, {Zhang}, {Zhou},
  {Tang}, {Xue}, {Issac}, {Soam}, \&
  {{\'A}lvarez-Guti{\'e}rrez}}]{2021MNRAS.505.2801L}
{Liu} H.-L. {et~al.}, 2021, \mnras, 505, 2801

\bibitem[{{Liu}, {Stutz} \& {Yuan}(2019){Liu}, {Stutz}, \&
  {Yuan}}]{2019MNRAS.487.1259L}
{Liu} H.-L., {Stutz} A., {Yuan} J.-H., 2019, \mnras, 487, 1259

\bibitem[{{Liu} {et~al}\mbox{.}(2022{\natexlab{a}}){Liu}, {Tej}, {Liu},
  {Goldsmith}, {Stutz}, {Juvela}, {Qin}, {Xu}, {Bronfman}, {Evans}, {Saha},
  {Issac}, {Tatematsu}, {Wang}, {Li}, {Zhang}, {Baug}, {Dewangan}, {Wu},
  {Zhang}, {Lee}, {Liu}, {Zhou}, \& {Soam}}]{2022MNRAS.511.4480L}
{Liu} H.-L. {et~al.}, 2022{\natexlab{a}}, \mnras, 511, 4480

\bibitem[{{Liu} {et~al}\mbox{.}(2022{\natexlab{b}}){Liu}, {Tej}, {Liu},
  {Issac}, {Saha}, {Goldsmith}, {Wang}, {Zhang}, {Qin}, {Wang}, {Li}, {Soam},
  {Dewangan}, {Lee}, {Li}, {Liu}, {Zhang}, {Ren}, {Juvela}, {Bronfman}, {Wu},
  {Tatematsu}, {Chen}, {Li}, {Stutz}, {Zhang}, {Viktor Toth}, {Luo}, {Xu},
  {Li}, {Liu}, {Zhou}, {Zhang}, {Tang}, {Zhang}, {Baug}, {Mannfors}, {Chakali},
  \& {Dutta}}]{2022MNRAS.510.5009L}
{Liu} H.-L. {et~al.}, 2022{\natexlab{b}}, \mnras, 510, 5009

\bibitem[{{Liu} {et~al}\mbox{.}(2023){Liu}, {Tej}, {Liu}, {Sanhueza}, {Qin},
  {He}, {Goldsmith}, {Garay}, {Pan}, {Morii}, {Li}, {Stutz}, {Tatematsu}, {Xu},
  {Bronfman}, {Saha}, {Issac}, {Baug}, {Toth}, {Dewangan}, {Wang}, {Zhou},
  {Lee}, {Yang}, {Luo}, {Shen}, {Zhang}, {Wu}, {Ren}, {Liu}, {Soam}, {Zhang},
  \& {Luo}}]{2023MNRAS.522.3719L}
{Liu} H.-L. {et~al.}, 2023, \mnras, 522, 3719

\bibitem[{{Liu} {et~al}\mbox{.}(2020){Liu}, {Evans}, {Kim}, {Goldsmith}, {Liu},
  {Zhang}, {Tatematsu}, {Wang}, {Juvela}, {Bronfman}, {Cunningham}, {Garay},
  {Hirota}, {Lee}, {Kang}, {Li}, {Li}, {Mardones}, {Qin}, {Ristorcelli}, {Tej},
  {Toth}, {Wu}, {Wu}, {Yi}, {Yun}, {Liu}, {Peng}, {Li}, {Li}, {Lee}, {Shen},
  {Baug}, {Wang}, {Zhang}, {Issac}, {Zhu}, {Luo}, {Soam}, {Liu}, {Xu}, {Wang},
  {Zhang}, {Ren}, \& {Zhang}}]{2020MNRAS.496.2790L}
{Liu} T. {et~al.}, 2020, \mnras, 496, 2790

\bibitem[{{Liu} {et~al}\mbox{.}(2016){Liu}, {Zhang}, {Kim}, {Wu}, {Lee},
  {Goldsmith}, {Li}, {Liu}, {Chen}, {Tatematsu}, {Wang}, {Lee}, {Qin},
  {Mardones}, \& {Cho}}]{2016ApJ...824...31L}
{Liu} T. {et~al.}, 2016, \apj, 824, 31

\bibitem[{{Lo} {et~al}\mbox{.}(2009){Lo}, {Cunningham}, {Jones}, {Bains},
  {Burton}, {Wong}, {Muller}, {Kramer}, {Ossenkopf}, {Henkel}, {Deragopian},
  {Donnelly}, \& {Ladd}}]{2009MNRAS.395.1021L}
{Lo} N. {et~al.}, 2009, \mnras, 395, 1021

\bibitem[{{Lu} {et~al}\mbox{.}(2018){Lu}, {Zhang}, {Liu}, {Sanhueza},
  {Tatematsu}, {Feng}, {Smith}, {Myers}, {Sridharan}, \&
  {Gu}}]{2018ApJ...855....9L}
{Lu} X. {et~al.}, 2018, \apj, 855, 9

\bibitem[{{Lumsden} {et~al}\mbox{.}(2013){Lumsden}, {Hoare}, {Urquhart},
  {Oudmaijer}, {Davies}, {Mottram}, {Cooper}, \& {Moore}}]{2013ApJS..208...11L}
{Lumsden} S.~L., {Hoare} M.~G., {Urquhart} J.~S., {Oudmaijer} R.~D., {Davies}
  B., {Mottram} J.~C., {Cooper} H.~D.~B., {Moore} T.~J.~T., 2013, \apjs, 208,
  11

\bibitem[{{Mai} {et~al}\mbox{.}(2024){Mai}, {Liu}, {Liu}, {Zhu}, {Garay},
  {Goldsmith}, {Juvela}, {Liu}, {Mannfors}, {Tej}, {Sanhueza}, {Li}, {Xu},
  {Semadeni}, {Jiao}, {Peng}, {Baug}, {Yang}, {Dewangan}, {Bronfman},
  {G{\'o}mez}, {Palau}, {Lee}, {Qin}, {Tatematsu}, {Chibueze}, {Yang}, {Lu},
  {Luo}, {Gu}, {Issac}, {Zhang}, {Li}, {Zhang}, \&
  {T{\'o}th}}]{2024ApJ...961L..35M}
{Mai} X. {et~al.}, 2024, \apjl, 961, L35

\bibitem[{{Marsh}, {Whitworth} \& {Lomax}(2015){Marsh}, {Whitworth}, \&
  {Lomax}}]{2015MNRAS.454.4282M}
{Marsh} K.~A., {Whitworth} A.~P., {Lomax} O., 2015, \mnras, 454, 4282

\bibitem[{{Marsh} {et~al}\mbox{.}(2017){Marsh}, {Whitworth}, {Lomax}, {Ragan},
  {Becciani}, {Cambr{\'e}sy}, {Di Giorgio}, {Eden}, {Elia}, {Kacsuk},
  {Molinari}, {Palmeirim}, {Pezzuto}, {Schneider}, {Sciacca}, \&
  {Vitello}}]{2017MNRAS.471.2730M}
{Marsh} K.~A. {et~al.}, 2017, \mnras, 471, 2730

\bibitem[{{Mauch} {et~al}\mbox{.}(2003){Mauch}, {Murphy}, {Buttery}, {Curran},
  {Hunstead}, {Piestrzynski}, {Robertson}, \& {Sadler}}]{2003MNRAS.342.1117M}
{Mauch} T., {Murphy} T., {Buttery} H.~J., {Curran} J., {Hunstead} R.~W.,
  {Piestrzynski} B., {Robertson} J.~G., {Sadler} E.~M., 2003, \mnras, 342, 1117

\bibitem[{{McClure-Griffiths} {et~al}\mbox{.}(2005){McClure-Griffiths},
  {Dickey}, {Gaensler}, {Green}, {Haverkorn}, \&
  {Strasser}}]{2005ApJS..158..178M}
{McClure-Griffiths} N.~M., {Dickey} J.~M., {Gaensler} B.~M., {Green} A.~J.,
  {Haverkorn} M., {Strasser} S., 2005, \apjs, 158, 178

\bibitem[{{McKee} \& {Tan}(2003)}]{2003ApJ...585..850M}
{McKee} C.~F., {Tan} J.~C., 2003, \apj, 585, 850

\bibitem[{{Molinari} {et~al}\mbox{.}(2016){Molinari}, {Schisano}, {Elia},
  {Pestalozzi}, {Traficante}, {Pezzuto}, {Swinyard}, {Noriega-Crespo}, {Bally},
  {Moore}, {Plume}, {Zavagno}, {di Giorgio}, {Liu}, {Pilbratt}, {Mottram},
  {Russeil}, {Piazzo}, {Veneziani}, {Benedettini}, {Calzoletti}, {Faustini},
  {Natoli}, {Piacentini}, {Merello}, {Palmese}, {Del Grande}, {Polychroni},
  {Rygl}, {Polenta}, {Barlow}, {Bernard}, {Martin}, {Testi}, {Ali},
  {Andr{\'e}}, {Beltr{\'a}n}, {Billot}, {Carey}, {Cesaroni}, {Compi{\`e}gne},
  {Eden}, {Fukui}, {Garcia-Lario}, {Hoare}, {Huang}, {Joncas}, {Lim}, {Lord},
  {Martinavarro-Armengol}, {Motte}, {Paladini}, {Paradis}, {Peretto},
  {Robitaille}, {Schilke}, {Schneider}, {Schulz}, {Sibthorpe}, {Strafella},
  {Thompson}, {Umana}, {Ward-Thompson}, \& {Wyrowski}}]{2016A&A...591A.149M}
{Molinari} S. {et~al.}, 2016, \aap, 591, A149

\bibitem[{{Molinari} {et~al}\mbox{.}(2010){Molinari}, {Swinyard}, {Bally},
  {Barlow}, {Bernard}, {Martin}, {Moore}, {Noriega-Crespo}, {Plume}, {Testi},
  {Zavagno}, {Abergel}, {Ali}, {Andr{\'e}}, {Baluteau}, {Benedettini},
  {Bern{\'e}}, {Billot}, {Blommaert}, {Bontemps}, {Boulanger}, {Brand},
  {Brunt}, {Burton}, {Campeggio}, {Carey}, {Caselli}, {Cesaroni}, {Cernicharo},
  {Chakrabarti}, {Chrysostomou}, {Codella}, {Cohen}, {Compiegne}, {Davis}, {de
  Bernardis}, {de Gasperis}, {Di Francesco}, {di Giorgio}, {Elia}, {Faustini},
  {Fischera}, {Fukui}, {Fuller}, {Ganga}, {Garcia-Lario}, {Giard}, {Giardino},
  {Glenn}, {Goldsmith}, {Griffin}, {Hoare}, {Huang}, {Jiang}, {Joblin},
  {Joncas}, {Juvela}, {Kirk}, {Lagache}, {Li}, {Lim}, {Lord}, {Lucas},
  {Maiolo}, {Marengo}, {Marshall}, {Masi}, {Massi}, {Matsuura}, {Meny},
  {Minier}, {Miville-Desch{\^e}nes}, {Montier}, {Motte}, {M{\"u}ller},
  {Natoli}, {Neves}, {Olmi}, {Paladini}, {Paradis}, {Pestalozzi}, {Pezzuto},
  {Piacentini}, {Pomar{\`e}s}, {Popescu}, {Reach}, {Richer}, {Ristorcelli},
  {Roy}, {Royer}, {Russeil}, {Saraceno}, {Sauvage}, {Schilke},
  {Schneider-Bontemps}, {Schuller}, {Schultz}, {Shepherd}, {Sibthorpe},
  {Smith}, {Smith}, {Spinoglio}, {Stamatellos}, {Strafella}, {Stringfellow},
  {Sturm}, {Taylor}, {Thompson}, {Tuffs}, {Umana}, {Valenziano}, {Vavrek},
  {Viti}, {Waelkens}, {Ward-Thompson}, {White}, {Wyrowski}, {Yorke}, \&
  {Zhang}}]{2010PASP..122..314M}
{Molinari} S. {et~al.}, 2010, \pasp, 122, 314

\bibitem[{{Morii} {et~al}\mbox{.}(2024){Morii}, {Sanhueza}, {Zhang}, {Guido},
  {Sabatini}, {Morii}, {Lu}, {Tafoya}, {Nakamura}, {Izumi}, {Tatematsu}, \&
  {Li}}]{Morii}
{Morii} S. {et~al.}, 2024

\bibitem[{{Motte}, {Bontemps} \& {Louvet}(2018){Motte}, {Bontemps}, \&
  {Louvet}}]{2018ARA&A..56...41M}
{Motte} F., {Bontemps} S., {Louvet} F., 2018, \araa, 56, 41

\bibitem[{{Neelamkodan} {et~al}\mbox{.}(2021){Neelamkodan}, {Tokuda}, {Barman},
  {Kondo}, {Sano}, \& {Onishi}}]{2021ApJ...908L..43N}
{Neelamkodan} N., {Tokuda} K., {Barman} S., {Kondo} H., {Sano} H., {Onishi} T.,
  2021, \apjl, 908, L43

\bibitem[{{Olguin} {et~al}\mbox{.}(2023){Olguin}, {Sanhueza}, {Chen}, {Lu},
  {Oya}, {Zhang}, {Ginsburg}, {Taniguchi}, {Li}, {Morii}, {Sakai}, \&
  {Nakamura}}]{2023ApJ...959L..31O}
{Olguin} F.~A. {et~al.}, 2023, \apjl, 959, L31

\bibitem[{{Olguin} {et~al}\mbox{.}(2022){Olguin}, {Sanhueza}, {Ginsburg},
  {Chen}, {Zhang}, {Li}, {Lu}, \& {Sakai}}]{2022ApJ...929...68O}
{Olguin} F.~A., {Sanhueza} P., {Ginsburg} A., {Chen} H.-R.~V., {Zhang} Q., {Li}
  S., {Lu} X., {Sakai} T., 2022, \apj, 929, 68

\bibitem[{{Olguin} {et~al}\mbox{.}(2021){Olguin}, {Sanhueza}, {Guzm{\'a}n},
  {Lu}, {Saigo}, {Zhang}, {Silva}, {Chen}, {Li}, {Ohashi}, {Nakamura}, {Sakai},
  \& {Wu}}]{2021ApJ...909..199O}
{Olguin} F.~A. {et~al.}, 2021, \apj, 909, 199

\bibitem[{{Ossenkopf} \& {Henning}(1994)}]{1994A&A...291..943O}
{Ossenkopf} V., {Henning} T., 1994, \aap, 291, 943

\bibitem[{{Padoan} {et~al}\mbox{.}(2020){Padoan}, {Pan}, {Juvela},
  {Haugb{\o}lle}, \& {Nordlund}}]{2020ApJ...900...82P}
{Padoan} P., {Pan} L., {Juvela} M., {Haugb{\o}lle} T., {Nordlund} {\r{A}}.,
  2020, \apj, 900, 82

\bibitem[{{Palau} {et~al}\mbox{.}(2015){Palau}, {Ballesteros-Paredes},
  {V{\'a}zquez-Semadeni}, {S{\'a}nchez-Monge}, {Estalella}, {Fall}, {Zapata},
  {Camacho}, {G{\'o}mez}, {Naranjo-Romero}, {Busquet}, \&
  {Fontani}}]{2015MNRAS.453.3785P}
{Palau} A. {et~al.}, 2015, \mnras, 453, 3785

\bibitem[{{Pan}, {Liu} \& {Qin}(2024){Pan}, {Liu}, \&
  {Qin}}]{2024ApJ...960...76P}
{Pan} S., {Liu} H.-L., {Qin} S.-L., 2024, \apj, 960, 76

\bibitem[{{Peretto} {et~al}\mbox{.}(2013){Peretto}, {Fuller}, {Duarte-Cabral},
  {Avison}, {Hennebelle}, {Pineda}, {Andr{\'e}}, {Bontemps}, {Motte},
  {Schneider}, \& {Molinari}}]{2013A&A...555A.112P}
{Peretto} N. {et~al.}, 2013, \aap, 555, A112

\bibitem[{{Peretto}, {Hennebelle} \& {Andr{\'e}}(2007){Peretto}, {Hennebelle},
  \& {Andr{\'e}}}]{2007A&A...464..983P}
{Peretto} N., {Hennebelle} P., {Andr{\'e}} P., 2007, \aap, 464, 983

\bibitem[{{Peretto} {et~al}\mbox{.}(2023){Peretto}, {Rigby}, {Louvet},
  {Fuller}, {Traficante}, \& {Gaudel}}]{2023MNRAS.525.2935P}
{Peretto} N., {Rigby} A.~J., {Louvet} F., {Fuller} G.~A., {Traficante} A.,
  {Gaudel} M., 2023, \mnras, 525, 2935

\bibitem[{{Peters} {et~al}\mbox{.}(2017){Peters}, {Zhukovska}, {Naab},
  {Girichidis}, {Walch}, {Glover}, {Klessen}, {Clark}, \&
  {Seifried}}]{2017MNRAS.467.4322P}
{Peters} T. {et~al.}, 2017, \mnras, 467, 4322

\bibitem[{{Pineda} {et~al}\mbox{.}(2012){Pineda}, {Maury}, {Fuller}, {Testi},
  {Garc{\'\i}a-Appadoo}, {Peck}, {Villard}, {Corder}, {van Kempen}, {Turner},
  {Tachihara}, \& {Dent}}]{2012A&A...544L...7P}
{Pineda} J.~E. {et~al.}, 2012, \aap, 544, L7

\bibitem[{{Pollack} {et~al}\mbox{.}(1994){Pollack}, {Hollenbach}, {Beckwith},
  {Simonelli}, {Roush}, \& {Fong}}]{1994ApJ...421..615P}
{Pollack} J.~B., {Hollenbach} D., {Beckwith} S., {Simonelli} D.~P., {Roush} T.,
  {Fong} W., 1994, \apj, 421, 615

\bibitem[{{Potdar} {et~al}\mbox{.}(2022){Potdar}, {Das}, {Issac}, {Tej}, {Vig},
  \& {Chandra}}]{2022MNRAS.510..658P}
{Potdar} A., {Das} S.~R., {Issac} N., {Tej} A., {Vig} S., {Chandra} C.~H.~I.,
  2022, \mnras, 510, 658

\bibitem[{{Ragan} {et~al}\mbox{.}(2012){Ragan}, {Henning}, {Krause}, {Pitann},
  {Beuther}, {Linz}, {Tackenberg}, {Balog}, {Hennemann}, {Launhardt}, {Lippok},
  {Nielbock}, {Schmiedeke}, {Schuller}, {Steinacker}, {Stutz}, \&
  {Vasyunina}}]{2012A&A...547A..49R}
{Ragan} S. {et~al.}, 2012, \aap, 547, A49

\bibitem[{{Rathborne} {et~al}\mbox{.}(2011){Rathborne}, {Garay}, {Jackson},
  {Longmore}, {Zhang}, \& {Simon}}]{2011ApJ...741..120R}
{Rathborne} J.~M., {Garay} G., {Jackson} J.~M., {Longmore} S., {Zhang} Q.,
  {Simon} R., 2011, \apj, 741, 120

\bibitem[{{Rebolledo} {et~al}\mbox{.}(2020){Rebolledo}, {Guzm{\'a}n},
  {Contreras}, {Garay}, {Medina}, {Sanhueza}, {Green}, {Castro}, {Guzm{\'a}n},
  \& {Burton}}]{2020ApJ...891..113R}
{Rebolledo} D. {et~al.}, 2020, \apj, 891, 113

\bibitem[{{Redaelli} {et~al}\mbox{.}(2022){Redaelli}, {Bovino}, {Sanhueza},
  {Morii}, {Sabatini}, {Caselli}, {Giannetti}, \& {Li}}]{2022ApJ...936..169R}
{Redaelli} E., {Bovino} S., {Sanhueza} P., {Morii} K., {Sabatini} G., {Caselli}
  P., {Giannetti} A., {Li} S., 2022, \apj, 936, 169

\bibitem[{{Rigby} {et~al}\mbox{.}(2024){Rigby}, {Peretto}, {Anderson}, {Ragan},
  {Priestley}, {Fuller}, {Thompson}, {Traficante}, {Watkins}, \&
  {Williams}}]{2024MNRAS.528.1172R}
{Rigby} A.~J. {et~al.}, 2024, \mnras, 528, 1172

\bibitem[{{Rosen} {et~al}\mbox{.}(2019){Rosen}, {Li}, {Zhang}, \&
  {Burkhart}}]{2019ApJ...887..108R}
{Rosen} A.~L., {Li} P.~S., {Zhang} Q., {Burkhart} B., 2019, \apj, 887, 108

\bibitem[{{Rosolowsky} {et~al}\mbox{.}(2008){Rosolowsky}, {Pineda},
  {Kauffmann}, \& {Goodman}}]{2008ApJ...679.1338R}
{Rosolowsky} E.~W., {Pineda} J.~E., {Kauffmann} J., {Goodman} A.~A., 2008,
  \apj, 679, 1338

\bibitem[{{Roy} {et~al}\mbox{.}(2014){Roy}, {Andr{\'e}}, {Palmeirim}, {Attard},
  {K{\"o}nyves}, {Schneider}, {Peretto}, {Men'shchikov}, {Ward-Thompson},
  {Kirk}, {Griffin}, {Marsh}, {Abergel}, {Arzoumanian}, {Benedettini}, {Hill},
  {Motte}, {Nguyen Luong}, {Pezzuto}, {Rivera-Ingraham}, {Roussel}, {Rygl},
  {Spinoglio}, {Stamatellos}, \& {White}}]{2014A&A...562A.138R}
{Roy} A. {et~al.}, 2014, \aap, 562, A138

\bibitem[{{Saha} {et~al}\mbox{.}(2022){Saha}, {Tej}, {Liu}, {Liu}, {Issac},
  {Lee}, {Garay}, {Goldsmith}, {Juvela}, {Qin}, {Stutz}, {Li}, {Wang}, {Baug},
  {Bronfman}, {Xu}, {Zhang}, \& {Eswaraiah}}]{2022MNRAS.516.1983S}
{Saha} A. {et~al.}, 2022, \mnras, 516, 1983

\bibitem[{{Sanhueza} {et~al}\mbox{.}(2019){Sanhueza}, {Contreras}, {Wu},
  {Jackson}, {Guzm{\'a}n}, {Zhang}, {Li}, {Lu}, {Silva}, {Izumi}, {Liu},
  {Miura}, {Tatematsu}, {Sakai}, {Beuther}, {Garay}, {Ohashi}, {Saito},
  {Nakamura}, {Saigo}, {Veena}, {Nguyen-Luong}, \&
  {Tafoya}}]{2019ApJ...886..102S}
{Sanhueza} P. {et~al.}, 2019, \apj, 886, 102

\bibitem[{{Sanhueza} {et~al}\mbox{.}(2021){Sanhueza}, {Girart}, {Padovani},
  {Galli}, {Hull}, {Zhang}, {Cortes}, {Stephens}, {Fern{\'a}ndez-L{\'o}pez},
  {Jackson}, {Frau}, {Kock}, {Wu}, {Zapata}, {Olguin}, {Lu}, {Silva}, {Tang},
  {Sakai}, {Guzm{\'a}n}, {Tatematsu}, {Nakamura}, \&
  {Chen}}]{2021ApJ...915L..10S}
{Sanhueza} P. {et~al.}, 2021, \apjl, 915, L10

\bibitem[{{Sanhueza} {et~al}\mbox{.}(2017){Sanhueza}, {Jackson}, {Zhang},
  {Guzm{\'a}n}, {Lu}, {Stephens}, {Wang}, \& {Tatematsu}}]{2017ApJ...841...97S}
{Sanhueza} P., {Jackson} J.~M., {Zhang} Q., {Guzm{\'a}n} A.~E., {Lu} X.,
  {Stephens} I.~W., {Wang} K., {Tatematsu} K., 2017, \apj, 841, 97

\bibitem[{{Scalise}, {Rodriguez} \& {Mendoza-Torres}(1989){Scalise},
  {Rodriguez}, \& {Mendoza-Torres}}]{1989A&A...221..105S}
{Scalise}, E. J., {Rodriguez} L.~F., {Mendoza-Torres} E., 1989, \aap, 221, 105

\bibitem[{{Schneider} {et~al}\mbox{.}(2010){Schneider}, {Csengeri}, {Bontemps},
  {Motte}, {Simon}, {Hennebelle}, {Federrath}, \&
  {Klessen}}]{2010A&A...520A..49S}
{Schneider} N., {Csengeri} T., {Bontemps} S., {Motte} F., {Simon} R.,
  {Hennebelle} P., {Federrath} C., {Klessen} R., 2010, \aap, 520, A49

\bibitem[{{Schuller} {et~al}\mbox{.}(2009){Schuller}, {Menten}, {Contreras},
  {Wyrowski}, {Schilke}, {Bronfman}, {Henning}, {Walmsley}, {Beuther},
  {Bontemps}, {Cesaroni}, {Deharveng}, {Garay}, {Herpin}, {Lefloch}, {Linz},
  {Mardones}, {Minier}, {Molinari}, {Motte}, {Nyman}, {Reveret}, {Risacher},
  {Russeil}, {Schneider}, {Testi}, {Troost}, {Vasyunina}, {Wienen}, {Zavagno},
  {Kovacs}, {Kreysa}, {Siringo}, \& {Wei{\ss}}}]{2009A&A...504..415S}
{Schuller} F. {et~al.}, 2009, \aap, 504, 415

\bibitem[{{Seshadri} {et~al}\mbox{.}(2024){Seshadri}, {Vig}, {Ghosh}, \&
  {Ojha}}]{2024MNRAS.527.4244S}
{Seshadri} A., {Vig} S., {Ghosh} S.~K., {Ojha} D.~K., 2024, \mnras, 527, 4244

\bibitem[{{Shimajiri} {et~al}\mbox{.}(2017){Shimajiri}, {Andr{\'e}}, {Braine},
  {K{\"o}nyves}, {Schneider}, {Bontemps}, {Ladjelate}, {Roy}, {Gao}, \&
  {Chen}}]{2017A&A...604A..74S}
{Shimajiri} Y. {et~al.}, 2017, \aap, 604, A74

\bibitem[{{Shirley}(2015)}]{2015PASP..127..299S}
{Shirley} Y.~L., 2015, \pasp, 127, 299

\bibitem[{{Shirley} {et~al}\mbox{.}(2011){Shirley}, {Huard}, {Pontoppidan},
  {Wilner}, {Stutz}, {Bieging}, \& {Evans}}]{2011ApJ...728..143S}
{Shirley} Y.~L., {Huard} T.~L., {Pontoppidan} K.~M., {Wilner} D.~J., {Stutz}
  A.~M., {Bieging} J.~H., {Evans}, Neal~J. I., 2011, \apj, 728, 143

\bibitem[{{Shu}, {Adams} \& {Lizano}(1987){Shu}, {Adams}, \&
  {Lizano}}]{1987ARA&A..25...23S}
{Shu} F.~H., {Adams} F.~C., {Lizano} S., 1987, \araa, 25, 23

\bibitem[{{Smith} {et~al}\mbox{.}(2016){Smith}, {Glover}, {Klessen}, \&
  {Fuller}}]{2016MNRAS.455.3640S}
{Smith} R.~J., {Glover} S. C.~O., {Klessen} R.~S., {Fuller} G.~A., 2016,
  \mnras, 455, 3640

\bibitem[{{Smith} {et~al}\mbox{.}(2020){Smith}, {Tre{\ss}}, {Sormani},
  {Glover}, {Klessen}, {Clark}, {Izquierdo}, {Duarte-Cabral}, \&
  {Zucker}}]{2020MNRAS.492.1594S}
{Smith} R.~J. {et~al.}, 2020, \mnras, 492, 1594

\bibitem[{{Stutz} \& {Gould}(2016)}]{2016A&A...590A...2S}
{Stutz} A.~M., {Gould} A., 2016, \aap, 590, A2

\bibitem[{{Tokuda} {et~al}\mbox{.}(2019){Tokuda}, {Fukui}, {Harada}, {Saigo},
  {Tachihara}, {Tsuge}, {Inoue}, {Torii}, {Nishimura}, {Zahorecz}, {Nayak},
  {Meixner}, {Minamidani}, {Kawamura}, {Mizuno}, {Indebetouw}, {Sewi{\l}o},
  {Madden}, {Galametz}, {Lebouteiller}, {Chen}, \&
  {Onishi}}]{2019ApJ...886...15T}
{Tokuda} K. {et~al.}, 2019, \apj, 886, 15

\bibitem[{{Trevi{\~n}o-Morales} {et~al}\mbox{.}(2019){Trevi{\~n}o-Morales},
  {Fuente}, {S{\'a}nchez-Monge}, {Kainulainen}, {Didelon}, {Suri}, {Schneider},
  {Ballesteros-Paredes}, {Lee}, {Hennebelle}, {Pilleri},
  {Gonz{\'a}lez-Garc{\'\i}a}, {Kramer}, {Garc{\'\i}a-Burillo}, {Luna},
  {Goicoechea}, {Tremblin}, \& {Geen}}]{2019A&A...629A..81T}
{Trevi{\~n}o-Morales} S.~P. {et~al.}, 2019, \aap, 629, A81

\bibitem[{{Urquhart} {et~al}\mbox{.}(2007){Urquhart}, {Busfield}, {Hoare},
  {Lumsden}, {Oudmaijer}, {Moore}, {Gibb}, {Purcell}, {Burton}, \&
  {Marechal}}]{2007A&A...474..891U}
{Urquhart} J.~S. {et~al.}, 2007, \aap, 474, 891

\bibitem[{{Urquhart} {et~al}\mbox{.}(2018){Urquhart}, {K{\"o}nig}, {Giannetti},
  {Leurini}, {Moore}, {Eden}, {Pillai}, {Thompson}, {Braiding}, {Burton},
  {Csengeri}, {Dempsey}, {Figura}, {Froebrich}, {Menten}, {Schuller}, {Smith},
  \& {Wyrowski}}]{2018MNRAS.473.1059U}
{Urquhart} J.~S. {et~al.}, 2018, \mnras, 473, 1059

\bibitem[{{Urquhart} {et~al}\mbox{.}(2014){Urquhart}, {Moore}, {Csengeri},
  {Wyrowski}, {Schuller}, {Hoare}, {Lumsden}, {Mottram}, {Thompson}, {Menten},
  {Walmsley}, {Bronfman}, {Pfalzner}, {K{\"o}nig}, \&
  {Wienen}}]{2014MNRAS.443.1555U}
{Urquhart} J.~S. {et~al.}, 2014, \mnras, 443, 1555

\bibitem[{{V{\'a}zquez-Semadeni}, {G{\'o}mez} \&
  {Gonz{\'a}lez-Samaniego}(2023){V{\'a}zquez-Semadeni}, {G{\'o}mez}, \&
  {Gonz{\'a}lez-Samaniego}}]{2023arXiv230613846V}
{V{\'a}zquez-Semadeni} E., {G{\'o}mez} G.~C., {Gonz{\'a}lez-Samaniego} A.,
  2023, arXiv e-prints, arXiv:2306.13846

\bibitem[{{V{\'a}zquez-Semadeni} {et~al}\mbox{.}(2009){V{\'a}zquez-Semadeni},
  {G{\'o}mez}, {Jappsen}, {Ballesteros-Paredes}, \&
  {Klessen}}]{2009ApJ...707.1023V}
{V{\'a}zquez-Semadeni} E., {G{\'o}mez} G.~C., {Jappsen} A.~K.,
  {Ballesteros-Paredes} J., {Klessen} R.~S., 2009, \apj, 707, 1023

\bibitem[{{V{\'a}zquez-Semadeni}, {Gonz{\'a}lez-Samaniego} \&
  {Col{\'\i}n}(2017){V{\'a}zquez-Semadeni}, {Gonz{\'a}lez-Samaniego}, \&
  {Col{\'\i}n}}]{2017MNRAS.467.1313V}
{V{\'a}zquez-Semadeni} E., {Gonz{\'a}lez-Samaniego} A., {Col{\'\i}n} P., 2017,
  \mnras, 467, 1313

\bibitem[{{V{\'a}zquez-Semadeni} {et~al}\mbox{.}(2019){V{\'a}zquez-Semadeni},
  {Palau}, {Ballesteros-Paredes}, {G{\'o}mez}, \&
  {Zamora-Avil{\'e}s}}]{2019MNRAS.490.3061V}
{V{\'a}zquez-Semadeni} E., {Palau} A., {Ballesteros-Paredes} J., {G{\'o}mez}
  G.~C., {Zamora-Avil{\'e}s} M., 2019, \mnras, 490, 3061

\bibitem[{{V{\'a}zquez-Semadeni} {et~al}\mbox{.}(2024){V{\'a}zquez-Semadeni},
  {Palau}, {G{\'o}mez}, {Arroyo-Ch{\'a}vez}, {Alig}, {Ballesteros-Paredes},
  {Camacho}, {Gonz{\'a}lez-Samaniego}, \& {Burkert}}]{2024arXiv240810406V}
{V{\'a}zquez-Semadeni} E. {et~al.}, 2024, arXiv e-prints, arXiv:2408.10406

\bibitem[{{Vuong} {et~al}\mbox{.}(2003){Vuong}, {Montmerle}, {Grosso},
  {Feigelson}, {Verstraete}, \& {Ozawa}}]{2003A&A...408..581V}
{Vuong} M.~H., {Montmerle} T., {Grosso} N., {Feigelson} E.~D., {Verstraete} L.,
  {Ozawa} H., 2003, \aap, 408, 581

\bibitem[{{Wang} {et~al}\mbox{.}(2014){Wang}, {Zhang}, {Testi}, {van der Tak},
  {Wu}, {Zhang}, {Pillai}, {Wyrowski}, {Carey}, {Ragan}, \&
  {Henning}}]{2014MNRAS.439.3275W}
{Wang} K. {et~al.}, 2014, \mnras, 439, 3275

\bibitem[{{Wang} {et~al}\mbox{.}(2011){Wang}, {Zhang}, {Wu}, \&
  {Zhang}}]{2011ApJ...735...64W}
{Wang} K., {Zhang} Q., {Wu} Y., {Zhang} H., 2011, \apj, 735, 64

\bibitem[{{Wang} {et~al}\mbox{.}(2010){Wang}, {Li}, {Abel}, \&
  {Nakamura}}]{2010ApJ...709...27W}
{Wang} P., {Li} Z.-Y., {Abel} T., {Nakamura} F., 2010, \apj, 709, 27

\bibitem[{{Xu} {et~al}\mbox{.}(2024){Xu}, {Wang}, {Liu}, {Tang}, {Evans},
  {Palau}, {Morii}, {He}, {Sanhueza}, {Liu}, {Stutz}, {Zhang}, {Chen}, {Li},
  {G{\'o}mez}, {V{\'a}zquez-Semadeni}, {Li}, {Mai}, {Lu}, {Liu}, {Chen}, {Li},
  {Shi}, {Ren}, {Li}, {Garay}, {Bronfman}, {Dewangan}, {Juvela}, {Lee},
  {Zhang}, {Yue}, {Wang}, {Ge}, {Jiao}, {Luo}, {Zhou}, {Tatematsu}, {Chibueze},
  {Su}, {Sun}, {Ristorcelli}, \& {Toth}}]{2024ApJS..270....9X}
{Xu} F. {et~al.}, 2024, \apjs, 270, 9

\bibitem[{{Xu} {et~al}\mbox{.}(2023){Xu}, {Wang}, {Liu}, {Goldsmith}, {Zhang},
  {Juvela}, {Liu}, {Qin}, {Li}, {Tej}, {Garay}, {Bronfman}, {Li}, {Wu},
  {G{\'o}mez}, {V{\'a}zquez-Semadeni}, {Tatematsu}, {Ren}, {Zhang}, {Toth},
  {Liu}, {Yue}, {Zhang}, {Baug}, {Issac}, {Stutz}, {Liu}, {Fuller}, {Tang},
  {Zhang}, {Dewangan}, {Lee}, {Zhou}, {Xie}, {Jiao}, {Wang}, {Liu}, {Luo},
  {Soam}, \& {Eswaraiah}}]{2023MNRAS.520.3259X}
{Xu} F.-W. {et~al.}, 2023, \mnras, 520, 3259

\bibitem[{{Yang} {et~al}\mbox{.}(2023){Yang}, {Liu}, {Tej}, {Liu}, {Sanhueza},
  {Qin}, {Lu}, {Wang}, {Pan}, {Xu}, {V{\'a}zquez-Semadeni​}, {Li},
  {G{\'o}mez}, {Palau}, {Garay}, {Goldsmith}, {Juvela}, {Saha}, {Bronfman},
  {Lee}, {Tatematsu}, {Dewangan}, {Zhou}, {Zhang}, {Stutz}, {Eswaraiah},
  {Toth}, {Ristorcelli}, {Shen}, {Luo}, \& {Chibueze}}]{2023ApJ...953...40Y}
{Yang} D. {et~al.}, 2023, \apj, 953, 40

\bibitem[{{Young} \& {Evans}(2005)}]{2005ApJ...627..293Y}
{Young} C.~H., {Evans}, Neal~J. I., 2005, \apj, 627, 293

\bibitem[{{Yuan} {et~al}\mbox{.}(2018){Yuan}, {Li}, {Wu}, {Ellingsen},
  {Henkel}, {Wang}, {Liu}, {Liu}, {Zavagno}, {Ren}, \&
  {Huang}}]{2018ApJ...852...12Y}
{Yuan} J. {et~al.}, 2018, \apj, 852, 12

\bibitem[{{Zapata} {et~al}\mbox{.}(2009){Zapata}, {Ho}, {Schilke},
  {Rodr{\'\i}guez}, {Menten}, {Palau}, \& {Garrod}}]{2009ApJ...698.1422Z}
{Zapata} L.~A., {Ho} P. T.~P., {Schilke} P., {Rodr{\'\i}guez} L.~F., {Menten}
  K., {Palau} A., {Garrod} R.~T., 2009, \apj, 698, 1422

\bibitem[{{Zhang} {et~al}\mbox{.}(2023){Zhang}, {Zhu}, {Liu}, {Ren}, {Liu},
  {Wang}, {Wu}, {Zhang}, {Zhou}, {Tatematsu}, {Garay}, {Tej}, {Li}, {Xu},
  {Lee}, {Bronfman}, {Soam}, \& {Li}}]{2023MNRAS.520.3245Z}
{Zhang} C. {et~al.}, 2023, \mnras, 520, 3245

\bibitem[{{Zhang} {et~al}\mbox{.}(2015){Zhang}, {Wang}, {Lu}, \&
  {Jim{\'e}nez-Serra}}]{2015ApJ...804..141Z}
{Zhang} Q., {Wang} K., {Lu} X., {Jim{\'e}nez-Serra} I., 2015, \apj, 804, 141

\bibitem[{{Zhou} {et~al}\mbox{.}(2022){Zhou}, {Liu}, {Evans}, {Garay},
  {Goldsmith}, {G{\'o}mez}, {V{\'a}zquez-Semadeni}, {Liu}, {Stutz}, {Wang},
  {Juvela}, {He}, {Li}, {Bronfman}, {Liu}, {Xu}, {Tej}, {Dewangan}, {Li},
  {Zhang}, {Zhang}, {Ren}, {Tatematsu}, {Shing Li}, {Won Lee}, {Baug}, {Qin},
  {Wu}, {Peng}, {Zhang}, {Liu}, {Luo}, {Ge}, {Saha}, {Chakali}, {Zhang}, {Kim},
  {Ristorcelli}, {Shen}, \& {Li}}]{2022MNRAS.514.6038Z}
{Zhou} J.-W. {et~al.}, 2022, \mnras, 514, 6038

\bibitem[{{Zhou} {et~al}\mbox{.}(2023){Zhou}, {Wyrowski}, {Neupane},
  {Urquhart}, {Evans}, {V{\'a}zquez-Semadeni}, {Menten}, {Gong}, \&
  {Liu}}]{2023A&A...676A..69Z}
{Zhou} J.~W. {et~al.}, 2023, \aap, 676, A69

\end{thebibliography}



\appendix

\section{3~mm continuum residual map} \label{resd_3mm}
\begin{figure}
\centering
\includegraphics[scale=0.4]{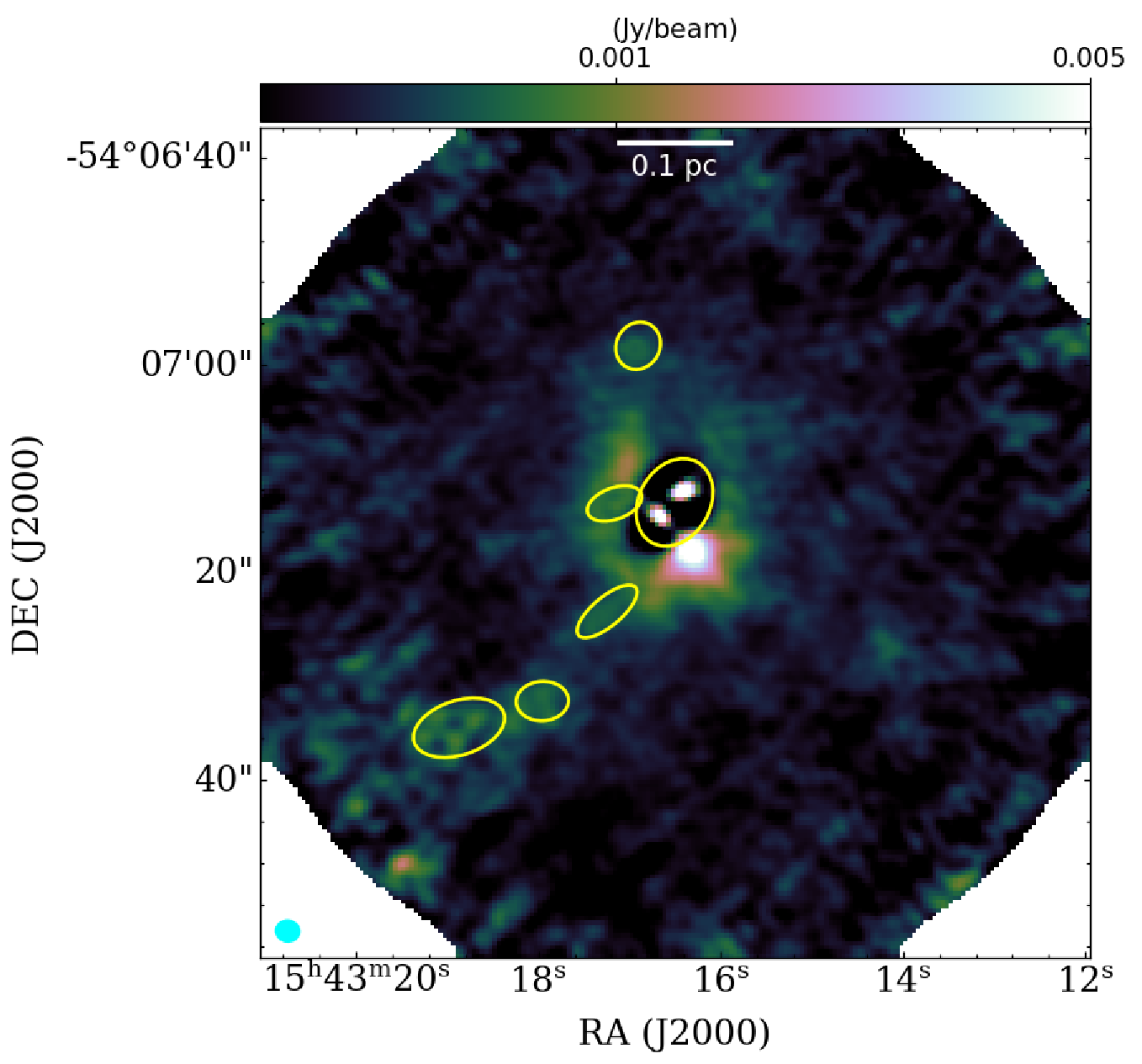}
\caption{The 3~mm continuum residual map retrived from CASA {\it imfit} task while obtaining the core properties (see Section \ref{3mm_core_ident}).  } 
\label{3mmcont_res_map}
\end{figure}


\bsp	
\label{lastpage}
\end{document}